\documentclass[a4paper, 12pt, reqno]{amsart}
\usepackage{amssymb, amsmath,amsfonts,amsthm,euscript,mathrsfs,  MnSymbol,verbatim, enumerate,multirow,bbding, slashed, multicol,
color,array,tikz, tikz-cd,tikz-3dplot, ytableau}
\usetikzlibrary{arrows,positioning,decorations.pathmorphing,
  decorations.markings, matrix, decorations.text} 
\usepackage[all]{xy}
\usepackage[english]{babel}
\usepackage{epsf}
\usepackage{graphicx}
\csname @addtoreset\endcsname{equation}{section}
\newcolumntype{C}{>{\centering\arraybackslash$}p{\linewidth}<{$}}
\newtheorem{theorem}{Theorem}[section]

\theoremstyle{definition}

\theoremstyle{remark}

\DeclareFontFamily{U}{rsf}{} \DeclareFontShape{U}{rsf}{m}{n}{ <5> <6> rsfs5 <7> <8> <9> rsfs7 <10-> rsfs10}{}
\DeclareMathAlphabet\Scr{U}{rsf}{m}{n}
\newcommand{\bIVeleven}[1]{
\ytableausetup{nosmalltableaux, boxsize=#1 em}
\begin{ytableau}
*(cyan!25)  & *(blue) & *(blue) &  *(blue) &  *(blue) \\
 \none & *(yellow!25) & *(yellow) &  *(yellow)  &   *(yellow)\\
 \none &    \none & *(yellow!25) &  *(yellow)&*(yellow) \\
  \none &  \none &  \none &  *(yellow!25) &*(yellow) \\
    \none &  \none &  \none &\none  &*(yellow!25) \\
\end{ytableau}
}
\newcommand{\bIIInine}[1]
{
\ytableausetup{nosmalltableaux, boxsize=#1em}
\begin{ytableau}
*(cyan!25)  & *(blue) & *(blue) &  *(blue) &  *(blue) \\
 \none & *(cyan!25) & *(blue) &  *(yellow)  &   *(yellow)\\
 \none &    \none & *(yellow!25) &  *(yellow)&*(yellow) \\
  \none &  \none &  \none &  *(yellow!25) &*(yellow) \\
    \none &  \none &  \none &\none  &*(yellow!25) \\
\end{ytableau}      
}
\newcommand{\bIIIeleven}[1]
{
\ytableausetup{nosmalltableaux, boxsize=#1em}
\begin{ytableau}
*(cyan!25)  & *(blue) & *(blue) &  *(blue) &  *(blue) \\
 \none & *(cyan!25) & *(yellow) &  *(yellow)  &   *(yellow)\\
 \none &    \none & *(yellow!25) &  *(yellow)&*(yellow) \\
  \none &  \none &  \none &  *(yellow!25) &*(yellow) \\
    \none &  \none &  \none &\none  &*(yellow!25) \\
\end{ytableau}   
}
\newcommand{\bIInine}[1]
{
\ytableausetup{nosmalltableaux, boxsize=#1em}
\begin{ytableau}
*(cyan!25)  & *(blue) & *(blue) &  *(blue) &  *(blue) \\
 \none & *(cyan!25) & *(blue) &  *(yellow)  &   *(yellow)\\
 \none &    \none & *(cyan!25) &  *(yellow)&*(yellow) \\
  \none &  \none &  \none &  *(yellow!25) &*(yellow) \\
    \none &  \none &  \none &\none  &*(yellow!25) \\
\end{ytableau}      
}
\newcommand{\bIIten}[1]
{
\ytableausetup{nosmalltableaux, boxsize=#1em}
\begin{ytableau}
*(cyan!25)  & *(blue) & *(blue) &  *(blue) &  *(yellow) \\
 \none & *(cyan!25) & *(blue) &  *(yellow)  &   *(yellow)\\
 \none &    \none & *(cyan!25) &  *(yellow)&*(yellow) \\
  \none &  \none &  \none &  *(yellow!25) &*(yellow) \\
    \none &  \none &  \none &\none  &*(yellow!25) \\
\end{ytableau}      
}
\newcommand{\bIIthirteen}[1]
{
\ytableausetup{nosmalltableaux, boxsize=#1em}
\begin{ytableau}
*(cyan!25)  & *(blue) & *(blue) &  *(yellow) &  *(yellow) \\
 \none & *(cyan!25) & *(blue) &  *(yellow)  &   *(yellow)\\
 \none &    \none & *(cyan!25) &  *(yellow)&*(yellow) \\
  \none &  \none &  \none &  *(yellow!25) &*(yellow) \\
    \none &  \none &  \none &\none  &*(yellow!25) \\
\end{ytableau}
}
\newcommand{\bIIIfour}[1]
{
\ytableausetup{nosmalltableaux, boxsize=#1em}
\begin{ytableau}
*(cyan!25)  & *(blue) & *(blue) &  *(blue) &  *(blue) \\
 \none & *(cyan!25) & *(blue) &  *(blue)  &   *(blue)\\
 \none &    \none & *(yellow!25) &  *(yellow)&*(yellow) \\
  \none &  \none &  \none &  *(yellow!25) &*(yellow) \\
    \none &  \none &  \none &\none  &*(yellow!25) \\
\end{ytableau}   
}
\newcommand{\bIIIseven}[1]
{
\ytableausetup{nosmalltableaux, boxsize=#1em}
\begin{ytableau}
*(cyan!25)  & *(blue) & *(blue) &  *(blue) &  *(blue) \\
 \none & *(cyan!25) & *(blue) &  *(blue)  &   *(yellow)\\
 \none &    \none & *(yellow!25) &  *(yellow)&*(yellow) \\
  \none &  \none &  \none &  *(yellow!25) &*(yellow) \\
    \none &  \none &  \none &\none  &*(yellow!25) \\
\end{ytableau}     
}
\newcommand{\bIIIeight}[1]
{
\ytableausetup{nosmalltableaux, boxsize=#1em}
\begin{ytableau}
*(cyan!25)  & *(blue) & *(blue) &  *(blue) &  *(yellow) \\
 \none & *(cyan!25) & *(blue) &  *(blue)  &   *(yellow)\\
 \none &    \none & *(yellow!25) &  *(yellow)&*(yellow) \\
  \none &  \none &  \none &  *(yellow!25) &*(yellow) \\
    \none &  \none &  \none &\none  &*(yellow!25) \\
\end{ytableau}   
}\newcommand{\bIIsix}[1]
{
\ytableausetup{nosmalltableaux, boxsize=#1em}
\begin{ytableau}
*(cyan!25)  & *(blue) & *(blue) &  *(blue) &  *(yellow) \\
 \none & *(cyan!25) & *(blue) &  *(blue)  &   *(yellow)\\
 \none &    \none & *(cyan!25) &  *(blue)&*(yellow) \\
  \none &  \none &  \none &  *(yellow!25) &*(yellow) \\
    \none &  \none &  \none &\none  &*(yellow!25) \\
\end{ytableau}     

}\newcommand{\bIIeight}[1]
{
\ytableausetup{nosmalltableaux, boxsize=#1em}
\begin{ytableau}
*(cyan!25)  & *(blue) & *(blue) &  *(blue) &  *(yellow) \\
 \none & *(cyan!25) & *(blue) &  *(blue)  &   *(yellow)\\
 \none &    \none & *(cyan!25) &  *(yellow)&*(yellow) \\
  \none &  \none &  \none &  *(yellow!25) &*(yellow) \\
    \none &  \none &  \none &\none  &*(yellow!25) \\
\end{ytableau}   
}\newcommand{\bIsix}[1]
{
\ytableausetup{nosmalltableaux, boxsize=#1em}
\begin{ytableau}
*(cyan!25)  & *(blue) & *(blue) &  *(blue) &  *(yellow) \\
 \none & *(cyan!25) & *(blue) &  *(blue)  &   *(yellow)\\
 \none &    \none & *(cyan!25) &  *(blue)&*(yellow) \\
  \none &  \none &  \none &  *(cyan!25) &*(yellow) \\
    \none &  \none &  \none &\none  &*(yellow!25) \\
\end{ytableau}  
}
  \usepackage[pdftex]{hyperref}
\begin{document}

\begin{titlepage}
~\\
\begin{center}
\baselineskip=14pt{\LARGE
Singularities and Gauge Theory Phases II  \\
}
\vspace{2 cm}
{\large  Mboyo Esole,$^{\spadesuit,\heartsuit}$  Shu-Heng Shao,$^{\heartsuit}$  and Shing-Tung Yau$^{\spadesuit,\clubsuit}$ } \\
\vspace{1 cm}
${}^\spadesuit$Department of Mathematics, \ Harvard University, Cambridge, MA 02138, U.S.A.\\
${}^\heartsuit$Jefferson Physical Laboratory, Harvard University, Cambridge, MA 02138, U.S.A.\\
${}^\clubsuit$Taida Institute for Mathematical Science, National Taiwan University, 
 Taipei, 
Taiwan.
\end{center}

\vfill
\begin{center}

{\bf Absract}
\vspace{.3 cm}
\end{center}

{\small
We present  a simple algebraic construction for all the small resolutions of the  $SU(5)$ Weierstrass model. Each resolution corresponds to a subchamber  
on the Coulomb branch of the five-dimensional $\mathcal{N}=1$ $SU(5)$ gauge theory 
with matter fields in the fundamental and two-index antisymmetric representations. 
This construction unifies all previous resolutions found in the literature in a single framework.
\vfill

Email:{\tt    \{esole , yau\} at  math.harvard.edu, shshao  at  physics.harvard.edu
}
}

\end{titlepage}
\addtocounter{page}{1}
 \tableofcontents{}

\parskip=12 pt 
\section{Introduction}

In this paper we describe a correspondence between two mathematical objects that are of great physical interest in the string/M-theory context. On the one hand, given a Lie algebra $\frak{g}$ and a  (not necessarily irreducible) representation $\mathbf{R}$, we consider the partitioning of the fundamental Weyl chamber of $\frak{g}$ by certain codimension one interior walls that we will define later. The Weyl chamber has the interpretation as the Coulomb branch of certain supersymmetric quantum field theories, and the interior walls are the Higgs branch roots where the Coulomb branch and the mixed branch intersect. 
On the other hand, 
we consider the  network of small resolutions for the  Weierstrass model of type $\frak{g}$ over a base $B$ of complex dimension two or three. The partitioning of the fundamental Weyl chamber has an exact correspondence with the network of resolutions. 
This correspondence can be most easily motivated from M-theory compactifications when the total space is an elliptic Calabi-Yau threefold, but the correspondence is much more general.

\subsection{Coulomb branches in M-theory compactifications}

M-theory compactified on a Calabi-Yau threefold leads to a five-dimensional supergravity theory with eight supercharges coupled to 
  vector multiplets and hypermultiplets \cite{CCDF,FKM, AFT, Wi}. 
We focus on the five-dimensional supersymmetric quantum field theories with eight supercharges discussed in \cite{MS,IMS}.

 The Coulomb branch is the vacuum moduli space where the gauge symmetry algebra $\frak{g}$ is completely  broken to its Cartan subalgebra $\frak{h}$ by the vacuum expectation value (vev) of the real adjoint scalar $\phi$ in the vector multiplet. 
 After modding out the Weyl group of $\frak{g}$,  we can take the real scalar field $\phi$ to be in the fundamental Weyl chamber of the Cartan subalgebra. Namely, we have $(\phi, \alpha_i)\ge0$ for all the simple roots $\alpha_i$. 
 
The hypermultiplet transforms in a given representation $\mathbf{R}$ of the gauge symmetry algebra $\frak{g}$. At a generic point on the Coulomb branch, the hypermultiplet fields are all massive due to the nonzero vevs of the vector multiplet scalars $\phi$. However, over some special codimension one loci on the Coulomb branch, some hypermultiplet fields become massless and one can turn on their vevs to go to the mixed Coulomb-Higgs branches. 
An  hypermultiplet field with weight $w_i$ in the representation $\mathbf{R}$ becomes massless over the codimension one  \textit{interior walls} defined by 
$$(\phi, w_i)=0.$$ We use the word ``interior" to remind the readers that these walls are in the bulk of the Coulomb branch where the hypermulitplet fields become massless, as opposed to the boundaries of the Coulomb branch where $W$-bosons become massless. These codimension one interior walls are sometimes called the Higgs branch roots on the Coulomb branch in the literature.

Given a gauge algebra $\frak{g}$ and a representation $\mathbf{R}$, the codimension one interior walls partition the fundamental Weyl chamber into several different subchambers.  Each subchamber of the Coulomb branch is called a {\it phase} of the Coulomb branch and is uniquely defined by a collection of signs of the scalar products $(\phi, w_i)$ for all the weights $w_i$ of the representation $\mathbf{R}$.

\subsection{Incidence geometries from Lie algebra representations}
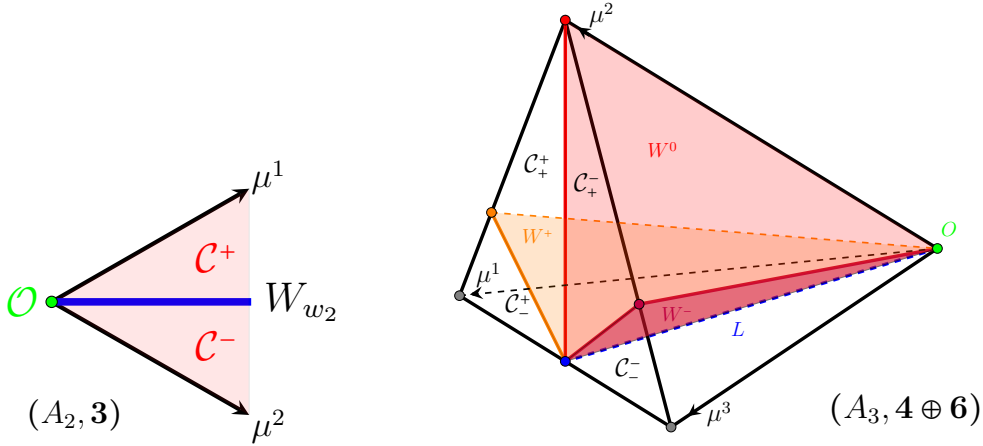
\begin{figure}

\begin{tabular}{ccc}

\raisebox{0cm}{
\begin{tikzpicture}[scale=1.5]
\coordinate (O) at (0,0);
\coordinate (A1)  at (30:2);
\coordinate (A2)  at (-30:2);
\draw[ultra thick, -stealth,black] (O)--(A1);
\draw[ultra thick, -stealth,black] (O)--(A2);
\draw[line width=1mm,blue] (O)--(0:1.75);
\draw[fill=red, opacity=.1] (O)--(A1)--(A2);
\draw[fill=green, opacity=1] (0,0) circle (.05);
\node at (-.2,0) {\color{green!} \scalebox{1.3}{$\mathcal{O}$~}};
\node at (15:1.5) {\color{red} \scalebox{1.3}{$\mathcal{C}^+$}};
\node at (-15:1.5) {\color{red} \scalebox{1.3}{$\mathcal{C}^-$}};
\node at (30:2.2) {\color{black} \scalebox{1.1}{$\mu^1$}  };
\node at (-30:2.2) {\color{black} \scalebox{1.1}{$\mu^2$}};
\node at (0:1.9) { \scalebox{1.3}{~~~~~$W_{w_2}$}};
\node at (.2,-1) {$(A_2, {\bf 3})$};
\end{tikzpicture}}

\phantom{~~~~~}

\scalebox{.8}{
\begin{tikzpicture}[scale=4]
\coordinate (O) at (0,0,0);
\coordinate   (P1) at (0.707107,0,0.707107);
\coordinate (P2) at (-0.707107,0,-0.707107);
\coordinate (P3) at (0 , 1.414,0);
\coordinate (w-) at (0.493561 ,0.427,0.493561);
\coordinate (w+) at (-0.493561 ,0.427,-0.493561 );

\coordinate (C) at (17:1.6);

\draw [black, ultra thick] (P1)--(P2)--(P3)--(P1);
\draw [orange,ultra thick](O)--(w+);
\draw [purple, ultra thick](O)--(w-);
\draw [red,ultra thick](O)--(P3);
\draw[blue,ultra thick,dashed] (C)--(O);
\draw [ultra thick,black] (P1)--(C);
\draw [purple, ultra thick](w-)--(C);
\draw [orange,  thick,dashed] (w+)--(C);
\draw [black,ultra thick] (P3)--(C);
\draw[dashed, black, thick] (P2)--(C);
\draw[fill=orange, opacity=.2] (w+)--(O)--(C);
\draw[fill=red, opacity=.2] (O)--(P3)--(C);
\draw[fill=purple, opacity=.4] (w-)--(O)--(C);

 \draw[ultra thick,-stealth] (.1 , 1.351)--(.05, 1.384);
   \coordinate[label={\color{black} \scalebox{1}{$\mu^2$}}] (MU2) at ( 0.15 ,1.35);

    \draw[ultra thick,-stealth] (.738, 0,0.53)--(.730, 0,0.59);
   \coordinate[label={\color{black} \scalebox{1}{$\mu^3$}}] (MU3) at ( 0.937 ,0,0.787);

   \draw[ultra thick,-stealth] (-.66 , 0,   -0.724)--(-.67, 0,   -0.72);
   \coordinate[label={\color{black} \scalebox{1}{$\mu^1$}}] (MU3) at ( -0.587 ,0,-0.707);

\coordinate[label={\color{blue} \scalebox{.8}{}}] (ell) at ( 270 : .15);
\draw[fill=blue, opacity=1] (0 ,0) circle (.02);

\coordinate[label={\color{purple} \scalebox{.8}{}}] (p-) at ( 0.43 ,0.38,0.493561);
\draw[fill=purple, opacity=1] (0.493561 ,0.427,0.493561)
 circle (.02);
 
\coordinate[label={\color{orange} \scalebox{.8}{}}] (p+) at (-0.553561 ,0.427,-0.493561 );
\draw[fill=orange, opacity=1] (-0.493561 ,0.427,-0.493561 )
 circle (.02);

\coordinate[label={\color{red} \scalebox{.8}{}}] (p0) at (0 , 1.44,0);
\draw[fill=red, opacity=1] (0 ,1.414,0)
 circle (.02);

 \coordinate[label={\color{green} \scalebox{.8}{$O$}}]  (c) at (17:1.65);
\draw[fill=green, opacity=1] (17:1.6)
 circle (.02);

 \coordinate[label={\color{purple} \scalebox{.8}{$W^-$}}] (W-) at ( 0.65 ,0.32,0.493561);
 
 \coordinate[label={\color{orange} \scalebox{.8}{$W^+$}}] (W+) at (-0.303561 ,0.265,-0.493561 );
 
 \coordinate[label={\color{red} \scalebox{.8}{$W^0$}}] (W0) at (0.4 , .81,0);

  \coordinate[label={\color{blue} \scalebox{.8}{$L$}}] (L) at ( 0.9 ,0.26,0.493561);

 \draw[fill=gray, opacity=1] (0.707107,0,0.707107)
 circle (.02);
  \coordinate[label={\color{gray} \scalebox{.8}{}}] (ell-) at ( 0.757 ,0,   .99);

 \draw[fill=gray, opacity=1] (-0.707107,0, -0.707107)
 circle (.02);
  \coordinate[label={\color{gray} \scalebox{.8}{}}] (ell+) at ( -0.705 ,0,   -.5);

     \coordinate[label={\color{black} \scalebox{1}{$\mathcal{C}^+_-$}}] (C+-) at ( -0.337 ,0,-0.377);
     
      \coordinate[label={\color{black} \scalebox{1}{$\mathcal{C}^+_+$}}] (C++) at (-0.303561 ,0.535,-0.493561 );
      
        \coordinate[label={\color{black} \scalebox{1}{$\mathcal{C}^-_+$}}] (C-+) at (-0.1 ,0.455,-0.493561 );

        \coordinate[label={\color{black} \scalebox{1}{$\mathcal{C}^-_-$}}] (C--) at ( 0.45 ,0.07,0.493561);
 \node at (1.4,-.2) { \scalebox{1.3}{$(A_3, {\bf 4}\oplus{\bf 6})$}};
\end{tikzpicture}
}

\end{tabular}

\caption{Left: The $(A_2,{\bf3})$ incidence geometry \cite{ESY}, or equivalently, the Coulomb branch for $SU(3)$ gauge theory with matter in the $\bf3$ representation. The Weyl chamber is spanned  by the two vectors $\mu^1$ and $\mu^2$, and is divided by the interior wall $W_{w_2}$ into two subchambers $\mathcal{C}^\pm$. The interior wall $W_{w_2}$ is the Higgs branch root where  matter fields become massless. The two boundary walls are the lines generated by $\mu^1$ and $\mu^2$ where the $W$-bosons become massless. Right: The $(A_3,{\bf 4\oplus\bf6})$ incidence geometry \cite{ESY}. The Weyl chamber is the three-dimensional cone spanned  by the vectors $\mu^1$, $\mu^2$, $\mu^3$. The three interior walls are $W^+$, $W^0$, $W^-$ where some matter fields become massless. The Coulomb branch is partitioned into four subchambers $\mathcal{C}^\pm_\pm$ by the three interior walls, which further intersect at the line $L$ lying at the bottom of the Weyl chamber. The three boundary walls are spanned by any pair of the three $\mu^i$'s.  }\label{SU34incidence}

\end{figure}

 The mathematics behind the determination of the number of phases of the Coulomb branch can be expressed as an enumerative problem for the  incidence geometry defined by a Lie algebra $\frak{g}$ and a  representation $\mathbf{R}$ of $\frak{g}$.  Let  $\frak{h}$ be the  Cartan subalgebra  of $\frak g$.  We recall that roots of $\frak{g}$ and weights of $\mathbf{R}$ are elements of $\frak{h}^\star$ and each  element  of  $\frak h^\star$ defines through its kernel a hyperplane in $\frak{h}$. 
 We call the hyperplanes defined by the weights of the representation $\mathbf{R}$  the {\it interior walls} and the hyperplanes defined by the roots of $\frak{g}$  the  {\it boundary walls}. 
The interior walls further intersect at some codimension two loci, and so on. The collection of interior and boundary walls and their successive  intersections give the Weyl chamber   an incidence structure. We will denote this incidence geometry by 
$$(\frak{g},\mathbf{R}).$$ For example, the incidence geometry $(A_2,{\bf3})$ and $(A_3, {\bf4\oplus\bf 6})$ studied in \cite{ESY} for the $SU(3)$ and $SU(4)$ model are shown in Figure \ref{SU34incidence}.

Let $n_d$ be the number of $d$-dimensional loci from the intersections of interior walls. We define a  polynomial
$$P_{\frak{g} , \mathbf{R}}(t)=n_0+n_1t+n_2 t^2+\cdots +n_r t^r.$$
We always have $n_0=1$ since all the interior walls   intersect at a unique point: the origin of the Weyl chamber. The integer $n_r$ is the number of subchambers partitioned by the interior walls from the representation $\mathbf{R}$ and $n_{r-1}$ is the number of interior walls.

In the two example shown in Figure \ref{SU34incidence}, we have two chambers $\mathcal{C}^\pm$ separated by one interior wall $W_{w_2}$ for  $(A_2,{\bf 3})$. The resulting generating polynomial is  $P_{A_2, \bf3}(t)=1+t + 2t^2$. In the $(A_3,{\bf4\oplus \bf6})$ example, we have four subchambers $\mathcal{C}^\pm_\pm$, $n_3=4$, three interior walls $W^\pm,W^0$, $n_2=3$, one line $L$, $n_1=1$, and one point $O$, $n_0=1$ giving $P_{A_3, \bf4\oplus \bf6}(t)=1+t+3 t^2+4t^3$.

\subsection{Coulomb branches and resolutions of elliptic fibrations} 
The correspondence we will describe in the current paper is that for certain choices of $\frak{g}$ and $\mathbf{R}$, the incidence geometry  $(\frak{g}, \mathbf{R})$ can be obtained from an  elliptic fibration with singular fibers in codimension one that collide in codimension two.

The types of singularities of an elliptic fibration over a codimension one locus are classified by Kodaira \cite{Kodaira}.  
Given a base $B$, the elliptic fibration with a given singular fiber over a codimension one locus $D$ of $B$  can be engineered systematically using the Weierstrass model with coefficients having appropriate vanishing orders over $D$. 
This is a direct consequence of the Tate's algorithm \cite{Tate} and for that reason  such Weierstrass models are said to be in  the {\it Tate forms} \cite{KMSNS}. Tate forms  were  first introduced in the physics literature in the context of F-theory  \cite{Vaf,MV1, MV2,BIKMSV}.  For recent works on elliptic Calabi-Yau threefolds see \cite{Ca,DKW,JT,Cat}. Many interesting questions on elliptic fibrations can also be discussed using different starting points than a Weierstrass model 
\cite{BKMT,AE2,EFY,MP} or considering the weak coupling limit \cite{Sen,CDE,CDW,AE1,AE2,ES}. 

A Weierstrass model in the Tate form is labeled by a choice of a Kodaira fiber and from that input a  Lie algebra $\frak{g}$ and a representation $\mathbf{R}$ can be uniquely deduced \cite{BIKMSV,GM}. 
The representation is due to the presence of singularities in  codimension two and can be understood in string theory from the picture of branes intersecting at angles: gauge fields live on the brane  (codimension one singularities) while matter fields are   localized  at the intersection of these branes. In the simplest case, the representation can be determined by the  Katz-Vafa method \cite{KV}. More generally, the representation can be determined by studying the intersection numbers 
between curves in the fiber after resolving the singularities \cite{MN,MT}.
Geometrically, these matter fields come from the collisions  of codimension one  singular fibers \cite{Mir1,AG,BJ} and the dictionary between collisions of singularities and the assignment of representations have been shown to be compatible with anomaly cancellation \cite{BJ,GM}. 

For the $SU(N)$ Weierstrass model {in the Tate form} I$_N^s$ with \textit{general} coefficients $a_{i,j}$, the codimension two rank-one fiber enhancements are\footnote{The type I$_2^s$ is special. The fiber enhancements are I$_2^s\rightarrow$I$_3^s$ and I$_2^s\rightarrow$III. Hence on the representation theory side we only get the fundamental representation $\bf2$.}
\begin{equation}
\begin{split}
&A_{N-1}\rightarrow  A_N,\\
&A_{N-1}\rightarrow D_N,
\end{split}
\end{equation}
which gives rise to the fundamental (\,\scalebox{.4}{\ydiagram{1}}\,) and the two-index antisymmetric representations (\scalebox{.4}{\raisebox{3pt}{\,\ydiagram{1,1}\,}}). This will be the case we focus on.

Given the Lie algebra $\frak{g}$ and the representation $\mathbf{R}$ arising from the rank-one enhancements described above, we consider all the small resolutions of the Weierstrass model in the Tate form  corresponding to the gauge algebra  $\frak{g}$. This network of small resolutions fits in nicely into the incidence geometry $(\frak{g},\mathbf{R})$. In particular, the number of subchambers $n_r$ is equal to the number of different small resolutions. This correspondence was recently discussed in \cite{HLN,ESY,HLM, CGKP}. The deformation side of the story was recently discussed in \cite{GHS1,GHS2}.

From the M-theory compactification point of view, this correspondence comes as no surprise. At each step of the blow up, one introduces an ambient projective space whose size is part of the K\"ahler moduli of the internal Calabi-Yau manifold. In M-theory compactification, the real scalars in the vector multiplet parametrize the K\"ahler moduli space of the internal Calabi-Yau manifold \cite{CCDF}. Therefore, the Coulomb branch, which is mathematically described by the incidence geometry $(\frak{g},\mathbf{R})$, should match with the network of blow ups. In particular, a sequence of blow ups on the geometry side should correspond to a trajectory on the Coulomb branch from the origin to one of the subchambers. Thus, the number of distinct resolutions should correspond to number of subchambers on the Coulomb branch.

In a previous paper \cite{ESY}, we have studied explicitly this correspondence for  Weierstrass models of type   I$_2^s$, I$_3^s$, and I$^s_4$.   The corresponding gauge theories have $SU(2)$ gauge symmetry with fundamental representation $\bf2$, $SU(3)$ gauge symmetry with matter in the fundamental representation $\bf3$ \footnote{Note that the two-index antisymmetric representation for $SU(3)$ is the same as the anti-fundamental representation, which defines the same interior walls as the fundamental representation. Therefore it suffices to consider the fundamental representation only.}, and $SU(4)$ gauge symmetry with matter in both the fundamental $\bf4$ and the antisymmetric representations $\bf6$.

In the current paper we consider the phenomenologically interesting $SU(5)$ model, namely, the Weierstrass model of the type I$_5^s$, with  fundamental representation $\bf5$ and the antisymmetric representation $\bf10$ from the rank-one enhancements.   
The network of small resolutions is much more complicated and richer than the lower rank cases.
From the two representations we obtain nine interior walls that  define  twelve subchambers in the Weyl chamber.  
The adjacency of subchambers of the incidence geometry $(A_4, \mathbf{5}\oplus\mathbf{10})$  is represented in Figure \ref{SU5Coulomb1}, which was first obtained in \cite{HLN}. The twelve subchambers are divided by nine interior walls, which intersect at nine planes. The nine planes further intersect  along four lines, which all intersect at a  the origin of the Weyl chamber. We summarize the incidence geometries in the following table:
$$
\begin{tabular}{|c|c|c|l|}
\hline
Weierstrass model & $\frak{g}$ & $\mathbf{R}$ & Polynomial $ P_{\frak{g}, {\mathbf{R}}} (t)$\\
\hline 
I$_2^s$ & $A_1$ & $\mathbf{2}$ & $1+t$\\
\hline 
I$_3^s$ & $A_2$ & $\mathbf{3}$& $1+t+2 t^2$\\
\hline 
I$_4^s$ & $A_3$ & $\mathbf{4}\oplus\mathbf{6}$ & $1+t+3t^2+4 t^3$\\
\hline
I$_5^s$ & $A_4$ & $\mathbf{5}\oplus\mathbf{10}$ & $1+4t+9t^2+9t^3+12t^4$\\
\hline
\end{tabular}
$$
 The connection between the Weyl groups with algebraic varieties and their birational transformations has been discussed in \cite{Mat1}. See also \cite{Nik,Slo,KM}.  
 
\subsection{Small resolutions of the $SU(5)$ model}
The $SU(5)$ model is usually defined by small resolutions of a singular Weierstrass model \cite{EY}. From a purely mathematical point of view, the $SU(5)$ model is interesting because it describes an elliptic fibration with a rich structure of fiber enhancements and flop transitions. 
 The first example of explicit algebraic  crepant resolutions was constructed in \cite{EY} where six different small resolutions connected by flop transitions were obtained.

Soon after \cite{EY}, another approach was introduced in \cite{KMW} to describe the resolutions of the $SU(5)$ model. The authors use a toric description following  \cite{BIKMSV}. The resolutions obtained this way are  usually called ``toric resolutions" of type I, II, and III, labeled by their Stanley-Reisner ideal. More recently, a new type of blow up were introduced to discuss the properties of the $SU(5)$ model \cite{HLN}. As a by-product of our analysis, we will clarify how these different resolutions are related to each other in a unified fashion. Other resolutions of singularities of elliptic fibrations are studied in \cite{LN, MCPRT,TW}.

\subsection{Summary of results}

We give a simple construction for all the small resolutions of the $SU(5)$ model. Our main result is the network of resolutions in Figure \ref{network}, which gives eighteen  resolutions, and the weighted blow ups in Section \ref{section:toric5} for the other two resolutions $\mathscr{B}_{2,3}^1, \mathscr{B}_{3,2}^1$. 
After identifying isomorphic resolutions, there are in total twelve resolutions shown in Figure \ref{SU5Res}. The number of resolutions exactly matches with the number of subchambers $n_4$ of the incidence geometry $(A_4, {\bf5}\oplus {\bf 10})$. In the physics language, the twelve resolutions correspond to the twelve subchambers on the Coulomb branch of the $SU(5)$ gauge theory with fundamental and two-index antisymmetric representations.

As a by-product, we obtained the codimension three fibers for the twelve resolutions in Table \ref{Table.Fo12}, \ref{type123}, \ref{Table.Fplus2minus}, \ref{Table.B132}. The codimension three fibers for the new resolutions $\mathscr{B}^1_{1,3}$ and $\mathscr{B}^2_{1,3}$ match with those of the phases (6.I) and (6.II) from the box graphs \cite{HLM}.

Flops between different resolutions are manifest from the ramifications of branches in the network of resolutions in Figure \ref{network}. At each step of blow ups, different choices of the blow up center  result in different branches of the network of resolutions.  Partial resolutions, corresponding to the branching points in the network of resolutions, has  conifold-like singularities fibered over certain subvarieties of the base.

We also clarify some confusions of the $SU(5)$ resolutions in the literature. 
We show that the  toric models of type  I and  II can not only be realized as sequences of blow ups, but are already obtained in the  resolutions of  \cite{EY}. More precisely, the toric model of type I is isomorphic to $\mathscr{B}_{1,3}$ and the toric model of type II is isomorphic to 
$\mathscr{B}_{2,3}$. As to the type III model, it can be obtained by a sequence of weighted blow ups as in Section \ref{section:toric5}.

\subsection{Organization of the paper}

\begin{figure}[htb]
\begin{center}
\includegraphics[scale=.8]{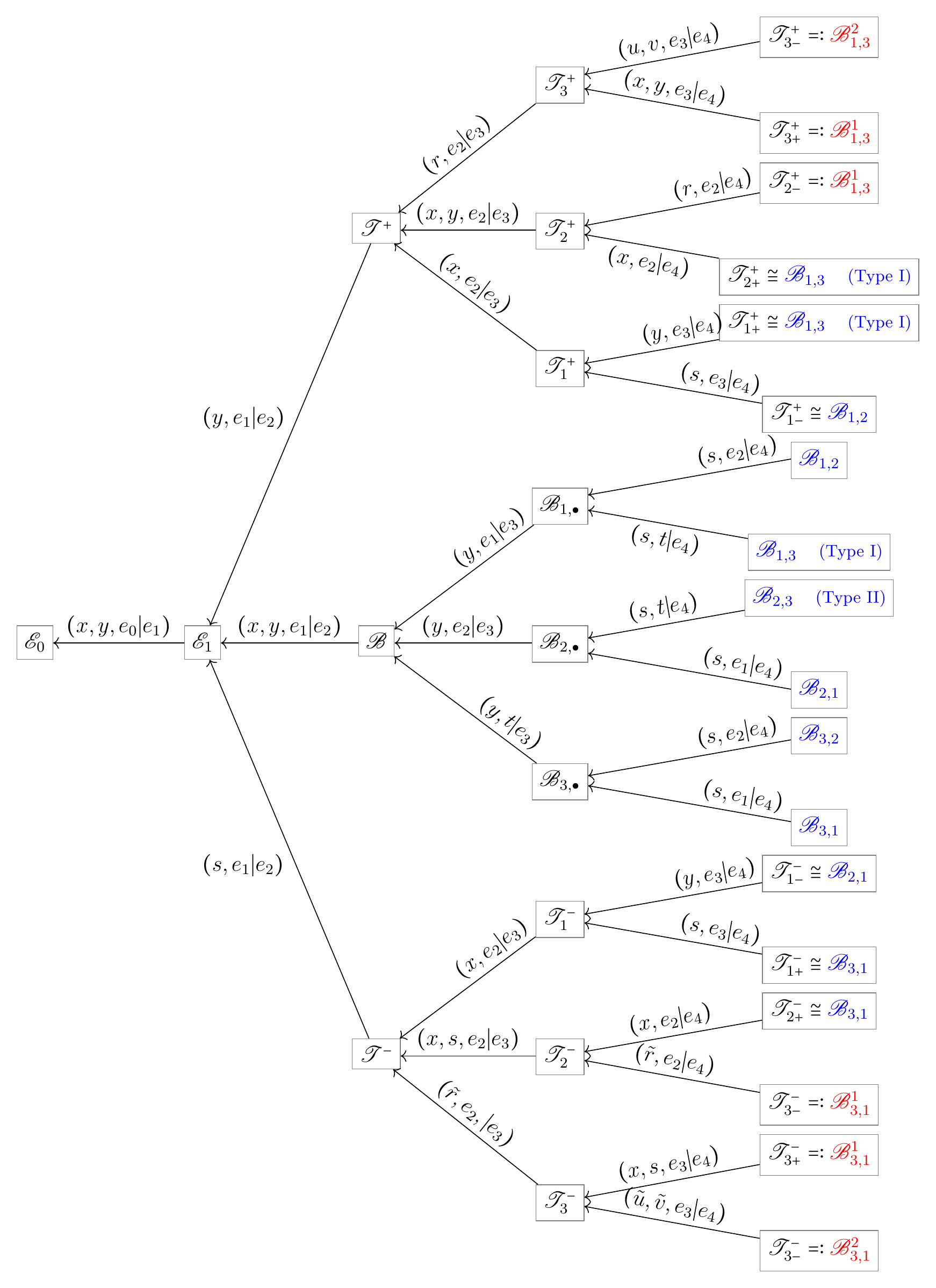}
\end{center}
\caption{The network of resolutions of the $SU(5)$ model.    
Resolutions found in \cite{EY} are in blue while new resolutions are in red. Some of the resolutions are isomorphic to each other and are therefore denoted by the same name. For example, it can be shown that $\mathscr{T}^+_{3+}\cong \mathscr{T}^+_{2-}$ and will therefore both be denoted by $\mathscr{B}_{1,3}^1$.  In the above network we have not exhausted all the possible partial resolutions,  but it is already sufficient to obtain ten out of the twelve resolutions. There are two more resolution $\mathscr{B}_{2,3}^1,\mathscr{B}_{3,2}^1$ that can be obtained by weighted blow ups described in Section \ref{section:toric5}. The variables $s,t,r,u,v$  are defined in \eqref{st}, \eqref{r}, \eqref{u}, \eqref{v}. 
The variables $\tilde{r}, \tilde{u},\tilde{v}$ are obtained from $r,u,v$ by replacing $y$ with $-s$, {\it i.e.} their Mordell-Weil duals. 
}\label{network}
\end{figure}

We summarize the main result in the network of resolutions in Figure \ref{network}. In Section \ref{section:coulomb}, we describe the Coulomb branch of an $SU(5)$ gauge theory with matters in the representations $\bf5$ and $\bf10$. The mathematical description is the incidence geometry $(A_4, {\bf5\oplus\bf10})$. In Section \ref{section:su5}, we define the $SU(5)$ model and present the first two blow ups. In Section \ref{EYdescription}, we give explicit constructions for $\mathscr{B}_{i,j}$ in terms of sequences of blow ups. In Section \ref{section:toric}, we construct sequences of (weighted) blow ups for the three toric resolutions. In Section \ref{section:T23}, we explore the remaining resolutions from the partial resolutions $\mathscr{T}^+_2$ and $\mathscr{T}^+_3$. In Section \ref{section:iso}, we briefly summarize all the isomorphism between resolutions.

\textbf{Note added}
After this paper appeared on arXiv, a related paper \cite{BSN} by Braun and Sch\"afer-Nameki  came out. The authors discussed  resolutions of the $SU(5)$ model using weighted blow ups in the special case of singular Calabi-Yau hypersurfaces in compact toric varieties.  In particular,  they reproduced the fan diagrams presented in Section \ref{section:toric4} of the current paper.

\clearpage

 \section{The Coulomb Branch of the $SU(5)$ Model}\label{section:coulomb}

\subsection{Coulomb branches and incidence geometries}

For definiteness we will consider the Coulomb branch of the five-dimensional $\mathcal{N}=1$ supersymmetric gauge theory, while a similar structure can also be found in the classical Coulomb branch of the three-dimensional $\mathcal{N}=2$ supersymmetric gauge theory.

In this section we consider the incidence structure of the Coulomb branch of the five-dimensional $\mathcal{N}=1$ supersymmetric quantum field theory with gauge symmetry $\frak{g}$ of rank $r$ and hypermultiplets in the representation $\mathbf{R}$. The precise multiplicity, as long as nonzero, for each hypermultiplet  will not affect our analysis. The mathematical description of the Coulomb branch is the Weyl chamber of $\frak{g}$ partitioned by several codimension one loci that we will introduce in a moment.

Let $\alpha_i$, $i=1,\cdots,r$, be the simple roots of $\frak{g}$ and $\mu_i$ be the dual basis of the simple roots $\alpha_i$ in the Cartan subalgebra $\frak{h}$:
\begin{align}
(\mu_i, \alpha_j ) = \delta_{ij}.
\end{align}
Let $\phi$ be a vector in the Weyl chamber. Physically it is the real scalar field in the vector multiplet. We can expand $\phi$ in the Weyl chamber as
\begin{align}
\phi = \phi_i \, \mu_i,~~~~\phi_i \ge 0.
\end{align}
The Weyl chamber is then parametrized by $(\phi_1, \cdots,\phi_r)\in \mathbb{R}_{\ge0}^r$.

\subsubsection{Boundary and interior walls}

The \textit{boundary wall} on the Coulomb branch labeled by the simple root $\alpha_i$  is defined by the loci where
\begin{align}
\phi_i = (\phi, \alpha_i) =0
\end{align}
for $i=1,\cdots,r$. Physically, this is the codimension one loci
where the $W$-boson becomes massless and the gauge symmetry is enhanced. 

 On the other hand, the \textit{interior wall} on the Coulomb branch labeled by the weight $w$ is defined by
 \begin{align}
 (\phi , w)=0.
 \end{align}
 Physically, this is the codimension one loci where the hypermultiplet field with weight $w$ becomes massless. In the physics literature, the interior walls are called the Higgs branch roots where one can turn on the vevs for the hypermultiplet fields to go to the mixed Coulomb-Higgs branch.

\subsubsection{Phases of the Coulomb branch}

The Coulomb branch is divided into several different \textit{subchambers} or \textit{phases} by the interior walls. 
Each subchamber is defined by a particular sign assignment for $(\phi , w_i)$ for all the weights $w_i$ in the representations $\mathbf{R}$. The boundary walls, on the other hand, are literally the boundary of the Coulomb branch so they do not divide the Coulomb branch. Moving on to higher codimensions, the intersections between the codimension one interior walls define several codimension two loci on the Coulomb branch \textit{etc}. The successive intersections between these singular loci in each codimension then define the incidence geometry $(\frak{g},
\mathbf{R})$ mentioned in the Introduction. 

One of the main goals of the current paper is to match the  Coulomb branch, described by the incidence geometry $(\frak{g},\mathbf{R})$, with the network of resolutions Figure \ref{network} in the case of the $SU(5)$ model.

\subsubsection{Choice of the representation $\mathbf{R}$}

As can be seen from the current paper or \cite{ESY}, for Weierstrass model of type $SU(N)$ \footnote{To be more specific, we consider the type I$_N^s$ Weierstrass model.} with \textit{general} coefficients $a_{i,j}$, the codimension two fiber enhancements are\footnote{For $N=2$ we do not have the latter enhancement to $SO(4)$. It follows that on the representation theory side we only get the fundamental representation $\bf2$ for the $SU(2)$ model.} either $SU(N)\rightarrow SU(N+1)$ or $SU(N)\rightarrow SO(2N)$. From the Katz-Vafa method \cite{KV}, the corresponding representations are the fundamental representation (\,\scalebox{.4}{\ydiagram{1}}\,) and the two-index antisymmetric representation (\scalebox{.4}{\raisebox{3pt}{\,\ydiagram{1,1}\,}}). Therefore, in the context of M-theory compactified on elliptic Calabi-Yau threefolds, the five-dimensional theories have $SU(N)$ gauge group with matters in the fundamental and two-index antisymmetric representation from the geometric data.  

Surprisingly, as shown in \cite{IMS} from a purely quantum field theory analysis, a UV complete $\mathcal{N}=1$ supersymmetric quantum field theory in five dimension with gauge group $SU(N)$ can only allow for matters in the fundamental and two-index antisymmetric representations, but no other representations! The supersymmetric quantum field theories and elliptic fibrations secretly put the same constraint on the possible representations from two completely different analyses.

\subsubsection{Quantum Coulomb branch}

The above description is completely classical and only relies on the representation theory. One might question whether the quantum Coulomb branch has a the same incidence structure as the classical one. In the five-dimensional $\mathcal{N}=1$ supersymmetric quantum field theory, the exact prepotential $\mathcal{F}(\phi)$ takes the following form \cite{IMS}
\begin{align}
\mathcal{F}(\phi) = {1\over 2} m_0 \text{Tr}(\phi^2) + {c_{cl}\over 6} \text{Tr}(\phi^3)+{1\over12}\left(
\sum_{\alpha:\text{roots} } | (\phi,\alpha )|^3 -\sum_{w:\text{weights}}|(\phi,w)|^3
\right).
\end{align}
Note that in particular the derivatives of the prepotential is not continuous at the boundary walls (where $(\phi,\alpha)=0$) and at the interior walls (where $(\phi,w)=0$). Therefore the quantum Coulomb branch also has singularities at the boundary and the interior walls. It follows that the incidence structure of the quantum Coulomb branch is still described by the incidence geometry $(\frak{g},\mathbf{R})$.

\subsection{The $SU(5)$ Coulomb Branch}

From the codimension two fiber enhancements of the $SU(5)$ model, the matters can be seen to be in the fundamental $\bf5$ and the two-index antisymmetric representations $\bf10$ by the Katz-Vafa method \cite{KV}    or by a direct computation of the weights from the resolved geometry \cite{MN}. The precise multiplicities for the hypermultiplets in each representation are not important for the Coulomb branch structure, as long as they are nonzero. The classical Coulomb branch of the $SU(5)$ gauge theory is given by the Weyl chamber of $A_4$. 

\begin{figure}[htb]
\scalebox{.8}{
\begin{tikzcd}[column sep=2.5cm, ampersand replacement=\&]
w_1^{\bf5}= \boxed{(1,0,0,0)}  \arrow[rightarrow]{d}{ \textstyle{-\alpha_1}}   \\
 w_2^{\bf 5}=\boxed{(-1,1,0,0) }\arrow[rightarrow]{d}{ \textstyle{-\alpha_2}} \\
 w_4^{\bf 5}=\boxed{(0,-1,1,0)} \arrow[rightarrow]{d}{ \textstyle{-\alpha_3}} \\
  w_5^{\bf 5}=\boxed{(0,0,-1,1)} \arrow[rightarrow]{d}{ \textstyle{-\alpha_4}} \\
  w_5^{\bf 5}=\boxed{(0,0,0,-1)}
  \end{tikzcd}}
  \quad 
  \scalebox{.8}{
  \begin{tikzcd}[column sep=.3cm, ampersand replacement=\&]
\&w_1^{\bf 10} =\boxed{(0,1,0,0)}\arrow[rightarrow]{d}{ \textstyle{-\alpha_2}} \& \\
\& w_2^{\bf 10}=\boxed{(1,-1,1,0)} \arrow[rightarrow]{dr}{ \textstyle{-\alpha_3}}  \arrow[rightarrow]{dl}[above]{ \textstyle{-\alpha_1}}\& \\
w_5^{\bf 10} =\boxed{(-1,0,1,0)} \arrow[rightarrow]{d}[left]{ \textstyle{-\alpha_3}}\& \&
 w_3^{\bf 10}=\boxed{(1,0,-1,1)}\arrow[rightarrow]{d}{ \textstyle{-\alpha_4}}  \arrow[rightarrow]{dll}{ \textstyle{-\alpha_1}}   \\
w_6^{\bf 10}=\boxed{(-1,1,-1,1)}\arrow[rightarrow]{d}[left]{ \textstyle{-\alpha_2}}  \arrow[rightarrow]{drr}{ \textstyle{-\alpha_4}} \& \&  
w_4^{\bf 10}=\boxed{(1,0,0,-1)}\arrow[rightarrow]{d}{ \textstyle{-\alpha_1}}\\
w_8^{\bf 10}=\boxed{(0,-1,0,1)} \arrow[rightarrow]{dr}[left]{ \textstyle{-\alpha_4}}\& \& 
w_7^{\bf 10}=\boxed{(-1,1,0,-1)} \arrow[rightarrow]{dl}{ \textstyle{-\alpha_2}}\\
\& w_9^{\bf 10}=\boxed{(0,-1,1,-1)} \arrow[rightarrow]{d}{ \textstyle{-\alpha_3}}\& \\
\& w_{10}^{\bf10}=\boxed{(0,0,-1,0)}\& 
  \end{tikzcd}}
  \caption{Weight diagrams for  the fundamental representation and antisymmetric representation of $A_4$. \label{HasseFA}}
\end{figure}
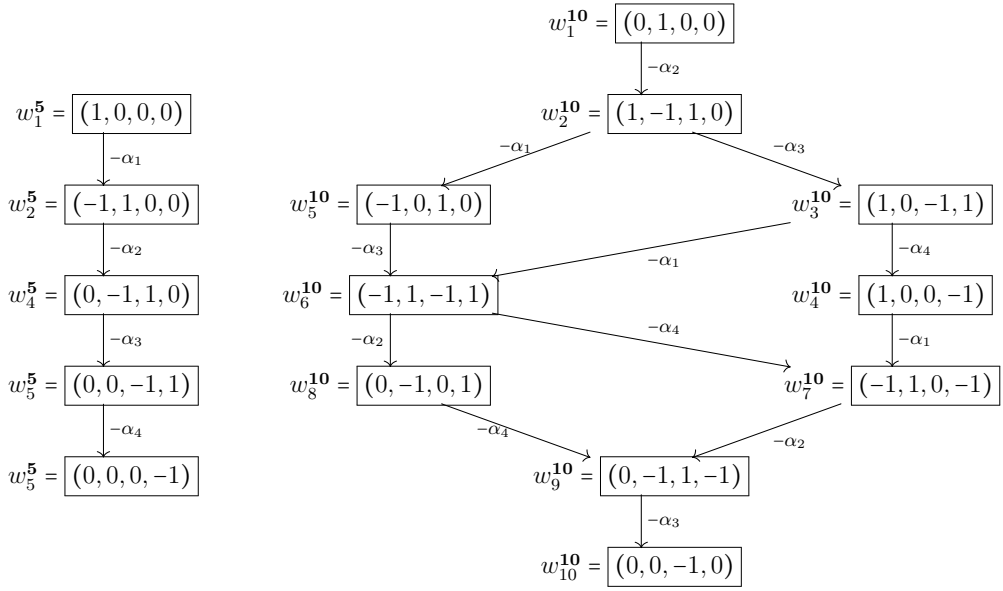

We will use the following conventions for the representation theory. The Dynkin labels for the simple roots of $A_4$ are given by:
\begin{equation}
\alpha_1=( 2 , -1 , 0 , 0), \quad 
\alpha_2=(-1, 2, -1, 0), \quad \alpha_3=(
 0 ,-1, 2 , -1)
,\quad \alpha_4=(
 0, 0,-1, 2)
\end{equation}
Let $w^5_i$ with $i=1,\cdots, 5$ be the weights in the fundamental representation of $A_4$. Due to the traceless condition of $A_4$, we have
\begin{align}
\sum_{i=1}^5 w_i^{\bf5}=0.
\end{align}
The simple roots can be expressed as 
\begin{equation}
\alpha_i=w^{\bf5}_i-w^{\bf5}_{i+1}, \quad i=1,2,3,4.
\end{equation}
The Cartan matrix is given by  the scalar products $(\alpha_i, \alpha_j)$: 
\begin{equation}
A_{ij}=\begin{pmatrix}
2& -1& 0 & 0 \\
-1 & 2 & -1& 0 \\
0& -1& 2& -1\\
0 & 0 & -1& 2
\end{pmatrix}
\end{equation}
where we have normalized $(\alpha_i,\alpha_i)=2$. 
The weights of the fundamental and antisymmetric representations of $A_4$ are given in Figure \ref{HasseFA}.

\begin{figure}[bht]
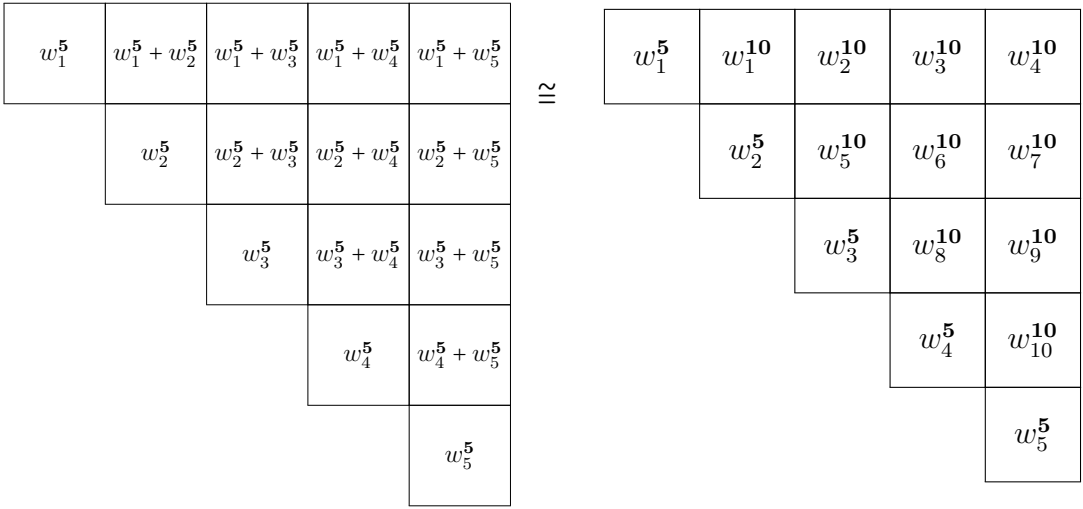

\begin{center}
\begin{tabular}{ccc}
\scalebox{.8}{\ytableausetup{smalltableaux, boxsize=4em}
\begin{ytableau}
w^{\bf5}_1 & w^{\bf5}_1+w^{\bf5}_2  &   w^{\bf5}_1+w^{\bf5}_3    &  w^{\bf5}_1+w^{\bf5}_4  &w^{\bf5}_1+w^{\bf5}_5  \\
 \none & w^{\bf5}_2   &    w^{\bf5}_2+w^{\bf5}_3   & w^{\bf5}_2+w^{\bf5}_4  &    w^{\bf5}_2+w^{\bf5}_5  \\
 \none  &   \none &  w^{\bf5}_3   &  w^{\bf5}_3+w^{\bf5}_4  &  w^{\bf5}_3+w^{\bf5}_5  \\
  \none &  \none &  \none  &w^{\bf5}_4 &  w^{\bf5}_4+w^{\bf5}_5 \\
  \none &  \none &  \none  &\none  & w^{\bf5}_5 \\
\end{ytableau}}
& \large
 $\cong$ & 
\scalebox{1}{
\ytableausetup{nosmalltableaux, boxsize=3em}
\begin{ytableau}
w_1^{\bf 5} & w_{1}^{\bf10}  &   w_{2}^{\bf10}    &    w_{3}^{\bf10}  &  w_{4}^{\bf10}  \\
 \none &w_2^{\bf 5}   &    w_{5}^{\bf10}   & w_{6}^{\bf10}  &    w_{7}^{\bf10}  \\
 \none  &   \none &  w_{3}^{\bf5}   &   w_{8}^{\bf10}  &  w_{9}^{\bf10}  \\
  \none &  \none &  \none  & w_{4}^{\bf 5}   &  w_{10}^{\bf10} \\
  \none &  \none &  \none  &\none  &  w_{5}^{\bf5} \\
\end{ytableau}}
\end{tabular}
\end{center}
\caption{The box graph  \cite{HLM} for the weights of the fundamental and antisymmetric representation. 
A given phase is characterized by specific signs assignment to each box of the box graph. 
 \label{fig.tableaux}}
\end{figure}

The relevant representations here are the $\bf 5$ and $\bf 10$ representations. 
The weights in $\bf 5$ are
\begin{align}
w^{\bf 5}_1 = [1~0~0~0],~w^{\bf 5}_2 = [-1~1~0~0],~
w^{\bf 5}_3=[0 -1~1~0],~w^{\bf 5}_4= [0~0 -1~1],
~w^{\bf 5}_5=[0 ~ 0~0 -1].
\end{align}
The weights in $\bf10$ are given by $w_i^{\bf5}+w_j^{\bf5}$ with $1\leq i <j\leq 5$. Explicitly they are
\begin{align}
\begin{split}
&w^{\bf 10}_1 = [ 0~1~0~0],~w^{\bf 10}_2 = [1 -1 ~1~0] ,~w^{\bf 10}_3= [1~0 -1~1],~w^{\bf 10}_4 = [1~0~0 -1],\\
&w^{\bf 10}_5 = [-1 ~0 ~1~0],~ w^{\bf 10}_6 = [-1~1 -1~1],~w^{\bf 10}_7 =[-1~1~0-1],\\
&w^{\bf 10}_8=[0-1~0~1],~w^{\bf 10}_9 = [0 -1 ~1 -1],~w^{\bf 10}_{10} = [0~0 -1~0].
\end{split}
\end{align}
 The weights of these representations can be organized inside the \textit{box graph} in Figure \ref{fig.tableaux}.

There are nine nontrivial interior walls\footnote{We use the same notation $w^{\bf5}_i$ or $w^{\bf10}_i$ to denote both the weight itself and the interior wall it defines in the Weyl chamber.} in the Weyl chamber $(\phi_1, \phi_2, \phi_3,\phi_4)\in \mathbb{R}_{\ge0}^4$:
\begin{align}
\begin{split}
&  w_2^{\bf 5}:~
 -\phi_1 + 3\phi_2 + 2\phi_3 + \phi_4=0,~~
  w_3^{\bf 5}:~
 -\phi_1   - 2 \phi_2 + 2\phi_3 + \phi_4=0,~~
  w_4^{\bf 5}:~ 
-\phi_1 - 2\phi_2 -3\phi_3 +\phi_4=0,\\
&  w_3^{\bf 10}:~
~3\phi_1 + \phi_2 - \phi_3 +2 \phi_4=0,~~
 w_4^{\bf 10}:~ 
~3\phi_1 + \phi_2 - \phi_3 -3 \phi_4=0,~~
 w_5^{\bf 10}:~ 
-2\phi_1 + \phi_2 +4 \phi_3 +2 \phi_4=0,\\
& w_6^{\bf 10}:~ 
-2\phi_1 + \phi_2 - \phi_3 +2 \phi_4=0,~~
 w_7^{\bf 10}:~ 
-2\phi_1 + \phi_2 - \phi_3 -3\phi_4=0,~~
 w_8^{\bf 10}:~ 
-2\phi_1 -4\phi_2 - \phi_3 +2 \phi_4=0.
\end{split}
\end{align}
The walls defined by other weights only intersect the Weyl chamber at the origin so do not divide the Weyl chamber into different subchambers. These nine interior walls divide the Coulomb branch into twelve different subchambers in Table \ref{match1}. The sign assignment for each subchamber is encoded in the box graph with blue (yellow) being the positive (negative) weights.

\newpage

\clearpage

The nine interior walls intersect at the following nine planes (Figure \ref{SU5Coulomb2}):
\begin{align}
\begin{split}
& P^1= w_3^{\bf5} \cap w_4^{\bf 10} \cap w_6^{\bf 10} :~ \phi_1 -\phi_4=\phi_2- \phi_3=0,\\
&P^2= w_2^{\bf5} \cap w_3^{\bf5} \cap w_5^{\bf10} :~ \phi_2 = -\phi_1 +2\phi_3 +\phi_4=0,\\
&P^3= w_3^{\bf5}\cap w_4^{\bf5} \cap w_8^{\bf 10}:~ \phi_3 = -\phi_1 -2 \phi_2 +\phi_4=0,\\
&P^4=w_3^{\bf10}\cap w_6^{\bf10}:~\phi_1= \phi_2 -\phi_3 +2\phi_4=0,\\
&P^5=w_4^{\bf10} \cap w_7^{\bf10}:~\phi_1=\phi_2 -\phi_3 -3 \phi_4=0,\\
&P^6=w_6^{\bf10}\cap w_8^{\bf10}:~\phi_2=-2\phi_1-\phi_3+2\phi_4=0,\\
&P^7=w_5^{\bf10}\cap w_6^{\bf10}:~\phi_3 = -2\phi_1+\phi_2+2\phi_4=0,\\
&P^8=w_3^{\bf10}\cap w_4^{\bf10}:~\phi_4= 3\phi_1 +\phi_2 -\phi_3=0,\\
&P^9=w_6^{\bf10}\cap w_7^{\bf10}:~\phi_4= -2\phi_1 +\phi_2-\phi_3=0.
\end{split}
\end{align}
The nine planes further intersect along the following four lines (Figure \ref{SU5Coulomb3}):
\begin{align}
\begin{split}
&L^1= P^1\cap P^2\cap P^3\cap P^6\cap P^7:~\phi_2=\phi_3= \phi_1-\phi_4=0,\\
&L^2= P^1\cap P^4\cap P^5\cap P^8\cap P^9:~\phi_1=\phi_4=\phi_2-\phi_3=0,\\
&L^3=P^4 \cap P^6:~ \phi_1=\phi_2 =  -\phi_3 +2\phi_4=0,\\
&L^4 = P^7\cap P^9:~ \phi_3=\phi_4= -2\phi_1+\phi_2=0.
\end{split}
\end{align}
Finally, the four lines intersect at the origin of the Weyl chamber (Figure \ref{SU5Coulomb4}):
\begin{align}
O=L^1\cap L^2\cap L^3\cap L^4.
\end{align}
The incidence structure of the $SU(5)$ Coulomb branch with matters in $\bf 5$ and $\bf10$ representations, or equivalently, the incidence geometry $(A_4,{\bf5\oplus\bf10})$, is presented in Figure \ref{SU5Coulomb1} to Figure \ref{SU5Coulomb4}.

\subsection{Identification with the network of resolutions}

We will now identify a subset of the incidence geometry $(A_4,{\bf 5\oplus 10})$ with the network of resolutions we constructed in the present paper. It would be interesting to find all the small \textit{partial} resolutions of the $SU(5)$ model to complete the match with the Coulomb branch incidence structure $(A_4,{\bf 5\oplus 10})$. Even though we do not have the complete network of all the small partial resolutions, the network in Figure \ref{network} is sufficient to obtain all twelve final resolutions.\footnote{The sequences of weighted blow ups for $\mathscr{B}_{2,3}^1$ and its dual are not presented in Figure \ref{network}. They are discussed in Section \ref{section:toric5}.}

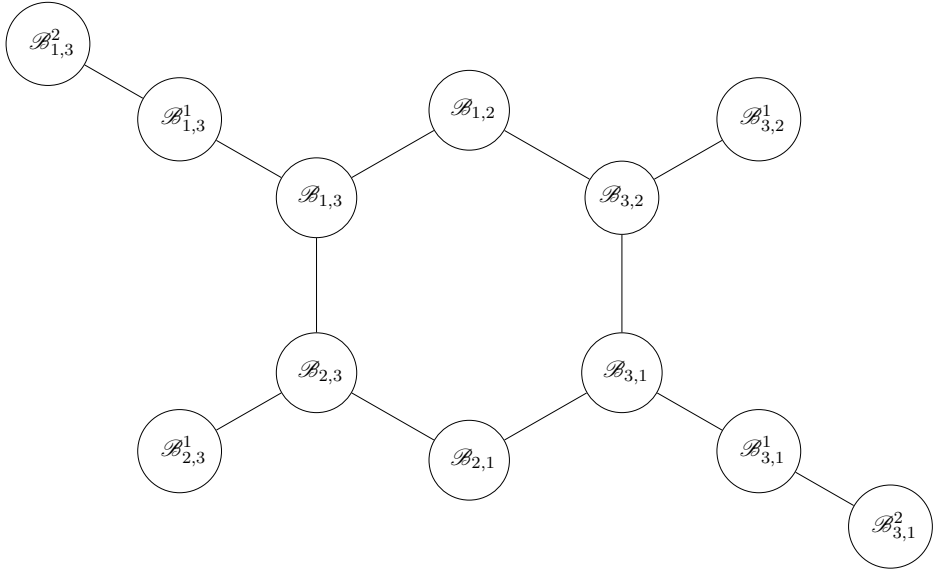
\begin{figure}
\scalebox{.8}{
\begin{tikzpicture}[every node/.style={circle,draw, minimum size= 10 mm},  scale=1]
\node (P11) at (1*30+60*0:2.9cm)  {\small $  \mathscr{B}_{3,2}$};
\node (P3)  at (1*30+60*1:2.9cm)  {\small $\  \mathscr{B}_{1,2}$};
\node (P4) at (1*30+60*2:2.9cm) {\small $ \ \mathscr{B}_{1,3}$};
\node (P2) at (1*30+60*3:2.9cm) {\small $  \ \mathscr{B}_{2,3}$};
\node (P10) at (1*30+60*4:2.9cm) {\small $\  \mathscr{B}_{2,1}$};
\node (P9) at (1*30+60*5:2.9cm) {\small $\ \mathscr{B}_{3,1}$};
\node (P5) at (1*30+60*2:5.5cm){\small $\ \mathscr{B}^1_{1,3}$};
\node (P6) at (1*30+60*2:8cm){\small $\  \mathscr{B}^2_{1,3}$};
\node (P8) at (180+1*30+60*2:5.5cm){\small $\  \mathscr{B}^1_{3,1}$};
\node (P7) at (180+1*30+60*2:8cm){\small $\  \mathscr{B}^2_{3,1}$};
\node (P1) at (120+1*30+60:5.5cm){\small  $\  \mathscr{B}_{2,3}^1$};
\node (P12) at (120+180+1*30+60:5.5cm){\small $\  \mathscr{B}_{3,2}^1$};
\draw
(P1)--(P2);
\draw
(P2)--(P10);
\draw
(P10)--(P9);
\draw (P11)--(P12);
\draw 
(P4)--(P3);
\draw
(P3)--(P11);
\draw  
(P6)--(P5);
\draw
(P5)--(P4);
\draw 
(P8)--(P7);
\draw
(P9)--(P8);
\draw (P4)--(P2);
\draw (P9)--(P11);
\end{tikzpicture}}
\caption{\small Network of  small resolutions of the  $SU(5)$ model. Each node corresponds to a resolution of the $SU(5)$ model. It has a perfect match with the intersections of subchambers of the $SU(5)$ Coulomb branch with representations $\bf5\oplus 10$ in Figure \ref{SU5Coulomb1}.
\label{SU5Res}}
 \end{figure}

\begin{figure}
\scalebox{.9}{
\begin{tikzpicture}[every node/.style={circle,draw, minimum size= 10 mm},  scale=1]
\node (P11) at (1*30+60*0:2.9cm) {\small $ 11$};
\node (P3)  at (1*30+60*1:2.9cm)  {\small $3$};
\node (P4) at (1*30+60*2:2.9cm) {\small $4$};
\node (P2) at (1*30+60*3:2.9cm) {\small $2$};
\node (P10) at (1*30+60*4:2.9cm) {\small $10$};
\node (P9) at (1*30+60*5:2.9cm) {\small $9$};
\node (P5) at (1*30+60*2:5.5cm){\small $5$};
\node (P6) at (1*30+60*2:8cm){\small $6$};
\node (P8) at (180+1*30+60*2:5.5cm){\small $8$};
\node (P7) at (180+1*30+60*2:8cm){\small $7$};
\node (P1) at (120+1*30+60:5.5cm){\small  $1$};
\node (P12) at (120+180+1*30+60:5.5cm){\small $12$};
\draw[postaction={decoration={raise=1ex,text along path,  text={{$w_{3}^{\bf 10}$}{}},text align={center}},decorate}] (P1)--(P2);
\draw[postaction={decoration={raise=1ex,text along path,  text={{$w_{4}^{\bf10}$}{}},text align={center}},decorate}] (P2)--(P10);
\draw[postaction={decoration={raise=1ex,text along path,  text={{$w_{3}^{\bf 5}$}{}},text align={center}},decorate}] (P10)--(P9);
\draw[postaction={decoration={raise=1ex,text along path,  text={{$w_{7}^{\bf 10}$}{}},text align={center}},decorate}] (P11)--(P12);
\draw[postaction={decoration={raise=1ex,text along path,  text={{$w_{3}^{\bf 5}$}{}},text align={center}},decorate}] (P4)--(P3);
\draw[postaction={decoration={raise=1ex,text along path,  text={{$w_{4}^{\bf 10}$}{}},text align={center}},decorate}] (P3)--(P11);
\draw[postaction={decoration={raise=1ex,text along path,  text={{$w_{4}^{\bf 5}$}{}},text align={center}},decorate}] (P6)--(P5);
\draw[postaction={decoration={raise=1ex,text along path,  text={{$w_{8}^{\bf 10}$}{}},text align={center}},decorate}] (P5)--(P4);
\draw[postaction={decoration={raise=1ex,text along path,  text={{$w_{2}^{\bf 5}$}{}},text align={center}},decorate}] (P8)--(P7);
\draw[postaction={decoration={raise=1ex,text along path,  text={{$w_{5}^{\bf 10}$}{}},text align={center}},decorate}] (P9)--(P8);
\draw (P4)--(P2);
\draw (P9)--(P11);
 \coordinate (M42) at ($($(P4)!+0.5!(P2)$)-(.4,0)$); \node[draw=none]  at (M42) {$w_6^{\bf 10}$};
  \coordinate (M911) at ($($(P9)!+0.5!(P11)$)+(.4,0)$); \node[draw=none]  at (M911) {$w_6^{\bf 10}$};
\end{tikzpicture}}
\caption{\small Intersections of subchambers in the $(A_4,{\bf5}\oplus {\bf 10})$ incidence geometry, namely, the $SU(5)$ Coulomb branch with representation $\mathbf{5}\oplus\mathbf{10}$.
  Each circle represents a subchamber and  each edge corresponds to a common interior wall between two adjacent subchambers. Physically, the interior walls labeled by a weight $w_i$ (the lines in the figure) correspond to the Higgs branch roots where the matter fields with weight $w_i$ become massless.
\label{SU5Coulomb1}}
 \end{figure}
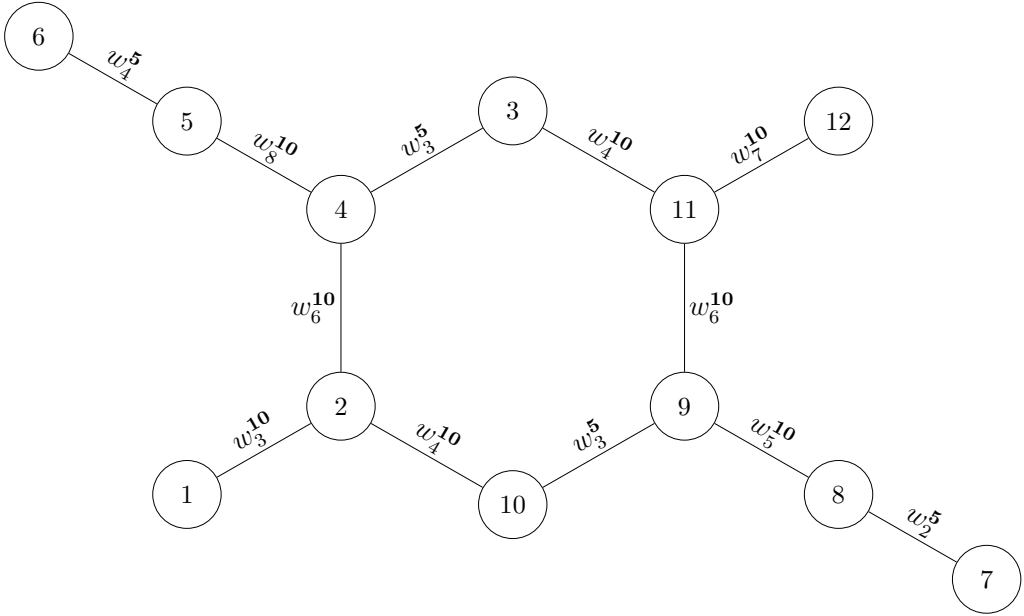

Let us start with the identification in codimension zero, namely, matching the subchambers with the final resolutions. This is demonstrated in Table \ref{match1} and can also be seen by comparing Figure \ref{SU5Res} with \ref{SU5Coulomb1}.

\begin{figure}
\begin{center}
\includegraphics[scale=1]{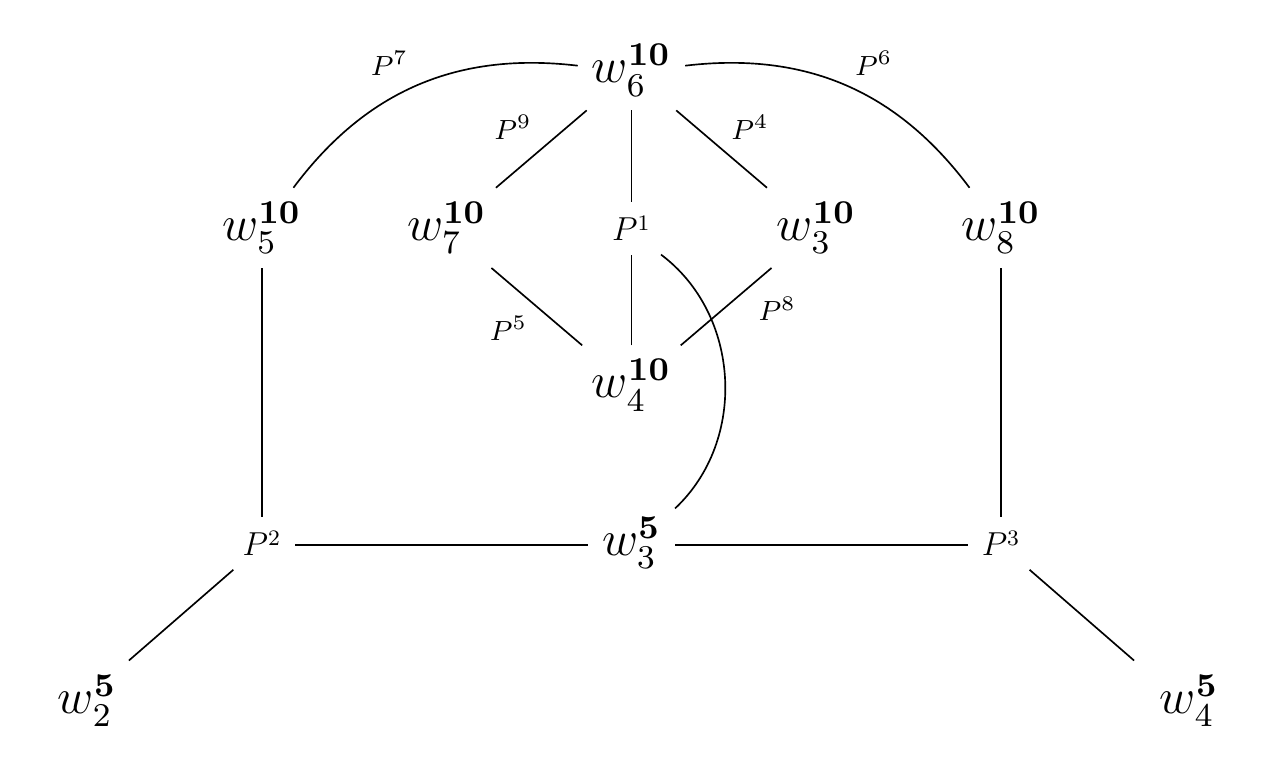}
\end{center}

\caption{Intersections of the nine (codimension one) interior walls $w_i^{\bf 5}$ and $w^{\bf10}_i$ at the nine planes $P^i$ in the $(A_4,{\bf5}\oplus {\bf 10})$ incidence geometry. }
\label{SU5Coulomb2}
\end{figure}

In codimension one, the partial resolutions in Figure \ref{network} are identified as the interior walls in Figure \ref{SU5Coulomb2}:
\begin{align}
\begin{split}
&w_3^{\bf 5} = (\mathscr{T}^+_1 \cong \mathscr{B}_{1,\bullet}) \cup (\mathscr{T}^-_1\cong\mathscr{B}_{\bullet, 1}),\\
&w_8^{\bf10}=\mathscr{T}_2^+,~~w_5^{\bf10}=\mathscr{T}^-_2,\\
&w_4^{\bf5}=\mathscr{T}^+_3,~~w_2^{\bf5}=\mathscr{T}^-_3,\\
&w_4^{\bf10}=\mathscr{B}_{2,\bullet}\cup\mathscr{B}_{\bullet, 2},\\
&w_6^{\bf10}=\mathscr{B}_{3,\bullet}\cup\mathscr{B}_{\bullet, 3}.
\end{split}
\end{align}
The $\cup$ sign for the interior wall $w_3^{\bf5}$ means that it is divided into two components by other interior walls, with one component corresponding to $\mathscr{T}^+_1\cong\mathscr{B}_{1,\bullet}$ and the other corresponding to 
$ \mathscr{T}^-_1\cong\mathscr{B}_{\bullet, 2}$. Similarly for $w_4^{\bf10}$ and $w_6^{\bf10}$. The isomorphism between $\mathscr{T}^+_1$ and $\mathscr{B}_{1,\bullet}$ is explained in Section \ref{section:iso}.

\begin{figure}
\begin{center}
\includegraphics[scale=1.2]{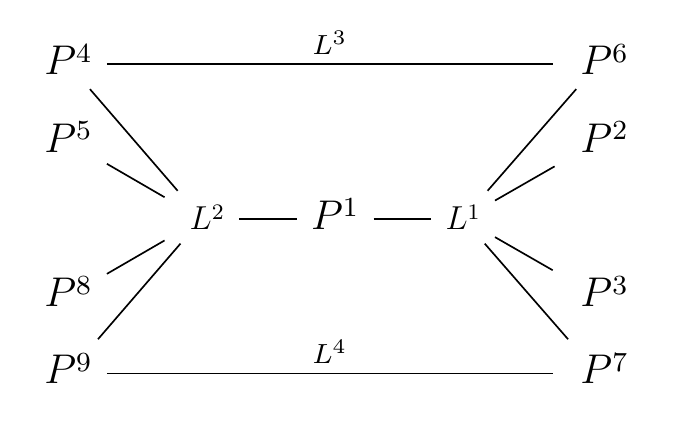}
\end{center}

\caption{Intersections of the nine (codimension two) planes in the  incidence geometry $(A_4,{\bf5\oplus 10})$.}
\label{SU5Coulomb3}
\end{figure}

In codimension two, the partial resolutions in Figure \ref{network} are identified as the planes in Figure \ref{SU5Coulomb3}:
\begin{align}
P^3 = \mathscr{T}^+,~~P^2 = \mathscr{T}^-,~~P^1=\mathscr{B}.
\end{align}

\begin{figure}
\begin{center}
\includegraphics[scale=1.3]{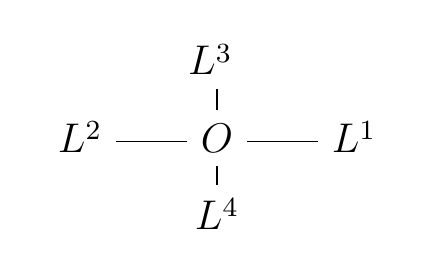}
\end{center}
\caption{Intersections of the four (codimension three) lines in the  incidence geometry $(A_4,{\bf5\oplus 10})$.}
\label{SU5Coulomb4}
\end{figure}

In codimension three, the partial resolution $\mathscr{E}_1$ is identified as one of the lines in Figure \ref{SU5Coulomb4}:
\begin{align}
L^1=\mathscr{E}_1. 
\end{align}
Finally, the origin $O$ of the Coulomb branch is identified as the original singular Weierstrass model $\mathscr{E}_0$:
\begin{align}
O=\mathscr{E}_0.
\end{align}
Thus we have successfully embedded the network in Figure \ref{network} into the incidence geometry $(A_4,{\bf 5\oplus 10})$. In the physics language, we identify the network of resolutions with the Coulomb branch incidence structure. 

In the rest of the paper we will construct the network in Figure \ref{network} by explicit blow ups.

\begin{table}[htb]
\scalebox{.9}{
{
\renewcommand{\arraystretch}{1.5}
\begin{tabular}{|c|c|c|c|c|c|c|}
\hline 
Phases & $\alpha_1$ & $\alpha_2$ & $\alpha_3$ & $\alpha_4$ & Box graph & \small Resolution \\ 
\hline 
\hline
 1 (13.II) & $\bullet$ &  $\bullet$& & $\bullet$ &  \bIIthirteen{.7} &  $\mathscr{B}_{2,3}^1$ \\
\hline 
 2 (10.II) & &  $\bullet$ & &   &  \bIIten{.7} &   $   \mathscr{B}_{2,3}$  \cite{EY}\\
 \hline 

 3 (8.III)
 & $\bullet$& & $\bullet$& & 
 \bIIIeight{.7} &  $\mathscr{B}_{1,2}$  \cite{EY} \\
 \hline 
 4 (8.II) & $\bullet$  & & & &  \bIIeight{.7} & $\mathscr{B}_{1,3}$  \cite{EY}\\
 \hline 
 5 (6.II) & $\bullet$ & $\bullet$& & &  \bIIsix{.7}& $\mathscr{B}^1_{1,3}$ \\
 \hline 
  6 (6.I) &   $\bullet$& $\bullet$ & $\bullet$ & &  \bIsix{.7}  &$\mathscr{B}^2_{1,3}$  \\
 \hline 
 7 (11.IV) & &  $\bullet$  & $\bullet$& $\bullet$  &\bIVeleven{.7}  & $\mathscr{B}^2_{3,1}$  \\
 \hline 
 8 (11.III) & &&   $\bullet$& $\bullet$&  \bIIIeleven{.7}  &  $\mathscr{B}^1_{3,1}$ \\
 \hline 
 9 (9.III) && & & $\bullet$ & \bIIInine{.7} & $\mathscr{B}_{3,1}$   \cite{EY} \\
   \hline 
  10 (9.II)
   & & $\bullet$& &$\bullet$ &
  \bIInine{.7}  &   $\mathscr{B}_{2,1}$  \cite{EY}\\
 \hline 
  11 (7.III)&   &  & $\bullet$  &  & \bIIIseven{.7} &  $\mathscr{B}_{3,2}$  \cite{EY}   \\
 \hline 
 12 (4.III)&  $\bullet$& &  $\bullet$ &$\bullet$ &\bIIIfour{.7}   & $\mathscr{B}_{3,2}^1$   \\
 \hline 
\end{tabular}}}
\caption{ \small Subchambers in the incidence geometry $(A_4, {\bf5\oplus \bf10})$.  A blue (yellow) box means the corresponding weight is positive (negative) in that phase. 
A dot under $\alpha_i$ indicates that the subchamber has the boundary wall corresponding to $\alpha_i$ as one of the components of its boundary. Physically that means the $W$-boson with root $\alpha_i$ can become massless in that phase.  Six of the twelve subchambers have been identified in \cite{EY}.}\label{match1}
\end{table} \clearpage

 \section{ Definition of the $SU(5)$ model}\label{section:su5}

A Weierstrass model $\mathscr{E}\rightarrow B$  over a base $B$ is an elliptic fibration defined as a hypersurface cut by a section of 
 the line bundle $\mathscr{O}(3)\otimes \pi^\star \mathscr{L}^6$ 
 in the projective bundle $\mathbb{P}(\mathscr{O}_B\oplus \mathscr{L}^2\oplus \mathscr{L}^3)\rightarrow B$  {\cite{Formulaire,Nakayama.Global, Nakayama.Local,MSu}}. Here $\mathscr{L}$ is a section on the base $B$, $\mathscr{O}(1)$ is the dual of the tautological line bundle of the projective bundle, and $\mathscr{O}(n)$ is its $n$-th tensor product. 
Projective coordinates of the $\mathbb{P}^2$ projective bundle are denoted by  $[x:y:z]$, where  $x$ is a section of $\mathscr{O}(1) \otimes \pi^\star \mathscr{L}^2$, $y$ is a section of $\mathscr{O}(1) \otimes \pi^\star \mathscr{L}^3$, and $z$ is a section of $\mathscr{O}(1)$. In terms of these projective coordinates, the Weierstrass model is defined by
\begin{equation}
\mathscr{E}:\quad z y^2 + a_1 x y z + a_3 y z^2 -(x^3 +a_2 x^2 z + a_4 x z^2 + a_6 z^3)=0.
\end{equation}
The coefficients $a_i$ are sections of $\mathscr{L}^i$.

Each fiber is an elliptic curve with the neutral element of  the Mordell-Weil group given by the point $z=x=0$. This defines a rational section of the elliptic fibration. The rational section of the elliptic fibration defines the  Mordell-Weil group of the elliptic fibration. On each fiber, the opposite of a point $[x:y:z]$ with respect to the Mordell-Weil group, is the point 
$[x:-y - a_1 x - a_3 z:z]$. This defines a fiberwise involutive automorphism of $\mathscr{E}$ \cite{Tate.Book}:
\begin{equation}
\iota:\mathscr{E}\rightarrow \mathscr{E}: [x:y:z]\mapsto [x:-y - a_1 x  - a_3 z:z].
\end{equation}
We will see that this involution induces flops between different small resolutions 
of $\mathscr{E}_0$.  We will call this involution the {\it Mordell-Weil involution} of the  elliptic fibration. When extended to the non-singular fibers it defines a birational map between different resolutions of the Weierstrass model \cite{EY}.

Given a Weierstrass model, at any stage of the resolution process, we denote by $s$ the proper transform of $(y+a_1 x+a_3)$ and by $t$ the proper transform of the cubic  $(x^3+a_2 x^2 + a_4 x + a_6)$: 
\begin{align}
&s:=\text{ Proper transform of }\   (y+a_1 x+a_3)\\
&t:=\text{ Proper transform of }\   (x^3+a_2 x^2 + a_4 x + a_6).
\end{align}
When we need to be more precise, we use the notation $s_k$ and $t_k$ to mean the proper transform of $s$ and $t$ after the $k$-th blow up. 
The  Weierstrass mosdel can then be rewritten as 
\begin{equation}
\mathscr{E}_0:\quad y s-t=0.
\end{equation}
In this new notation, the  Mordell-Weil involution  is simply \begin{equation}
(y,s)\mapsto (-s, -y).
\end{equation}

The 
$SU(5)$ model is  usually defined by the following singular Weierstrass model directly  from Tate's algorithm (in the $z=1$ patch of the $\mathbb{P}^2$) \cite{BIKMSV}:
\begin{equation} 
\mathscr{E}_0:\quad y(y + a_1 x + a_{3,2} e_0^2)-(x^3 + a_{2,1} e_0  x^2 +a_{4,3} e_0^3 x +a_{6,5} e_0^5)=0.
\end{equation}
Here
\begin{align}
e_0=0
\end{align}
 is a codimension one locus in the base $B$ where the fiber of $\mathscr{E}_0$ becomes singular.  Throughout this paper we will assume $a_{i,j}$ to be {\it general} sections on the base. Many results in the current paper do \textit{not} hold for special choices of $a_{i,j}$.

The discriminant locus of $\mathscr{E}_0$ is
\begin{align}\label{discriminant}
\Delta= e_0^5 \left[ -a_1^4 P + \mathcal{O}(e_0)\right],
\end{align}
where 
\begin{align}
P:= a_{2,1} a_{3,2}^2 -a_1 a_{3,2} a_{4,3} +a_1^2 a_{6,5}^2.
\end{align}

\subsection{Resolution of singularities}
A {\em resolution of singularities} is a map $\pi: X'\rightarrow X$ between a nonsingular variety $X'$ and a singular variety $X$ such that the following conditions are satisfied:
\begin{enumerate}
\item $X'$ is a nonsingular variety 
\item $\pi$ is a surjective birational map 
\item $\pi$ is a proper map 
\item $\pi$  is  an isomorphism away from the singular locus of $X$ 
\end{enumerate}
We will require our resolutions to be crepant and the projection of the elliptic fibration $X'\rightarrow B$ to the base $B$ should define a flat morphism. Since we work over $\mathbb{C}$, a morphism $\pi: Y\rightarrow B$  between irreducible varieties $Y$ and $B$ with  $Y$ Cohen-Macaulay and $B$ regular is  {\em  flat } if and only if  the fibers are all equidimensional. 
In our case, flatness will mean that the resolved elliptic fibrations admit fibers that are always one-dimensional.

A birational map  is   said to be {\em small } when  the exceptional locus is of codimension two or higher. 
 A birational map is  said to be {\em crepant} when  $X$ is normal and $\pi$  preserves the canonical class, that is $\pi^\star K_X=K_{X'}$.
A small resolution is always crepant, but a crepant resolution is not necessary small. 
One way to construct a small resolution is to give a sequence of blow ups with centers that are non-Cartier  Weil divisors.

\subsubsection{Notations for  blow ups.}

If we blow up $X$ along the  ideal generated by   $g_i$ to obtain $X'$, we write: 
$$
X\xleftarrow{(g_1, \cdots, g_n |e)}X',$$
 where $e$ defines a generator of the  principal ideal corresponding to the exceptional locus of the blow up. 
Such a blow up is induced by the following replacements
 $$g_k\rightarrow e g_k, \quad k=1,\cdots, n.$$ 
where   $e$ is a section of $\mathscr{O}(E)$. The $g_k$ on the lefthand side are the original generators of the ideal while on those on the  righthand side are now the projective coordinates of the  projective bundle generated by the blow up. In this economic notation we do not have to introduce an extra set of variables to denote the new projective coordinates. On the other hand, the original generators of the ideal are now expressed as $eg_k$.

\par Since we will often need successive blow ups, we will denote by  $E_k$ the exceptional divisor of the $k$-th blow up and by $e_k$ a rational section of $\mathscr{O}(E_k)$. For the resolutions discussed in the current paper,  the total transform  of the divisor $e_0=0$ in the base $B$ is $e_0e_1e_2e_3e_4$, which we will denote by $w$:
\begin{align}
w:=e_0e_1e_2e_3e_4.
\end{align}

As we will see, \textit{all} twelve resolutions of $\mathscr{E}_0$ have the following universal fibers over the following loci:
\begin{align}
\begin{split}
&\text{affine}~A_4~\text{over} ~w=0,\\
&\text{affine}~A_5 ~\text{over}~ w= P=0,\\
&\text{affine}~D_5~\text{over}~ w=a_1=0,
\end{split}
\end{align}
The codimension two fiber then determines the relevant matter representations to be $\bf 5$ and $\bf10$ by the Katz-Vafa method. 
In codimension three, we will have more complicated (generally non-Kodaira) fibers.

\subsection{First blow up of the $SU(5)$ model and conifold singularities}

The total space of the Weierstrass model $\mathscr{E}_0$ is singular at
\begin{align}
x=y=e_0.
\end{align}
Therefore we first blow up the ideal $(x,y,e_0)$ to obtain the partial resolution $\mathscr{E}_1$:
\begin{equation}
\begin{tikzcd}[column sep=huge]
\mathscr{E}_0 \arrow[leftarrow]{r}{ \textstyle{( x,y,e_0 |e_1 )}} & 
\mathscr{E}_1 
\end{tikzcd}
\end{equation}
the proper transform of $\mathscr{E}_0$ is 
\begin{equation}
\mathscr{E}_1: y (y+a_1 x+a_{3,2} e_1 e_0^2)-e_1 (x^3+ a_{2,1} e_0x^2 
+ a_{4,3} e_0^3 e_1 x + a_{6,5}e_0^5 e_1^2)=0.
\end{equation}
The blow up introduces  a $\mathbb{P}^2$ bundle (in the ambient space) with  projective coordinates  $[x:y:e_0]$. Altogether after the first blow up the ambient space has projective coordinates
\begin{equation}
[e_1x:e_1y:z][x: y: e_0].
\end{equation}
   The proper transform $\mathscr{E}_1$ is a partial resolution that will be the common ancestor of all our resolutions with the exception of $\mathscr{B}_{2,3}^1$ and $\mathscr{B}_{3,2}^1$.

In order to understand the next possible steps, it is useful to rewrite $\mathscr{E}_1$ as follows
\begin{equation}
\mathscr{E}_1: y s -e_1 t=0,
\end{equation}
where  $s$ and $t$ are defined as:
 \begin{equation}\label{st}
s=y+a_1 x+a_{3,2} e_1 e_0^2\quad \text{ and}\quad   t =x^3+ e_0 Q(x,e_1 e_0^2)
\end{equation}
with
\begin{equation} 
 Q(x_1,x_2):=a_{2,1} x^2_1+a_{4,3}  x_1 x_2 +a_{6,5} x_2^2.
\end{equation}
In this form, it is clear that $\mathscr{E}_1$ has a conifold singularity at 
\begin{equation}
y=s=e_1=t=0,
\end{equation}
which is  
 \begin{equation}
e_1=y=a_1 x =x^2(x+a_{2,1}  e_0)=0.
\end{equation}
This locus has two components: 
 \begin{equation}
e_1=y= x =0, \quad \text{and}\quad  e_1=a_1=y=x+a_{2,1} e_0=0,
\end{equation}
 in codimension one and in codimension two, respectively. 
Blowing up $e_1=x=y=0$  gives the binomial geometry leading to the six small resolutions discussed in \cite{EY}. Alternatively, we could also blow up $y=e_1=0$ or $s=e_1=0$. These are the two small resolutions of $ys-e_1 t=0$ and are connected to each other by a flop. The flop is induced by the  Mordell-Weil involution $(y,s)\rightarrow (-s, -y)$ which is an involutive symmetry of the partial resolution $\mathscr{E}_1$. 

Before doing the second blow up, let us take a closer look at the fiber structure of $\mathscr{E}_1$. This will give a geometric guidance for the possible options of the following blow ups.  The singular fiber is located at the total transform of the original divisor $e_0=0$, which is $e_0 e_1=0$. 
The component $e_0=0$  gives the node $C_0$ that contains the section of the elliptic fibration. 
This node is the proper transform of the original singular fiber:
\begin{equation}
C_0:\quad e_0= y (y+a_1 x)-e_1 x^3=0. 
\end{equation}
The other component $e_1=0$   defines two nodes $C_{1\pm}$ in the fiber: 
\begin{equation}
C_{1+}: e_1=y=0\quad \text{ and } \quad C_{1-}:e_1=s=0.
\end{equation} 
These two components are exchanged by the Mordell-Weil involution of the elliptic fiber $(y,s)\rightarrow (-s,-y)$.   The two nodes $C_{1\pm}$  intersect only over the codimension two locus $e_0e_1=a_1=0$ in the base $B$. 
The intersection of $C_{1\pm}$ is $x=y=e_1=0$, which is  contained in the center of the conifold $ys=e_1 t$.   The ideal $(x,y,e_1)$ is the center of the second blow up of \cite{EY} that gives the binomial variety $\mathscr{B}$. Alternatively, blowing up $C_\pm$ themselves  give new partial resolutions $\mathscr{T}^\pm$ that we will study in the following. 
We summarize these three options for the second blow ups in the following tree diagram:
\begin{equation}
\begin{tikzcd}[column sep=huge]
& & \mathscr{T}^+  \arrow[bend left=75, leftrightarrow, dashed]{dd}[right] {\quad \text{\large flop}}\\
\mathscr{E}_0\arrow[leftarrow]{r}[near end]{\textstyle{(x,y,e_0)}}& \mathscr{E}_1\arrow[leftarrow]{r}[near end]{ \textstyle{(x,y,e_1)}}\arrow[leftarrow]{rd}[sloped,below, near end]{\textstyle{(s,e_1)}}\arrow[leftarrow]{ru}[sloped,above,near end] {\textstyle{(y,e_1)}}& \mathscr{B}\\
& & \mathscr{T}^-
\end{tikzcd}
\end{equation}
A few comments are in order:
\begin{itemize}
\item $\mathscr{T}^+$ is obtained by   blowing up the non-Cartier divisor $C_{1+}:(y,e_1)$.
\item  $\mathscr{T}^-$ is obtained by   blowing up the non-Cartier divisor  $C_{1-}:(s,e_1)$.
\item $\mathscr{B}$ is obtained by blowing up the ideal $(x,y,e_1)$. This is the second blow up in   \cite{EY} with the ideal being the intersection of $C_{1\pm}$. 
\item $\mathscr{T}^{\pm}$ are mapped into each other under  the Mordell-Weil involution $(y,s)\mapsto (-s,-y)$ (which is now a birational map)  while $\mathscr{B}$ is invariant. This involution is the origin of the $\mathbb{Z}_2$ 
symmetry  in the full network of resolutions in Figure \ref{network}. 
\end{itemize}

In the following sections we will explore the full network of resolutions from the three partial resolutions $\mathscr{B},\mathscr{T}^\pm$.

\section{ Esole-Yau  small resolutions \label{EYdescription}}

In this section we will reproduce the six resolutions of the $SU(5)$ model that were previously obtained in  \cite{EY} from the partial resolution $\mathscr{B}$ discussed in the last section:

\subsection{The binomial structure of the $\mathscr{B}$ branch}

The partial resolution $\mathscr{B}$ is obtained by
\begin{equation}
\begin{tikzcd}[column sep=huge]
\mathscr{E}_0 \arrow[leftarrow]{r}{ \textstyle{( x,y,e_0 |e_1 )}} & 
\mathscr{E}_1 \arrow[leftarrow]{r}{ \textstyle{( x,y,e_1 |e_2)}} & 
\mathscr{B}
\end{tikzcd}
\end{equation}
\begin{equation}
\mathscr{B}:\quad y(y + a_1 x + a_{3,2} e_1 e_0^2 )= e_1e_2(e_2 x^3 + a_{2,1} e_0 x^ 2 +a_{4,3} e_1 e_0^3 x +a_{6,5} e_1^2  e_0^5),
\end{equation}
in the ambient space with projective coordinates
\begin{equation}
[e_2x:e_2y: e_0]\quad [x:y: e_1].
\end{equation}
Note that $\mathscr{B}$ can be embedded into a hypersurface in $\mathbb{A}^5$ as follows:
\begin{equation}
\begin{tikzcd}[column sep=normal]
\mathscr{B} \arrow[hookrightarrow]{r} & V(y s-v_1 v_2 v_3)\subset \mathbb{A}^5,
\end{tikzcd}
\end{equation}
with
\begin{align}
&s=y + a_1 x + a_{3,2} e_1 e_0^2, \quad\\
& v_1=e_1, \quad v_2=e_2, \quad v_3=e_2 x^3 + a_{2,1} e_0 x^ 2 +a_{4,3} e_1 e_0^3 x +a_{6,5} e_1^2  e_0^5=:t.
\end{align}
The network of resolutions from the $\mathscr{B}$ branch immediately follows from the binomial structure of $V(ys-v_1v_2v_3)$. There are six resolutions obtained as follows:
\begin{equation}
\begin{tikzcd}[column sep=huge]
\mathscr{B} \arrow[leftarrow]{r}{ \textstyle{( y,v_i |\ell_1 )}} & 
\mathscr{B}_{i,\bullet}  \arrow[leftarrow]{r}{ \textstyle{( s,v_j |\ell_2)}} & 
\mathscr{B}_{i,j}  
\end{tikzcd} \quad\quad i\neq j.
\end{equation}
We could also obtain the same six resolutions by exchanging the order of the blow ups: 
\begin{equation}
\begin{tikzcd}[column sep=huge]
\mathscr{B} \arrow[leftarrow]{r}{ \textstyle{( s,v_j |\ell_2 )}} & 
\mathscr{B}_{\bullet,j}  \arrow[leftarrow]{r}{ \textstyle{( y,v_i |\ell_1)}} & 
\mathscr{B}_{i,j}  
\end{tikzcd} \quad\quad i\neq j.
\end{equation}
We will present the explicit formulas for these resolutions in a moment.

In summary, the branch coming out of the partial resolution $\mathscr{B}$ consists of the following  14 (partial) resolutions:
\begin{itemize}
\item the partial resolutions $\mathscr{E}_1$ and $\mathscr{B}$
\item  the three partial resolutions $\mathscr{B}_{i,\bullet}$
\item the three partial resolutions $\mathscr{B}_{\bullet,j}$
\item the  six resolutions $\mathscr{B}_{i,j}$
\end{itemize}

These different  resolutions $\mathscr{B}_{i,j}$ are connected to each other by a network of flop transitions  induced by    automorphisms of the binomial variety $\mathscr{B}$.   In particular

\begin{itemize}
\item the involution $(y,s)\mapsto (-s,-y)$ is exactly the Mordell-Weil involution sending a point on the elliptic fiber to its inverse under the Mordell-Weil group. Interestingly, it induces a flop: 
\begin{equation}
\begin{tikzcd}\mathscr{B}_{i,j}\arrow[rightarrow, dashed] {r} &\mathscr{B}_{j,i}\end{tikzcd}.
\end{equation}
 \item the involution $(j \leftrightarrow k)$ connects  $\mathscr{B}_{i,j}$  and $\mathscr{B}_{i,k}$  (resp.  $\mathscr{B}_{j,i}$  and $\mathscr{B}_{k,i}$) through the flop induced by contracting to the partial resolution $\mathscr{B}_{i,\bullet}$  (resp.  $\mathscr{B}_{\bullet,i}$)
 \end{itemize}

The six resolutions $\mathscr{B}_{i,j}$ can be described by the following complete intersection:
\begin{equation}
\mathscr{B}_{i,j}
\begin{cases}
 v_i\alpha_+ -y\beta_+ =0\\
 v_j\alpha_- -s\beta_- =0\\
\alpha_+\alpha_--\beta_+\beta_- v_k=0
\end{cases}\quad \text{where}\quad   (i,j,k) \quad \text{is a permutation of $(1,2,3)$}
\end{equation}
where $[\alpha_+:\beta_+][\alpha_-:\beta_-]$ are the projective coordinates introduced for the last two steps of blow ups. 
Let us rearrange the equations in the following suggestive form: 
\begin{equation}
\mathscr{B}_{i,j}
\begin{cases}
 v_i\alpha_+ -y\beta_+ =0\\
(\alpha_+)\alpha_--(\beta_+ v_k) \beta_-=0\\
 v_j\alpha_- -\beta_- s=0
\end{cases}
\end{equation}  This is to emphasize that $\mathscr{B}_{i,j}$ can also be obtained from  $\mathscr{B}$ by first  blow up at $(y, v_i)$ and then blowing up  $(\alpha_+, \beta_+ v_k)$. 
Since $\alpha_+$ and $\beta_+$ are projective coordinates of a $\mathbb{P}^1$, they  cannot vanish at the same time. It follows that the ideal  $(\alpha_+, \beta_+ v_k)$ is the same as $(\alpha_+, v_k)$. Therefore, the blow up  with center $(\alpha_+ ,\beta_+ v_k)$   can be equivalently described as the blow up with center $(\alpha_+, v_k)$. This gives:
\begin{equation}
\mathscr{B}_{i,j}
\begin{cases}
 v_i\alpha_+ -y\beta_+ =0\\
\alpha_+ \alpha_- - v_k \beta_-=0\\
 s \beta_-  -\beta_+\alpha_- v_j=0
\end{cases}
\end{equation}
We recognize the $\mathscr{B}_{i,j}$ as the blow up $(1,i)(1,k)$  introduced in  \cite{HLN}. In the same way, we can also show that the resolution $(2,j)(2,k)$ is also the resolution $\mathscr{B}_{i,j}$.  Hence we arrive at the following theorem:
\begin{theorem}
Let   $(i,j,k)$ be  a permutation of $(1,2,3)$. 
The resolution $(1,i)(1,k)$ and $(2,j)(2,k)$ are isomorphic to  $\mathscr{B}_{i,j}$:
\begin{equation}
\mathscr{B}_{i,j}\cong (1i)(1k)\cong(2,j)(2,k).
\end{equation}
\end{theorem}
In other words, the resolutions $(1,i)(1,k)$ and $(2,j)(2,k)$ introduced in \cite{HLN} are reformulation of the resolutions $\mathscr{B}_{i,j}$ of \cite{EY}. This is our first result in providing a unifying picture for the known resolutions in the literature from direct blow up. 

In the remaining part of this section we will study the three resolutions $\mathscr{B}_{1,3},~\mathscr{B}_{2,3},~ \mathscr{B}_{1,2}$ in details, while the other three are trivially related by Mordell-Weil involution.

\subsection{Resolutions $\mathscr{B}_{1,3}$ and $\mathscr{B}_{2,3}$}\label{section:EY2}

In this subsection we will redo the blow up by explicit introducing the generator $e_k$ for the exceptional divisor. This will prove to be convenient when comparing with the other resolutions in the network.

The centers for the last two blow ups for $\mathscr{B}_{1,3}$ are 
\begin{align}
(y,e_1)(s,t).
\end{align}
Since the $\mathscr{B}_{1,\bullet}$ takes the conifold form $ys = (e_1 e_2) t$, we can equivalently replace the center of the blow up $(s,t)$ by $(y,e_1e_2)$. Furthermore, as noted previously, since $y$ and $e_1$ cannot vanish at the same time, the center $(y,e_1e_2)$ is the same as $(y,e_2)$. All in all, we can obtain $\mathscr{B}_{1,3}$ from $\mathscr{B}$ by performing two blow ups along the following centers
\begin{align}
(y,e_1)\quad (y,e_2).
\end{align}
The complete sequence of blow ups for $\mathscr{B}_{1,3}$ is
\begin{align}
 \mathscr{E}_0 \xleftarrow{(x,y,e_0|e_1)}\mathscr{E}_1 \xleftarrow{(x,y,e_1|e_2)} \mathscr{B}\xleftarrow{(y,e_1|e_3)}\mathscr{B}_{1,\bullet}\xleftarrow{(y,e_2|e_4)}\mathscr{B}_{1,3}
  \end{align}
The resolution can be written as
\begin{align}
\mathscr{B}_{1,3} : ~y (e_4e_3 y+a_1 x +a_{3,2} e_0^2 e_1 e_3) =e_1 e_2 (e_4e_2 x^3  + a_{2,1} e_0x^2 +a_{4,3} e_0^3 e_1 e_3 x +a_{6,5} e_0^5e_1^2 e_3^2),
\end{align}
in the following projective space
\begin{align}
\, [e_4^2e_3 e_2^2e_1 x: e_4^3e_3^2 e_2^2 e_1 y:z] \quad[e_4e_2 x:e_4^2e_3 e_2 y:e_0]  \quad[x:e_4e_3 y:e_3 e_1] \quad[e_4y:e_1] \quad[y:e_2].
\end{align}

Similarly, $\mathscr{B}_{2,3}$ can be obtained by the following sequence of blow ups:
\begin{align}
 \mathscr{E}_0 \xleftarrow{(x,y,e_0|e_1)}\mathscr{E}_1 \xleftarrow{(x,y,e_1|e_2)} \mathscr{B}\xleftarrow{(y,e_2|e_3)}\mathscr{B}_{2,\bullet} \xleftarrow{(y,e_1|e_4)}\mathscr{B}_{2,3}
  \end{align}
  $\mathscr{B}_{2,3}$ can be written as
  \begin{align}
  \mathscr{B}_{2,3}:~y( e_3e_4  y+a_1 x+a_{3,2}e_0^2e_1e_4)
  = e_1 e_2 ( e_3e_2 x^3 +a_{2,1} e_0x^2 +a_{4,3} e_0^3e_1e_4 x +a_{6,5}e_0^5 e_1^2e_4^2)
  \end{align}
  in the following projective space
  \begin{align}
  \, [e_4 e_3^2e_2^2e_1x:e_4^2 e_3^3 e_2^2 e_1 y:z] \quad[e_3 e_2 x:e_4e_3^2 e_2 y:e_0] \quad[x:e_3 e_4y: e_4e_1] \quad[e_4y:e_2] \quad[y:e_1].
  \end{align}
  
As we will see in Section \ref{section:toric}, $\mathscr{B}_{1,3}$ and $\mathscr{B}_{2,3}$ provide realizations of the toric resolutions  of type  I and type II in \cite{KMW}, respectively.  The identification will be explained in Section \ref{section:toric2}. We postpone the study of the fiber enhancements for $\mathscr{B}_{1,3}$ and $\mathscr{B}_{2,3}$ to Section \ref{section:toric1} when we discuss the toric resolutions.

\subsection{Resolution $\mathscr{B}_{1,2}\cong\mathscr{T}^+_{1-}$ }

The last resolution we want to study is $\mathscr{B}_{1,2}$, which can be obtained from the following sequence of blow ups:
\begin{align}
 \mathscr{E}_0 \xleftarrow{(x,y,e_0|e_1)}\mathscr{E}_1 \xleftarrow{(x,y,e_1|e_2)} \mathscr{B}\xleftarrow{(y,e_1|e_3)}\mathscr{B}_{1,\bullet}\xleftarrow{(s,e_2|e_4)}\mathscr{B}_{1,2}
  \end{align}
We can write $\mathscr{B}_{1,2}$ as follows: 
\begin{equation}
\mathscr{B}_{1,2}
\begin{cases}
e_3 y + a_1 x + a_{3,2} e_1 e_3 e_0^2=e_4 s, \\
ys - e_1e_2 (e_2e_4 x^3 + a_{2,1} e_0 x^ 2 +a_{4,3} e_1e_3 e_0^3 x +a_{6,5} e_1^2 e_3^2  e_0^5)=0.
 \end{cases}
\end{equation}
with projective coordinates 
\begin{equation}
[e_1 e_2^2 e_3 e_4^2 x:e_1 e_2^2e_3^2 e_4^2 y:z], \quad [e_2  e_4 x:e_2 e_3  e_4y: e_0],\quad[x: e_3y : e_3 e_1], \quad [y:e_1]\quad [s:e_2].
\end{equation}

After a direct computation, the fibers for $\mathscr{B}_{1,2}$ in various codimensions are obtained in Table \ref{Table.Fo12}.

In fact, one can show that the $\mathscr{B}_{1,2}$ is isomorphic to another resolution $\mathscr{T}^+_{1-}$ 
\begin{align}
\mathscr{T}^+_{1-}\cong \mathscr{B}_{1,2}.
\end{align}
This isomorphism will be further discussed in Section \ref{section:iso}. See also Appendix B.4.1 of \cite{ESY}.

Finally, since both $\mathscr{B}_{1,2}$ and $\mathscr{B}_{1,3}$ are obtained from the partial resolution $\mathscr{B}_{1,\bullet}$, they are related by a flop:
\begin{equation}
\begin{tikzcd}[column sep=huge]
& & & & \mathscr{B}_{1,2} \arrow[leftrightarrow, bend left, dashed]{dd}{\textstyle{flop}}\\
\mathscr{E}_0 \arrow[leftarrow]{r}{ \textstyle{( x,y,e_0 |e_1 )}} & 
\mathscr{E}_1 \arrow[leftarrow]{r}{ \textstyle{( x,y,e_1 |e_2)}} & 
\mathscr{B} \arrow[leftarrow]{r}{ \textstyle{( y,e_1 |e_3)}} & 
\mathscr{B}_{1,\bullet} 
\arrow[leftarrow]{ru}[ sloped,  near end]{\textstyle{( s,e_2 |e_4)}}[sloped,below]{\textstyle{or\,( y,t|e_4)}}
 \arrow[leftarrow]{rd}[sloped, near start]{ \textstyle{( s,t |e_4)}}[below,sloped]{\textstyle{or \,( y,e_2|e_4)}}& \\
& & & & \mathscr{B}_{1,3}
\end{tikzcd}
\end{equation} 

\begin{table}[!h]
\begin{center}
\scalebox{.85}{\begin{tabular}{|c|c|c|c|c|}
\hline
$w=0$& $w=P=0$ &  $w=a_1=0$ & $w=a_1=a_{3,2}=0$ & $w=a_1=a_{2,1}=0$ \\ 
\hline
& {\footnotesize $C_4\to C_{4a}+C_{4b}$ }&
\scalebox{.8}{ \begin{tabular}{l}
$C_1\to C_{13}$\\
$C_3\to C_{13}+C_{34}+C_3'$\\
$C_4\to C_{34}+C_4'$
\end{tabular} }
 & 
\scalebox{.8}{ \begin{tabular}{l}
$C_1\to C_{13}$\\
$C_2\to C_{24}$\\
$C_3\to C_{13}+C_{34}+C_3'$\\
$C_4\to C_{24}+C_{34}+C_{4a}''+C_{4b}''$
\end{tabular} }
 & 
\scalebox{.8}{\begin{tabular}{l}
$C_1\to C_{13}$\\
$C_3\to C_{34+}+2C_{34-}+C_{13}$\\
$C_4\to C_{34+}+C_{34-}+C_4'$
\end{tabular} }
 \\
\hline
\scalebox{.95}{\begin{tikzpicture}[every node/.style={circle,draw, minimum size= 5 mm}]
\node (C2)  at (18+72*5:1.4cm)  {\tiny $C_{3}$};
\node (C4) at (18+72*4:1.4cm) {\tiny $C_{2}$};
\node (C3) at (18+72*3:1.4cm) {\tiny $C_{4}$};
\node (C1) at (18+72*2:1.4cm) {\tiny $C_{1}$};
\node (C0) at (18+72*1:1.4cm) {\tiny $C_0$};
\draw (C1)--(C3)--(C4)--(C2)--(C0)--(C1);
\end{tikzpicture}}& 
\scalebox{.95}{\begin{tikzpicture}[every node/.style={circle,draw, minimum size= 5 mm}]
\node (C2b)  at (30+60*6:1.4cm)  {\tiny $C_{3}$};
\node (C2)  at (30+60*5:1.4cm)  {\tiny $C_{2}$};
\node (C4) at (30+60*4:1.4cm) {\scalebox{.7}{ $C_{4b}$}};
\node (C3) at (30+60*3:1.4cm) {\scalebox{.7}{$C_{4a}$}};
\node (C1) at (30+60*2:1.4cm) {\tiny $C_{1}$};
\node (C0) at (30+60*1:1.4cm) {\tiny $C_0$};
\draw (C1)--(C3)--(C4)--(C2)--(C2b)--(C0)--(C1);
\end{tikzpicture}}& 
\scalebox{1}{\begin{tikzpicture}[every node/.style={circle,draw, minimum size= 5 mm}, scale=.8]
\node (C2) at (0,0) {\tiny $2C_{13}$};
\node (C5) at (90:-1.8cm) { \scalebox{.7}{$2C_{34}$}};
\node (C0) at (90-35:1.8cm) { \tiny $C_3'$} ;
\node (C1) at (90+35:1.8cm) {\tiny $C_0$};
\node [yshift=-1.8cm] (C3) at  (-90-35:1.8cm){ \tiny $C_4'$};
\node [yshift=-1.8cm] (C4) at (-90+35:1.8cm) {\tiny $C_{2}$};
\draw (C2)--(C0);
\draw (C2)--(C1);
\draw (C2)--(C5);
\draw (C5)--(C3);
\draw (C5)--(C4);
\end{tikzpicture}}
& 
\scalebox{1} {\begin{tikzpicture}[every node/.style={circle,draw, minimum size= 4 mm}, scale=.8]
\node (C2) at (0,0) {\tiny $2C_{13}$};
\node (C5) at (90:-1.8cm) {\tiny $2C_{34}$};
\node (C0) at (90-35:1.8cm) {\tiny $C_3' $} ;
\node (C1) at (90+35:1.8cm) {\small $C_0$};
\node (C6) at (90:-2*1.8cm) { \tiny {$2C_{24}$}};
\node [yshift=-1.8*2cm] (C3) at  (-90-35:1.4cm){\tiny $C_{4a}''$};
\node [yshift=-1.8*2cm] (C4) at (-90+35:1.4cm) {\tiny $C_{4b}''$};
\draw (C2)--(C0);
\draw (C2)--(C1);
\draw (C2)--(C5);
\draw (C6)--(C3);
\draw (C6)--(C4);
\draw (C5)--(C6);
\end{tikzpicture}}

& 
\scalebox{1}{\begin{tikzpicture}[every node/.style={circle,draw, minimum size= 4 mm}, scale=.8]
\node (C2) at (0,0) {\tiny $2C_{13}$};
\node (C5) at (90:-1.8cm) {\scalebox{.65}{$3C_{34-}$}};
\node (C0) at (90:1.8cm) {\small $C_0$} ;
\node [xshift=-1.4cm] (C3) at  (90:-1.8cm){\tiny $C_4'$};
\node  (C6) at (90:-2*1.8cm) {\scalebox{.65}{$2C_{34+}$}};
\node  (C4) at (90:-3*1.8cm) {\tiny $C_2$};
\draw (C2)--(C0);
\draw (C2)--(C5)--(C6)--(C4);
\draw (C5)--(C3);
\end{tikzpicture}}

\\ 
\hline
\end{tabular}}
\end{center}
\caption{Fibers of the resolution $\mathscr{B}_{1,2}\cong\mathscr{T}^+_{1-}$ in various codimensions. $C_i$ is the node coming from the $e_i=0$ component of  $w=e_0e_1e_2e_3e_4=0$  on the base. $C_{ij}$ is node comes from the intersection of $e_i=0$ and $e_j=0$. Note that  $C_i$ are not ordered by their positions in the I$_5^s$ fiber. Here $P=
a_{2,1} a_{3,2}^2 - a_{4,3} a_1 a_{3,2}  +a_{6,5}a_1^2=0$. \label{Table.Fo12} }
\end{table}

\section{Models of  type I, II, and III}\label{section:toric}

Few months after \cite{EY}, another recipe was presented in \cite{KMW} to resolve the singularities of the $SU(5)$  model. This method is more combinatorial in nature and is inspired from the procedure outlined in \cite{BIKMSV} and toric resolutions of Calabi-Yau
 hypersurfaces as in \cite{CPR}.  The blow up is expected to be defined by  monoidal transformations (blow ups of regular primes).  For these reasons, the constructions of \cite{KMW} are usually called  {\em toric resolutions} in the F-theory literature.
 
In this section we will obtain all the toric resolutions by sequences of (weighted) blow ups. In particular, the type I and type II toric resolutions are identified as $\mathscr{B}_{1,3}$ and $\mathscr{B}_{2,3}$ obtained in Section \ref{EYdescription}.

Let us start with the definitions of toric resolutions.
There are three different toric resolutions in the $SU(5)$ model called  { type I}, { type II}, and { type III}. They are related by flops as will be explained in Section \ref{section:toric4}. These three types are supposed to come from  a succession of blow ups  that combine  to the  following  morphism:
\begin{align}\label{ToricTrans}
(x,y,e_0)\rightarrow (x e_1 e_4 e_2^2 e_3^2, y e_1 e_4^2 e_2^2 e_3^3,e_0 e_1 e_2 e_3 e_4).
\end{align}
However, the explicit sequences of blow ups are not known before the current paper. 
Under this morphism, the proper transform of the elliptic fibration is:
\begin{align}\label{ET}
\mathscr{E}_T:
y( e_3 e_4y + a_1 x  + a_{3,2}   e_0^2 e_1 e_4) 
=
e_1 e_2(x^3 e_2 e_3 + a_{2,1} x^2 e_0  + a_{4,3} x e_0^3 e_1  e_4 
+ a_{6,5} e_0^5 e_1^2 e_4^2).
\end{align}
The projective bundles produced by the sequence of blow ups leading to the morphism \eqref{ToricTrans} defines together with the equation of $\mathscr{E}_T$ a  Stanley-Reisner  (SR) ideal common to all the three cases:
\begin{equation}\label{SRIdeal}
\{xyz, z e_1 , z e_2 , ze_3, z  e_4 , x e_0 , y e_0,  x e_1, 
y e_1, y e_2 , xe_4, e_0e_2 \}.
\end{equation}
Each monomial of the Stanley-Reisner ideal indicates a set a variables that cannot vanish simultaneously. 
The three types of toric resolutions are then defined by the following additional elements to add to the SR ideal:

\begin{equation}\label{SRIdeal2}
\begin{tabular}{|r|l|}
\hline
Type\   I &  $e_0 e_3, \ e_1 e_3$ \\
\hline
Type II&  $e_0 e_3, \  e_2 e_4 $ \\
\hline
Type III &  $e_1 e_4, \  e_2 e_4 $\\
\hline  
\end{tabular}
\end{equation}

\subsection{Fiber structure of toric models of type I, II, and III}\label{section:toric1}

Before we give the explicit constructions for type I, II, and III, we can already determine their fiber structures from \eqref{ET} and their SR ideals \eqref{SRIdeal2}.

For any of the toric model I, II, and III, the fiber over the codimension one locus $e_0e_1e_2e_3e_4=0$ consists of the following five rational curves: 
\begin{align}
\begin{split}
& C_0: \quad e_0=e_3 e_4 y^2+ a_1 x y -e_1 e_2^2 e_3 x^3=0,\\
& C_1 : \quad e_1= e_3 e_4y + a_1 x=0 ,\\ 
& C_2:\quad  e_2= e_3 e_4y + a_1 x  + a_{3,2}   e_0^2 e_1 e_4=0 ,\\
& C_3: \quad e_3=  
y( a_1 x  + a_{3,2}   e_0^2 e_1 e_4) 
-
e_0 e_1 e_2( a_{2,1} x^2   + a_{4,3} x e_0^2 e_1  e_4 
+ a_{6,5} e_0^4 e_1^2 e_4^2)=0,\\
& C_4: \quad e_4=a_1 y    -
e_1 e_2 x(x e_2 e_3 + a_{2,1} e_0 )=0,
\end{split}
\end{align}
where we have used the elements $e_1 y, e_2 y , e_4 x$ of the SR ideal to  eliminate some components in  $C_1$, $C_2$ and $C_4$. 
Using the SR ideal, the intersections of the five nodes can be shown to be the affine $A_4$ Dynkin diagram in Figure \ref{I5fiber}. The component $C_0$ is the only one that touches the section $x=z=0$ thanks to the elements $\{ ze_1, z e_2, z_3, z_4 \}$ in the SR ideal. 

\begin{center}
\begin{figure}[htb]
\begin{tikzpicture}[every node/.style={circle,draw, minimum size= 5 mm}]
\node (C4)  at (18+72*5:1.8cm)  {\tiny $C_{4}$};
\node (C3) at (18+72*4:1.8cm) {\tiny $C_{3}$};
\node (C2) at (18+72*3:1.8cm) {\tiny $C_{2}$};
\node (C1) at (18+72*2:1.8cm) {\tiny $C_{1}$};
\node (C0) at (18+72*1:1.8cm) {\tiny $C_0$};
\draw (C1)--(C2)--(C3)--(C4)--(C0)--(C1);
\end{tikzpicture}
\caption{ Fiber of type I$_5^s$. Here we label the nodes by their positions in the $A_4$ Dynkin diagram. \label{I5fiber}}
\end{figure}
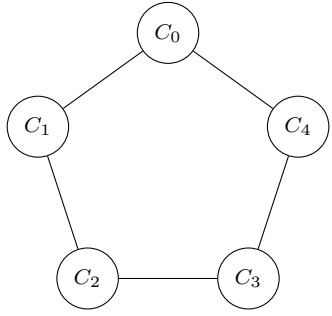
\end{center}

The different models distinguish themselves  by the splitting of their nodes in higher codimensions. Let us now study their codimension two fibers.

Let us start with the codimension two locus $e_0e_1e_2e_3e_4=P=0$, where
\begin{align}
P:= 
a_{2,1} a_{3,2}^2 - a_{4,3} a_1 a_{3,2}  +a_{6,5}a_1^2=0,
\end{align}
is one of the component of the discriminant locus $\Delta$ \eqref{discriminant}. 
 All three types of resolutions share the same fiber enhancement over $P=0$ where the third node $C_3$ splits into two
\begin{align}
C_3\rightarrow C_{3a}+C_{3b}.
\end{align}
The original affine $A_4$ Dynkin diagram now enhances to the affine $A_5$ Dynkin diagram. 

\begin{table}[htb]
\begin{tabular}{|c|c|}
\cline{2-2}
\multicolumn{1}{c|}{}& $C_3\rightarrow C_{3a}+C_{3b}$ \\
\hline
&\\
\scalebox{.9}{
\begin{tikzpicture}[every node/.style={circle,draw, minimum size= 5 mm}]
\node (C4)  at (18+72*5:1.8cm)  {\tiny $C_{4}$};
\node (C3) at (18+72*4:1.8cm) {\tiny $C_{3}$};
\node (C2) at (18+72*3:1.8cm) {\tiny $C_{2}$};
\node (C1) at (18+72*2:1.8cm) {\tiny $C_{1}$};
\node (C0) at (18+72*1:1.8cm) {\tiny $C_0$};
\draw (C1)--(C2)--(C3)--(C4)--(C0)--(C1);
\end{tikzpicture}}
& \scalebox{.9}{\begin{tikzpicture}[every node/.style={circle,draw, minimum size= 5 mm}]
\node (C2b)  at (30+60*6:1.8cm)  {\tiny $C_{4}$};
\node (C2)  at (30+60*5:1.8cm)   {\scalebox{.7}{ $C_{3b}$}};
\node (C4) at (30+60*4:1.8cm) {\scalebox{.7}{ $C_{3a}$}};
\node (C3) at (30+60*3:1.8cm) {\tiny $C_{2}$};
\node (C1) at (30+60*2:1.8cm) {\tiny $C_{1}$};
\node (C0) at (30+60*1:1.8cm) {\tiny $C_0$};
\draw (C1)--(C3)--(C4)--(C2)--(C2b)--(C0)--(C1);
\end{tikzpicture}}\\
 & \\
\hline
\end{tabular}
\caption{I$_5^s\rightarrow $I$_6^s$ enhancement from $w=0$ to $w=P=0$ for models of type I, II and III. Here $w=e_0e_1e_2e_3e_4$ and $P=
a_{2,1} a_{3,2}^2 - a_{4,3} a_1 a_{3,2}  +a_{6,5}a_1^2=0$.}\label{type123P}
\end{table}

Over the codimension two locus $e_0e_1e_2e_3e_4=a_1=0$, the nodes split as (we use the notation $C_{ab}: e_a=e_b=0$):
\begin{equation}
\begin{aligned}
C_0&\quad \to C_{03}, \quad  C_{0}':e_0=e_4 y^2 -e_1 e_2^2  x^3=0,\\
C_1 &\quad \to C_{13}, \quad C_{14},\\ 
C_2 &\quad \to C_{24}, \quad  C_2' :e_2=e_3 y   + a_{3,2}   e_0^2 e_1=0, \\
 C_3&\quad \to C_{03}, \quad C_{13},\quad C_3' :e_3=
y a_{3,2}   e_0 e_4 
- e_2( a_{2,1} x^2 + a_{4,3} x e_0^2 e_1  e_4 
+ a_{6,5} e_0^5 e_1^2 e_4^2)=0,\\
C_4 &\quad \to C_{14}, \quad C_{24},\quad  C_4' :e_4=x e_2 e_3 + a_{2,1} e_0=0.
\end{aligned}
\end{equation}
In the above splitting, we have \textit{not} imposed the SR ideal constraints. When specializing to a particular type of resolution, its SR ideal would forbid some of the $C_{ab}$ nodes on the right hand side to exist if $e_ae_b$ belongs to it. However, note that the nodes $C_0', C_2', C_3'$ and $C_4'$ are common to all the three types of resolutions.

Over  $e_0e_1e_2e_3e_4=a_1=a_{3,2}=0$, the node splits as
\begin{equation}
\begin{aligned}
C_0&\quad \to C_{03}, \quad  C_{0}':e_0=e_4 y^2 -e_1 e_2^2  x^3=0,\\
C_1 &\quad \to C_{13}, \quad C_{14},\\ 
C_2&\quad \to C_{24}, \quad  C_{23}, \\
 C_3&\quad \to C_{03}, \quad C_{13},\quad C_{23}, \quad C_3'' :e_3= a_{2,1} x^2   + a_{4,3} x e_0^2 e_1  e_4 
+ a_{6,5} e_0^4 e_1^2 e_4^2=0,\\
C_4 &\quad \to C_{14}, \quad C_{24},\quad  C_4' :e_4=x e_2 e_3 + a_{2,1} e_0=0.
\end{aligned}
\end{equation}
The node $C_3''$ is actually composed of two nodes  $C_3^{(1)}+C_3^{(2)}$ whose locations are given by solving the equation for $x$. 
However, these two nodes are not split in the sense that the discriminant is not a perfect square. If we work on a base of dimension three, this is not a problem as 
 all the coefficients are just in $k$. 

Finally, the fiber enhancements over $e_0e_1e_2e_3e_4=a_1=a_{2,1}=0$ are: 
\begin{equation}
\begin{aligned}
C_0&\quad \to C_{03}, \quad  C_{0}':e_0=e_4 y^2 -e_1 e_2^2  x^3=0,\\
 C_1 &\quad \to C_{13}, \quad C_{14}, \\ 
 C_2&\quad \to C_{24}, \quad  C_2':e_2=e_3 y   + a_{3,2}   e_0^2 e_1=0, \\
 C_3&\quad \to 2C_{03}, \quad C_{13}\quad C_{34}, \quad C_3''' :e_3=
y a_{3,2}  
- e_0e_1 e_2 ( a_{4,3} x  
+ a_{6,5} e_0^2 e_1 e_4)=0,\\
 C_4&\quad \to  C_{14}, \quad 2  C_{24},\quad  C_{34}.
\end{aligned}
\end{equation}

After imposing the SR ideal to the above fiber enhancements for three types of toric resolutions, we obtain their fiber structure from codimension one to codimension three in Table \ref{type123P} and Table \ref{type123}.

\begin{center}
\begin{tabular}{|c|c|c|c|}
\cline{2-4}
\multicolumn{1}{c|}{}& \multicolumn{3}{c|}{Fiber enhancements over $w=a_1=0$} \\
\hline 
Components & Type I & Type II & Type III \\
\hline 
$C_0$ & $C_0'$ & $C_0'$ & $C_0'+C_{03}$  \\
\hline
$C_1$ & $C_{14}$ & $C_{13}+C_{14}$  & $C_{13}$\\
\hline
$C_2$ & $C_{24}+C_2'$ & $C_2'$ & $C_2'$\\
\hline
$C_3$ & $C_3'$& $C_{13}+C_3'$ &$C_{03}+C_{13}+C_3'$  \\
\hline
$C_4$ &  $C_{14}+C_{24}+C_4'$ &$C_{14}+C_4'$  & $C_4'$ \\
\hline 
{
\begin{tikzpicture}[every node/.style={circle,draw, minimum size= 8 mm}, scale=.9]
\node (C4)  at (18+72*5:1.8cm)  {\tiny $C_{4}$};
\node (C3) at (18+72*4:1.8cm) {\tiny $C_{3}$};
\node (C2) at (18+72*3:1.8cm) {\tiny $C_{2}$};
\node (C1) at (18+72*2:1.8cm) {\tiny $C_{1}$};
\node (C0) at (18+72*1:1.8cm) {\tiny $C_0$};
\draw (C1)--(C2)--(C3)--(C4)--(C0)--(C1);
\end{tikzpicture}
}
& {\begin{tikzpicture}[every node/.style={circle,draw, minimum size= 4 mm}, scale=.8]
\node (C2) at (0,0) {\tiny $2C_{14}$};
\node (C5) at (90:-1.8cm) {\tiny $2C_{24}$};
\node (C0) at (45+90*0:1.8cm) {\small $C_0' $} ;
\node (C1) at (45+90*1:1.8cm) {\small $C_4'$};
\node [yshift=-1.8cm] (C3) at  (45+90*2:1.8cm){\small $C_2'$};
\node [yshift=-1.8cm] (C4) at (45+90*3:1.8cm) {\small $C_3'$};
\draw (C2)--(C0);
\draw (C2)--(C1);
\draw (C2)--(C5);
\draw (C5)--(C3);
\draw (C5)--(C4);
\end{tikzpicture}}
& 
{\begin{tikzpicture}[every node/.style={circle,draw, minimum size= 4 mm}, scale=.8]
\node (C2) at (0,0) {\tiny $2C_{14}$};
\node (C5) at (90:-1.8cm) {\tiny $2C_{13}$};
\node (C0) at (45+90*0:1.8cm) {\small $C_0' $} ;
\node (C1) at (45+90*1:1.8cm) {\small $C_4'$};
\node [yshift=-1.8cm] (C3) at  (45+90*2:1.8cm){\small $C_2'$};
\node [yshift=-1.8cm] (C4) at (45+90*3:1.8cm) {\small $C_3'$};
\draw (C2)--(C0);
\draw (C2)--(C1);
\draw (C2)--(C5);
\draw (C5)--(C3);
\draw (C5)--(C4);
\end{tikzpicture}}

&
{\begin{tikzpicture}[every node/.style={circle,draw, minimum size= 4 mm}, scale=.8]
\node (C2) at (0,0) {\tiny $2C_{03}$};
\node (C5) at (90:-1.8cm) {\tiny $2C_{13}$};
\node (C0) at (45+90*0:1.8cm) {\small $C_0' $} ;
\node (C1) at (45+90*1:1.8cm) {\small $C_{4}'$};
\node [yshift=-1.8cm] (C3) at  (45+90*2:1.8cm){\small $C_{2}'$};
\node [yshift=-1.8cm] (C4) at (45+90*3:1.8cm) {\small $C_3'$};
\draw (C2)--(C0);
\draw (C2)--(C1);
\draw (C2)--(C5);
\draw (C5)--(C3);
\draw (C5)--(C4);
\end{tikzpicture}} \\
\hline
\end{tabular}

\end{center}

\begin{center}
\begin{tabular}{|c|c|c|c|}
\cline{2-4}
\multicolumn{1}{c|}{}& \multicolumn{3}{c|}{Fiber enhancements over $w=a_1=a_{3,2}=0$} \\
\hline 
Components & Type I & Type II & Type III \\
\hline 
$C_0$ & $C_0'$ & $C_0'$ & $C_0'+C_{03}$  \\
\hline
$C_1$ & $C_{14}$ & $C_{13}+C_{14}$  & $C_{13}$\\
\hline
$C_2$ & $C_{24}+C_{23}$ & $C_{23}$ & $C_{23}$\\
\hline
$C_3$ & $C_{23}+C_3^{(1)}+C_3^{(2)}$& $C_3^{(1)}+C_3^{(2)}+C_{13} +C_{23}$ &$\substack{C_3^{(1)}+C_3^{(2)}\\+C_{03}+C_{13}+C_{23}\\~}$  \\
\hline
$C_4$ &  $C_{14}+C_{24}+C_4'$ &$C_{14}+C_4'$  & $C_{4}'$ \\
\hline 
{
\begin{tikzpicture}[every node/.style={circle,draw, minimum size= 8 mm}, scale=.9]
\node (C4)  at (18+72*5:1.8cm)  {\tiny $C_{4}$};
\node (C3) at (18+72*4:1.8cm) {\tiny $C_{3}$};
\node (C2) at (18+72*3:1.8cm) {\tiny $C_{2}$};
\node (C1) at (18+72*2:1.8cm) {\tiny $C_{1}$};
\node (C0) at (18+72*1:1.8cm) {\tiny $C_0$};
\draw (C1)--(C2)--(C3)--(C4)--(C0)--(C1);
\end{tikzpicture}
}
& {\begin{tikzpicture}[every node/.style={circle,draw, minimum size= 4 mm}, scale=.7]
\node (C2) at (0,0) {\tiny $2C_{14}$};
\node (C5) at (90:-1.8cm) {\tiny $2C_{24}$};
\node (C0) at (45+90*0:1.8cm) {\small $C_0' $} ;
\node (C1) at (45+90*1:1.8cm) {\small $C_4'$};
\node (C6) at (90:-2*1.8cm) {\tiny $2C_{23}$};
\node [yshift=-1.6*2cm] (C3) at  (45+90*2:1.6cm){\tiny $C_3^{(1)}$};
\node [yshift=-1.6*2cm] (C4) at (45+90*3:1.6cm) {\tiny $C_3^{(2)}$};
\draw (C2)--(C0);
\draw (C2)--(C1);
\draw (C2)--(C5);
\draw (C6)--(C3);
\draw (C6)--(C4);
\draw (C5)--(C6);
\end{tikzpicture}}
& 
{\begin{tikzpicture}[every node/.style={circle,draw, minimum size= 4 mm}, scale=.8]
\node (C2) at (0,0) {\tiny $2C_{14}$};
\node (C5) at (90:-1.8cm) {\tiny $2C_{13}$};
\node (C0) at (45+90*0:1.8cm) {\small $C_0' $} ;
\node (C1) at (45+90*1:1.8cm) {\small $C_4'$};
\node (C6) at (90:-2*1.8cm) {\tiny $2C_{23}$};
\node [yshift=-1.6*2cm] (C3) at  (45+90*2:1.6cm){\tiny $C_3^{(1)}$};
\node [yshift=-1.6*2cm] (C4) at (45+90*3:1.6cm) {\tiny $C_3^{(2)}$};
\draw (C2)--(C0);
\draw (C2)--(C1);
\draw (C2)--(C5);
\draw (C6)--(C3);
\draw (C6)--(C4);
\draw (C5)--(C6);
\end{tikzpicture}}
&
{\begin{tikzpicture}[every node/.style={circle,draw, minimum size= 4 mm}, scale=.7]
\node (C2) at (0,0) {\tiny $2C_{03}$};
\node (C5) at (90:-1.8cm) {\tiny $2C_{13}$};
\node (C0) at (45+90*0:1.8cm) {\small $C_0' $} ;
\node (C1) at (45+90*1:1.8cm) {\small $C_4'$};
\node (C6) at (90:-2*1.8cm) {\tiny $2C_{23}$};
\node [yshift=-1.6*2cm] (C3) at  (45+90*2:1.6cm){\tiny $C_3^{(1)}$};
\node [yshift=-1.6*2cm] (C4) at (45+90*3:1.6cm) {\tiny $C_3^{(2)}$};
\draw (C2)--(C0);
\draw (C2)--(C1);
\draw (C2)--(C5);
\draw (C6)--(C3);
\draw (C6)--(C4);
\draw (C5)--(C6);
\end{tikzpicture}}\\
\hline 
\end{tabular}
\end{center}

\begin{center}
\begin{tabular}{|c|c|c|c|}
\cline{2-4}
\multicolumn{1}{c|}{}& \multicolumn{3}{c|}{Fiber enhancements over $w=a_1=a_{2,1}=0$} \\
\hline 
Components & Type I & Type II & Type III \\
\hline 
$C_0$ & $C_0'$ & $C_0'$ & $C_0'+C_{03}$  \\
\hline
$C_1$ & $C_{14}$ & $C_{13}+C_{14}$  & $C_{13}$\\
\hline
$C_2$ & $C_{24}+C_2'$ & $C_2'$ & $C_2'$\\
\hline
$C_3$ & $C_{34}+C_3'''$& $C_3'''+C_{13}+C_{34}$ &$C_3'''+2C_{03}+C_{13}+C_{34}$  \\
\hline
$C_4$ &  $C_{14}+2C_{24}+C_{34}$ &$C_{14}+C_{34}$  & $C_{34}$ \\
\hline

{
\begin{tikzpicture}[every node/.style={circle,draw, minimum size= 8 mm}, scale=.9]
\node (C4)  at (18+72*5:1.8cm)  {\tiny $C_{4}$};
\node (C3) at (18+72*4:1.8cm) {\tiny $C_{3}$};
\node (C2) at (18+72*3:1.8cm) {\tiny $C_{2}$};
\node (C1) at (18+72*2:1.8cm) {\tiny $C_{1}$};
\node (C0) at (18+72*1:1.8cm) {\tiny $C_0$};
\draw (C1)--(C2)--(C3)--(C4)--(C0)--(C1);
\end{tikzpicture}
}
& 

{\begin{tikzpicture}[every node/.style={circle,draw, minimum size= 4 mm}, scale=.8]
\node (C2) at (0,0) {\tiny $2C_{14}$};
\node (C5) at (90:-1.8cm) {\tiny $3C_{24}$};
\node (C0) at (90:1.8cm) {\small $C_0' $} ;
\node [xshift=-1.8cm] (C3) at  (90:-1.8cm){\small $C_2'$};
\node  (C6) at (90:-2*1.8cm) {\tiny $2C_{34}$};
\node  (C4) at (90:-3*1.8cm) {\small $C_3'''$};
\draw (C2)--(C0);
\draw (C2)--(C5)--(C6)--(C4);
\draw (C5)--(C3);
\end{tikzpicture}}

& 
{\begin{tikzpicture}[every node/.style={circle,draw, minimum size= 4 mm}, scale=.8]
\node (C2) at (0,0) {\tiny $2C_{14}$};
\node (C5) [xshift=-1.8cm]  at (90:-1.8*0 cm) {\tiny $2C_{13}$};
\node (C0) at (90:1.8cm) {\small $C_0' $} ;
\draw (- 1.1cm,0)--+(0,-1.2cm);
\node [xshift=-1.8cm] (C3) at  (90:1.8cm){\small $C_2'$};
\node  (C6) [xshift=-.9 cm] at (90:-1.8cm) {\tiny $2C_{34}$};
\node [xshift=-.9 cm, yshift=-1.8cm] (C4) at (90:-1.8cm) {\small $C_3'''$};
\draw (C2)--(C0);
\draw (C4)--(C6);
\draw (C2)--(C5);
\draw (C5)--(C3);
\end{tikzpicture}}

&

{\begin{tikzpicture}[every node/.style={circle,draw, minimum size= 4 mm}, scale=.8]
\node (C2) at (0,0) {\tiny $2C_{13}$};
\node (C5) at (90:-1.8cm) {\tiny $3C_{03}$};
\node (C0) at (90:1.8cm) {\small $C_2' $} ;
\node [xshift=-1.8cm] (C3) at  (90:-1.8cm){\small $C_0'$};
\node  (C6) at (90:-2*1.8cm) {\tiny $2C_{34}$};
\node  (C4) at (90:-3*1.8cm) {\small $C_3'''$};
\draw (C2)--(C0);
\draw (C2)--(C5)--(C6)--(C4);
\draw (C5)--(C3);
\end{tikzpicture}}

\\
\hline
\end{tabular}

\end{center}

\begin{table}[htb]
\begin{center}
\scalebox{.9}{
\begin{tabular}{|>{\centering  $\displaystyle}c<{$}| >{\centering}m{30mm}<{}| >{\centering}m{30mm}<{}  | >{\centering}m{30mm}<{}| c }
\cline{1-4}
Locus & Type I & Type II & Type III &\\
\cline{1-4}
& & & &\\
w=a_1=0 &
\scalebox{1}{\begin{tikzpicture}[every node/.style={circle,draw, minimum size= 5 mm}, scale=.8]
\node (C2) at (0,0) {\tiny $2C_{14}$};
\node (C5) at (90:-1.8cm) { \tiny $2C_{24}$};
\node (C0) at (45+90*0:1.8cm) { $C_0' $} ;
\node (C1) at (45+90*1:1.8cm) { $C_4'$};
\node [yshift=-1.8cm] (C3) at  (45+90*2:1.8cm){ $C_2'$};
\node [yshift=-1.8cm] (C4) at (45+90*3:1.8cm) {$C_3'$};
\draw (C2)--(C0);
\draw (C2)--(C1);
\draw (C2)--(C5);
\draw (C5)--(C3);
\draw (C5)--(C4);
\end{tikzpicture}}
& 
\scalebox{1}{\begin{tikzpicture}[every node/.style={circle,draw, minimum size= 4 mm}, scale=.8]
\node (C2) at (0,0) {\tiny $2C_{14}$};
\node (C5) at (90:-1.8cm) {\tiny $2C_{13}$};
\node (C0) at (45+90*0:1.8cm) {\small $C_0' $} ;
\node (C1) at (45+90*1:1.8cm) {\small $C_4'$};
\node [yshift=-1.8cm] (C3) at  (45+90*2:1.8cm){\small $C_2'$};
\node [yshift=-1.8cm] (C4) at (45+90*3:1.8cm) {\small $C_3'$};
\draw (C2)--(C0);
\draw (C2)--(C1);
\draw (C2)--(C5);
\draw (C5)--(C3);
\draw (C5)--(C4);
\end{tikzpicture}}

&
\scalebox{1}{\begin{tikzpicture}[every node/.style={circle,draw, minimum size= 4 mm}, scale=.8]
\node (C2) at (0,0) {\tiny $2C_{03}$};
\node (C5) at (90:-1.8cm) {\tiny $2C_{13}$};
\node (C0) at (45+90*0:1.8cm) {\small $C_0' $} ;
\node (C1) at (45+90*1:1.8cm) {\small $C_{4}'$};
\node [yshift=-1.8cm] (C3) at  (45+90*2:1.8cm){\small $C_{2}'$};
\node [yshift=-1.8cm] (C4) at (45+90*3:1.8cm) {\small $C_3'$};
\draw (C2)--(C0);
\draw (C2)--(C1);
\draw (C2)--(C5);
\draw (C5)--(C3);
\draw (C5)--(C4);
\end{tikzpicture}} &\\
& & &&\\
\cline{1-4} 
& && &\\
w=a_1=a_{3,2}=0&
\scalebox{1} {\begin{tikzpicture}[every node/.style={circle,draw, minimum size= 4 mm}, scale=.8]
\node (C2) at (0,0) {\tiny $2C_{14}$};
\node (C5) at (90:-1.8cm) {\tiny $2C_{24}$};
\node (C0) at (45+90*0:1.8cm) {\small $C_0' $} ;
\node (C1) at (45+90*1:1.8cm) {\small $C_4'$};
\node (C6) at (90:-2*1.8cm) {\tiny $2C_{23}$};
\node [yshift=-1.8*2cm] (C3) at  (45+90*2:1.8cm){\scalebox{.7}{$C_3^{(1)}$}};
\node [yshift=-1.8*2cm] (C4) at (45+90*3:1.8cm){\scalebox{.7}{$C_3^{(2)}$}};
\draw (C2)--(C0);
\draw (C2)--(C1);
\draw (C2)--(C5);
\draw (C6)--(C3);
\draw (C6)--(C4);
\draw (C5)--(C6);
\end{tikzpicture}}
& 
\scalebox{1}{\begin{tikzpicture}[every node/.style={circle,draw, minimum size= 4 mm}, scale=.8]
\node (C2) at (0,0) {\tiny $2C_{14}$};
\node (C5) at (90:-1.8cm) {\tiny $2C_{13}$};
\node (C0) at (45+90*0:1.8cm) {\small $C_0' $} ;
\node (C1) at (45+90*1:1.8cm) {\small $C_4'$};
\node (C6) at (90:-2*1.8cm) {\tiny $2C_{23}$};
\node [yshift=-1.8*2cm] (C3) at  (45+90*2:1.8cm){\scalebox{.7}{$C_3^{(1)}$}};
\node [yshift=-1.8*2cm] (C4) at (45+90*3:1.8cm) {\scalebox{.7}{$C_3^{(2)}$}};
\draw (C2)--(C0);
\draw (C2)--(C1);
\draw (C2)--(C5);
\draw (C6)--(C3);
\draw (C6)--(C4);
\draw (C5)--(C6);
\end{tikzpicture}}
&
\scalebox{1}{\begin{tikzpicture}[every node/.style={circle,draw, minimum size= 4 mm}, scale=.8]
\node (C2) at (0,0) {\tiny $2C_{03}$};
\node (C5) at (90:-1.8cm) {\tiny $2C_{13}$};
\node (C0) at (45+90*0:1.8cm) {\small $C_0' $} ;
\node (C1) at (45+90*1:1.8cm) {\small $C_4'$};
\node (C6) at (90:-2*1.8cm) {\tiny $2C_{23}$};
\node [yshift=-1.8*2cm] (C3) at  (45+90*2:1.8cm){\scalebox{.7}{$C_3^{(1)}$}};
\node [yshift=-1.8*2cm] (C4) at (45+90*3:1.8cm) {\scalebox{.7}{ $C_3^{(2)}$}};
\draw (C2)--(C0);
\draw (C2)--(C1);
\draw (C2)--(C5);
\draw (C6)--(C3);
\draw (C6)--(C4);
\draw (C5)--(C6);
\end{tikzpicture}}&\\ & & && \\
\cline{1-4}
& & &&\\
 w=a_1=a_{2,1}=0 &

 {\begin{tikzpicture}[every node/.style={circle,draw, minimum size= 4 mm}, scale=.8]
\node (C2) at (0,0) {\tiny $2C_{14}$};
\node (C5) at (90:-1.8cm) {\tiny $3C_{24}$};
\node (C0) at (90:1.8cm) {\small $C_0' $} ;
\node [xshift=-1.8cm] (C3) at  (90:-1.8cm){\small $C_2'$};
\node  (C6) at (90:-2*1.8cm) {\tiny $2C_{34}$};
\node  (C4) at (90:-3*1.8cm) {\small $C_3'''$};
\draw (C2)--(C0);
\draw (C2)--(C5)--(C6)--(C4);
\draw (C5)--(C3);
\end{tikzpicture}}

& 
\scalebox{1}{\begin{tikzpicture}[every node/.style={circle,draw, minimum size= 4 mm}, scale=.8]
\node (C2) at (0,0) {\tiny $2C_{14}$};
\node (C5) [xshift=-1.8cm]  at (90:-1.8*0 cm) {\tiny $2C_{13}$};
\node (C0) at (90:1.8cm) {\small $C_0' $} ;
\draw (- 1.1cm,0)--+(0,-1.2cm);
\node [xshift=-1.8cm] (C3) at  (90:1.8cm){\small $C_2'$};
\node  (C6) [xshift=-.9 cm] at (90:-1.8cm) {\tiny $2C_{34}$};
\node [xshift=-.9 cm, yshift=-1.8cm] (C4) at (90:-1.8cm) {\small $C_3'''$};
\draw (C2)--(C0);
\draw (C4)--(C6);
\draw (C2)--(C5);
\draw (C5)--(C3);
\end{tikzpicture}}

&

\scalebox{1}{\begin{tikzpicture}[every node/.style={circle,draw, minimum size= 4 mm}, scale=.8]
\node (C2) at (0,0) {\tiny $2C_{13}$};
\node (C5) at (90:-1.8cm) {\tiny $3C_{03}$};
\node (C0) at (90:1.8cm) {\small $C_2' $} ;
\node [xshift=-1.8cm] (C3) at  (90:-1.8cm){\small $C_0'$};
\node  (C6) at (90:-2*1.8cm) {\tiny $2C_{34}$};
\node  (C4) at (90:-3*1.8cm) {\small $C_3'''$};
\draw (C2)--(C0);
\draw (C2)--(C5)--(C6)--(C4);
\draw (C5)--(C3);
\end{tikzpicture}}

&
\\
\cline{1-4}

\end{tabular}}
\end{center}
\caption{Fibers of toric resolutions of type I, II and III. Here $w=e_0 e_1 e_2 e_3 e_4$ the equation of the divisor over which we have the I$_5^s$ fiber. The fiber enhancement I$_5^s\rightarrow$I$_6^s$ over the codimension two locus $w=P=0$ is not presented here but in Table \ref{type123P} for all three types of toric resolutions. As we will see the following subsections, the type I can be realized as $\mathscr{T}^+_{1+}\cong\mathscr{T}^+_{2+}\cong\mathscr{B}_{1,3}$, type II as $\mathscr{B}_{2,3}$, and type III as $\mathscr{B}_{2,3}^1$.}\label{type123}
\end{table}
\clearpage

\subsection{Explicit constructions of models of type I and II}\label{section:toric2}

In this subsection we will present the explicit sequences of blow ups for the type I and type II toric resolutions, and identify them as the resolutions in our network in Figure \ref{network}. The construction for type III will be discussed in Section \ref{section:toric3}. The sequences of blow ups for these four resolutions are:

The relevant resolutions in Figure \ref{network} are  $\mathscr{T}^+_{1+}$, $\mathscr{T}^+_{2+}$, $\mathscr{B}_{1,3}$ and $\mathscr{B}_{2,3}$. We will relabel the $e_i$ to comply with the convention in the toric resolution \eqref{ET}.
\begin{enumerate} 
\item \begin{equation}
\begin{tikzcd}[column sep=huge]
\mathscr{E}_0 \arrow[leftarrow]{r}{ \textstyle{( x,y,e_0 |e_1 )}} & 
\mathscr{E}_1 \arrow[leftarrow]{r}{ \textstyle{( y,e_1 |e_4)}} & 
\mathscr{T}^+ \arrow[leftarrow]{r}{ \textstyle{( x,e_4|e_2 )}} & 
\mathscr{T}^+_{1} \arrow[leftarrow]{r}{ \textstyle{( y,e_2 |e_3 )}} & \mathscr{T}^+_{1+}
\end{tikzcd}
\end{equation}
The defining equation for $\mathscr{T}^+_{1+}$ is precisely \eqref{ET}. It lives in the following ambient space parametrized by the projective coordinates:
\begin{equation}\label{typeIambient1}
[e_1 e_4 e_2^2  e_3^2x:e_1  e_2^2 e_3^3 e_4^2 y:z]\   \    [e_2e_3x:e_2e_3^2 e_4 y:e_0]\  \    [e_3 y:e_1]\  \   [x:e_4  ]\  \   [y:e_2]
\end{equation}
From the ambient space parametrization, one can read off the relevant SR ideals $e_0e_3, e_1e_3$ in \eqref{SRIdeal2} that defines the type I resolution.

\item 
\begin{equation}
\begin{tikzcd}[column sep=huge]
\mathscr{E}_0 \arrow[leftarrow]{r}{ \textstyle{( x,y,e_0 |e_1 )}} & 
\mathscr{E}_1 \arrow[leftarrow]{r}{ \textstyle{( y,e_1 |e_4)}} & 
\mathscr{T}^+ \arrow[leftarrow]{r}{ \textstyle{( x,y,e_4|e_3 )}} & 
\mathscr{T}^+_2 \arrow[leftarrow]{r}{ \textstyle{( x,e_4 |e_2 )}} & \mathscr{T}^+_{2+}
\end{tikzcd}
\end{equation}
with projective coordinates 
\begin{align}\label{typeIambient2}
~~~~~~~~[e_1 e_2^2 e_3^2 e_4x:e_1   e_2^2  e_3^3 e_4^2 y:z]\ \  [ e_2 e_3 x:e_2 e_3^2 e_4 y:e_0]\  \   [ e_3  y:e_1] \  \  [e_2x: y:e_2 e_4]\   \ [x:e_4].
\end{align}
$\mathscr{T}^+_{2+}$ again shares the same defining equation \eqref{ET} and the same SR ideal as the type I resolution. Hence it is also a toric resolution of type  I. 

\item 

\begin{equation}
\begin{tikzcd}[column sep=huge]
\mathscr{E}_0 \arrow[leftarrow]{r}{ \textstyle{( x,y,e_0 |e_1 )}} & 
\mathscr{E}_1 \arrow[leftarrow]{r}{ \textstyle{( x,y,e_1 |e_2)}} & 
\mathscr{B} \arrow[leftarrow]{r}{ \textstyle{( y,e_1|e_4 )}} & 
\mathscr{B}_{1,\bullet} \arrow[leftarrow]{r}{ \textstyle{( y,e_2 |e_3 )}}[below] {\textstyle{(s, t|e_3)}} &\mathscr{B}_{1,3}
\end{tikzcd}
\end{equation}
with projective coordinates 
\begin{equation}\label{typeIambient3}
~~~~~~~~[e_1 e_2^2 e_3^2 e_4x:e_1   e_2^2  e_3^3 e_4^2 y:z] \ \ [e_2 e_3 x:e_2 e_3^2 e_4 y:e_0] \ \ [x:e_3 e_4 y:e_1e_4]\ \ [e_3 y:e_1] \ \ [y:e_2]
\end{equation}
$\mathscr{B}_{1,3}$ again shares the same defining equation \eqref{ET} and the same SR ideal as the type I resolution. Hence it is also the type I resolution as already mentioned in Section \ref{section:EY2}.

\item 

\begin{equation}\label{toric.res.2}
\begin{tikzcd}[column sep=huge]
\mathscr{E}_0 \arrow[leftarrow]{r}{ \textstyle{( x,y,e_0 |e_1 )}} & 
\mathscr{E}_1 \arrow[leftarrow]{r}{ \textstyle{( x,y,e_1 |e_2)}} & 
\mathscr{B} \arrow[leftarrow]{r}{ \textstyle{( y,e_2|e_3 )}} & 
\mathscr{B}_{3,\bullet} \arrow[leftarrow]{r}{ \textstyle{( y,e_1 |e_4 )}}[below]{\textstyle{or\, (s, t|e_3)}}& \mathscr{B}_{2,3}
\end{tikzcd}
\end{equation}
with projective coordinates 
\begin{equation}
~~~~~~~~[e_1 e_2^2 e_3^2 e_4x:e_1   e_2^2  e_3^3 e_4^2 y:z]\quad  [e_2 e_3 x:e_2 e_3^2 e_4 y:e_0]\quad [x:e_3 e_4 y:e_1e_4] \quad [e_4y:e_2]\quad [y:e_1]
\end{equation}
From the ambient space parametrization, one can read off the relevant SR ideals $e_0e_3, e_2e_4$ in \eqref{SRIdeal2} that defines the type II resolution.
 We notice that the first two blow ups are the same as those of \cite{EY}. The last two blow ups show that this resolution is exactly the same as the resolution denoted as  $(1,2)(1,1)$  in \cite{LN,HLN}. 
\end{enumerate}

The three resolutions $\mathscr{T}^+_{1+}, \mathscr{T}^+_{2+}, \mathscr{B}_{1,3}$ in Figure \ref{network} share the same defining equation $\eqref{ET}$ and the same SR ideal, but they live in seemingly different projective spaces parametrized by \eqref{typeIambient1}, \eqref{typeIambient2}, \eqref{typeIambient3}. In fact, the three ambient spaces \eqref{typeIambient1}, \eqref{typeIambient2}, \eqref{typeIambient3} can be shown to be isomorphic, therefore $\mathscr{T}^+_{1+}, \mathscr{T}^+_{2+}, \mathscr{B}_{1,3}$ should be identified as one single resolution in the network of resolutions in Figure \ref{network}:
\begin{align}
\text{type I}:~\mathscr{T}^+_{1+}\cong\mathscr{T}^+_{2+}\cong \mathscr{B}_{1,3}.
\end{align}
In addition to the above isomorphism, there are many more in the network of resolutions in Figure \ref{network}. We summarize all the isomorphisms between different sequences of resolutions in Section \ref{section:iso}. We do not present the detailed calculation of the isomorphism in the current paper as it is completely parallel to the $SU(4)$ example in Section B.4.1 of \cite{ESY}. The isomorphism between the ambient spaces \eqref{typeIambient1}, \eqref{typeIambient2}, \eqref{typeIambient3} is also a direct consequence of the toric diagram discussed in Section \ref{section:toric4}.

On the other hand, the type II toric resolution is identified  as the resolution $\mathscr{B}_{2,3}$, which is obtained by a sequence of blow ups:
\begin{align}
\text{type II}:~ \mathscr{B}_{2,3}.
\end{align}

In summary, we have provided explicit constructions for the toric resolutions of type I and type II in terms of sequences of blow ups. We have included the type I and type II resolutions into a unified framework of blow ups summarized in Figure \ref{network}. In particular, the type I and type II resolutions are \textit{not} new resolutions in addition to the six resolutions $\mathscr{B}_{i,j}$ found in \cite{EY}; rather, they can be realized as $\mathscr{B}_{1,3}$ and $\mathscr{B}_{2,3}$, respectively.

\subsection{Type III from a flop of $\mathscr{B}_{2,3}$}\label{section:toric3}

In this subsection we will construct the type III toric resolution from a flop of $\mathscr{B}_{2,3}$ (type II). Indeed, from the representation theory prediction Figure \ref{SU5Coulomb1}, the type III model (which corresponds to the subchamber 2) is related to $\mathscr{B}_{2,3}$ (which corresponds to the subchamber 1) by a flop. Later in Section \ref{section:toric5}, we will realize type III as a sequence of weighted blow up. Here we follow a similar idea in \cite{TY}.

The flop of $\mathscr{B}_{2,3}$ is obtained by first blowing up $\mathscr{B}_{2,3}$ and then blowing down to obtain type III. Recall that the resolution $\mathscr{B}_{2,3}$ (type II) is obtained by the following sequence of blow ups
\begin{equation}
\begin{tikzcd}[column sep=huge]
\mathscr{E}_0 \arrow[leftarrow]{r}{ \textstyle{( x,y,e_0 |e_1 )}} & 
\mathscr{E}_1 \arrow[leftarrow]{r}{ \textstyle{( x,y,e_1 |e_2)}} & 
\mathscr{B} \arrow[leftarrow]{r}{ \textstyle{( y,e_2|e_3 )}} & 
\mathscr{B}_{2,\bullet} \arrow[leftarrow]{r}{ \textstyle{( y,e_1 |e_4 )}} & \mathscr{B}_{2,3}.
\end{tikzcd}
\end{equation}
It can be written as
\begin{align}
\mathscr{B}_{2,3}:
y( e_3 e_4y + a_1 x  + a_{3,2}   e_0^2 e_1 e_4) 
=
e_1 e_2(x^3 e_2 e_3 + a_{2,1} x^2 e_0  + a_{4,3} x e_0^3 e_1  e_4 
+ a_{6,5} e_0^5 e_1^2 e_4^2),
\end{align}
in the ambient space parametrized by
\begin{equation}
[e_1 e_2^2 e_3^2 e_4x:e_1   e_2^2  e_3^3 e_4^2 y:z]\quad  [e_2 e_3 x:e_2 e_3^2 e_4 y:e_0]\quad [x:e_3 e_4 y:e_1e_4] \quad [e_4y:e_2]\quad [y:e_1].
\end{equation}

The locus we wish to blow up in $\mathscr{B}_{2,3}$ is the codimension two locus defined by  $C_{14}: e_1=e_4=a_1=0$. $C_{14}$ is parametrized by:
\begin{equation}
[0:0:z]\quad  [e_2 e_3 x:0:e_0]\quad [x:0:0] \quad [0:e_2]\quad [y: 0]
\end{equation}
in the ambient space. 

Let us now blow up $\mathscr{B}_{2,3}$ along $C_{14}$
\begin{align}
\mathscr{B}_{2,3} \xleftarrow { (e_1,e_4|e_5)} \hat{\mathscr{B}}
\end{align}
which gives
\begin{align}
\hat{\mathscr{B}}:
y( e_3 e_4 e_5 y + a_1 x  + a_{3,2}   e_0^2 e_1 e_4 e_5^2) 
=
e_1  e_5 e_2(x^3 e_2 e_3 + a_{2,1} x^2 e_0  + a_{4,3} x e_0^3 e_1  e_4 e_5^2
+ a_{6,5} e_0^5 e_1^2 e_4^2 e_5^4),
\end{align}
with projective coordinates
\begin{align}
\begin{split}
&[e_1 e_2^2 e_3^2 e_4 e_5^2 x:e_1   e_2^2  e_3^3 e_4^2  e_5^3 y:z]  [e_2 e_3 x:e_2 e_3^2 e_4 e_5 y:e_0] [x:e_3 e_4 e_5 y:e_1e_4 e_5^2 ] \\
&
 [e_4e_5 y:e_2] [y:e_1 e_5]  [e_1:e_4].
 \end{split}
\end{align}
The variety $\hat{\mathscr{B}}$ is nonsingular and isomorphic to $\mathscr{B}_{2,3}$ away from  the center of the blow up.  
Now the codimension two locus $C_{14}$ is replaced in  $\hat{\mathscr{B}}$  by the divisor 
\begin{equation}
C_{14}\rightarrow D_5:e_5=a_1=0.
\end{equation}
which is a $\mathbb{F}_p$-fibration over $w=a_1=0$ in the base for some integer $p$. The integer $p$ will be determined to be zero in a moment. Over that locus in the base the projective coordinates are:
\begin{equation}
[0:0:z]\quad[e_2e_3 x:0: e_0]\quad[x:0:0]\quad[0:e_2] \quad [y:0]\quad[e_1: e_4].
\end{equation}
If we blow down $\hat{\mathscr{B}}$ by contracting the new  $\mathbb{P}^1:[e_1:e_4]$ in $D_5$, it results in the  original type II model $\mathscr{B}_{2,3}$. 
 To obtain the flop of $\mathscr{B}_{2,3}$, we should therefore contract the other $\mathbb{P}^1:[e_2e_3x:0:e_0]$ in $D_5$.

Before performing the alternative blow down, let us first summarize the scalings for the variables of $\hat{\mathscr{B}}$: 
\begin{equation}
\begin{tabular}{|c|c|c|c|c|c|c|c|c|}
\cline{2-9}
\multicolumn{1}{c|}{}& $x$ & $y$ & $e_0$ & $e_1$ & $e_2$ & $e_3$ & $e_4$ & $e_5$\\
\hline 
$\ell_1$ & 1 & 1 & 1& -1 &  &  &  &  \\
\hline
$\ell_2$ & 1 & 1 &  &  1 & -1 &  &  &  \\
\hline
$\ell_3$ &  & 1 & &  & 1 & -1 & &  \\
\hline
$\ell_4$ &  & 1 & & 1 &  &  & -1 &  \\
\hline
$\ell_5$ &  &  & &  1 &  &  & 1 & -1 \\
\hline
\end{tabular}
\end{equation}
It is useful to replace the previous scalings $\ell_i$ by the following equivalent scalings $\ell'_i$: 
\begin{equation}
\begin{pmatrix}
\ell_1' \\
\ell_2'\\
\ell_3'\\
\ell_4'\\
\ell_5'
\end{pmatrix}=
\begin{pmatrix}
1& -1& -1& 1& 1\\
0& 1&1 & -2& 0\\
0& 0& 1& -1&0 \\
0&0 &0 & 1& 0\\
0& 0 & 0 & 0 &1 
\end{pmatrix}
\begin{pmatrix}
\ell_1 \\
\ell_2\\
\ell_3\\
\ell_4\\
\ell_5
\end{pmatrix}
\end{equation}
The scalings for each variable with respect to $\ell'_i$ are now:
   \begin{equation}
\begin{tabular}{|l|c|c|c|c|c|c|c|c|}
\cline{2-9}
\multicolumn{1}{c|}{}& $x$ & $y$ & $e_0$ & $e_1$ & $e_2$ & $e_3$ & $e_4$ & $e_5$\\
\hline 
$\ell_1'$& 0 & 0  & 1& 0 &0  & 1 & 0  & -1  \\
\hline
$\ell_2'$& 1 &0   &  0& -1 & 0 &-1  & 2 & 0 \\
\hline
$\ell_3'$ &  0& 0 & 0 &  -1& 1 & -1 & 1& 0 \\
\hline
$\ell_4'$ & 0 & 1 &0 & 1 &  0 & 0  & -1 &0   \\
\hline
$\ell_5'$ &  0& 0 & 0&  1 & 0 & 0 & 1 & -1 \\
\hline
\end{tabular}
\end{equation}
We are interested in the locus of $D_5$ where $x$, $y$, $e_2$ cannot be zero. Therefore we can rescale  $x=y=e_2=1$ by the scaling freedom $\ell_2', \ell_4', \ell_3'$, respectively.
 We are  then left with the scalings $\ell_1', \ell_5'$  under which $x$, $y$ and $e_2$ are uncharged:
\begin{equation}
\begin{tabular}{|l|c|c|c|c|c|c|c|c|}
\cline{2-9}
\multicolumn{1}{c|}{}& $x$ & $y$ & $e_0$ & $e_1$ & $e_2$ & $e_3$ & $e_4$ & $e_5$\\
\hline 
$\ell_1'$ & 0 & 0  & 1& 0 &0  & 1 & 0  & -1  \\
\hline
$\ell_5'$ &  0& 0 & 0&  1 & 0 & 0 & 1 & -1 \\
\hline
\end{tabular}
\end{equation}
Observe that $D_5 :e_5=a_1=0$ is parametrized by $(e_3,e_0, e_1, e_4)$ in this patch. From the scalings, we recognize $D_5$ to be a $\mathbb{F}_0\simeq \mathbb{P}^1\times \mathbb{P}^1$ in the fiber with projective coordinates $[e_3:e_0][e_1:e_4]$.

To obtain the type III model, we contract the $\mathbb{P}^1:[e_3:e_0]$ by considering the following projection map \cite{TY}:
\begin{equation}
([e_3:e_0] , [e_1: e_4], e_5)\longrightarrow ([e_1: e_4], \tilde{e}_3=e_3 e_5 ,\tilde{e}_0= e_0 e_5)
\end{equation}
The blow down variety, which we will denote by $\mathscr{B}_{2,3}^1$ since it is related to $\mathscr{B}_{2,3}$ by a flop, is then
\begin{align}\label{B231}
{\mathscr{B}}_{2,3}^1:
y( \tilde{e}_3 e_4 y + a_1 x  + a_{3,2}   \tilde{e}_0^2 e_1 e_4 ) 
-
e_1  e_2(x^3 e_2 \tilde{e}_3 + a_{2,1} x^2 \tilde{e}_0  + a_{4,3} x \tilde{e}_0^3 e_1  e_4 
+ a_{6,5} \tilde{e}_0^5 e_1^2 e_4^2 )=0,
\end{align}
with projective coordinates
  \begin{equation}\label{typeIII.eq1}
\begin{tabular}{|l|c|c|c|c|c|c|c|}
\cline{2-8}
\multicolumn{1}{c|}{}& $x$ & $y$ & $e_1$ & $e_2$ & $e_4$ & $\tilde{e}_3$ & $\tilde{e}_0$\\
\hline 
$\ell_2'$& 1 & 0  & -1& 0 & 2 & -1  & 0  \\
\hline
$\ell_3'$ &0   &  0& -1 & 1 &1  &-1 & 0 \\
\hline
$\ell_4'$ &  0& 1 & 1 &  0 & -1 &  0 & 0 \\
\hline
$\ell_5'$ & 0 & 0 & 1 &  0 & 1 & -1 &-1  \\
\hline
\end{tabular}
\end{equation}
Even though we performed the blow down in the patch $x=y=e_2=1$, it is obvious to take closure to the whole space as presented above. 

We notice that the defining equation \eqref{B231} is exactly the same as the defining equation for the toric resolution \eqref{ET}. It remains to check that the SR ideal of $\mathscr{B}_{2,3}^1$ contains $e_1e_4,e_2e_4$ for it be of type III (see \eqref{SRIdeal}). We can determine the SR ideal of $\mathscr{B}_{2,3}^1$ by following the sequence of blow up and blow down from $\mathscr{B}_{2,3}$.

To begin with, the SR ideal for $\hat{\mathscr{B}}$ contains
\begin{align}
e_0e_3,~e_2e_4,~e_1e_4
\end{align}
where the first two are inherited from $\mathscr{B}_{2,3}$ while $e_1e_4$ comes the blow up $\mathscr{B}_{2,3}\xleftarrow{(e_1,e_4|e_5)}\hat{\mathscr{B}}$. Next, in blowing down $\hat{\mathscr{B}}$ to $\mathscr{B}_{2,3}^1$, we define $\tilde e_3=e_3e_5, \tilde e_0=e_0e_5$ and ``forget" the original variables $e_3$ and $e_0$. Note that $\tilde e_3$ and $\tilde e_0$ can be zero at the same time by setting $e_5=0$, so $\tilde e_0\tilde e_3$ is \textit{not} part of the SR ideal of $\mathscr{B}_{2,3}^1$. It follows that the SR ideal of $\mathscr{B}_{2,3}^1$ contains
\begin{align}
e_2e_4,~e_1e_4,
\end{align}
and we conclude that 
\begin{align}
\text{type III}:~\mathscr{B}_{2,3}^1.
\end{align}
In summary, we constructed the type III model $\mathscr{B}_{2,3}^1$ from a flop from the type II model $\mathscr{B}_{2,3}$. Since our flop is the composition of an explicit blow up and a blow down, the projectivity of the resolution $\mathscr{B}_{2,3}^1$ is ensured. In Section \ref{section:toric5}, we will give a more direct construction of type III in terms of a sequence of weighted blow ups.

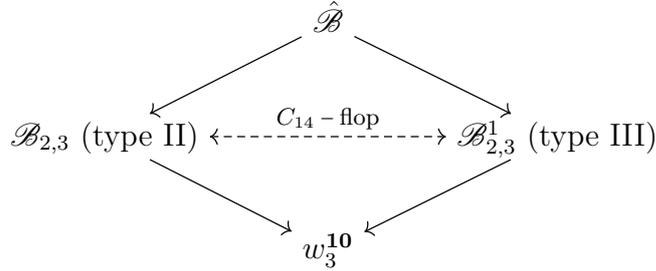
\begin{figure}[htb]
\begin{tikzcd}[column sep=normal]
& \hat{\mathscr{B}}   {\arrow[rightarrow]{ld}[above=2mm]{}  }{\arrow[rightarrow]{dr}[above=2mm]{}}& \\
\mathscr{B}_{2,3} ~\text{(type II)} \arrow[leftrightarrow, dashed]{rr}{ C_{14}-\text{flop}}& & \mathscr{B}^1_{2,3}~\text{(type III)} \\
& w_3^{\bf10} \arrow[leftarrow]{ul}  \arrow[leftarrow]{ur}&
\end{tikzcd}
\caption{Flop of a codimension two locus as the composition of a blow up and a blow down. The common blow down of type II ($\mathscr{B}_{2,3}$) and type III ($\mathscr{B}_{2,3}^1$) is a partial resolution corresponding to the wall $w_3^{\bf10}$ as can be seen from Figure \ref{SU5Coulomb1}.  }
\end{figure}

In the next subsection we will provide the toric description that unifies all the discussion of type I, II, and III models.

\subsection{Toric Descriptions of Type I, II, III}\label{section:toric4}

The Weierstrass models $\mathscr{E}_0$ has the description as a hypersurface in the $\mathbb{P}^2$-bundle with homogeneous coordinates $[x:y:z]$ over a base $B$. As always, we will restrict ourselves to the patch $z\neq 1$ where all the nontrivial fiber enhancements take place. If we further restrict to the normal direction of the divisor $e_0=0$ on the base $B$ parametrized $e_0$, locally the Weierstrass model $\mathscr{E}_0$ is described by a hypersurface in this ambient space $\mathbb{C}^3:(x,y,e_0)$. 

To resolve the singularity of $\mathscr{E}_0$, we successively blow up the ambient space $\mathbb{C}^3$ along certain ideals. The blow up of $\mathbb{C}^3$ has a concise toric geometry description. In particular, for the type I ($\mathscr{T}^+_{1+}\cong \mathscr{T}^+_{2+}\cong \mathscr{B}_{1,3}$), the type II ($\mathscr{B}_{2,3}$), and the type III ($\mathscr{B}_{2,3}^1$) resolutions, they are defined by the \textit{same} hypersurface equation \eqref{ET} in \textit{different} blow ups of the ambient $\mathbb{C}^3$. Their ambient spaces are related by toric flops as we will describe in the following.

\subsubsection{Fans}

Here we give a lightening review on the fan diagram for toric geometry \cite{Mirror}. 

Let $N\cong \mathbb{Z}^r$ be a lattice of rank $r$ and set $N_{\mathbb{R}} = N\otimes \mathbb{R}$. A cone $\sigma\subset N_{\mathbb{R}}$ is defined by the set
\begin{align}
\sigma = \left\{ a_1 v_1 +\cdots  +a_k v_k |a_i\ge0 \right\}
\end{align}
generated by a finite set of lattice vectors $v_i$ in $N$ such that $\sigma \cap (-\sigma)=\{0\}$.

A collection $\Sigma$ of cones in $N_{\mathbb{R}}$ is called a fan if
\begin{enumerate}
  \item each face of a cone in $\Sigma$ is also a cone in $\Sigma$, and
  \item the intersection of two cones in $\Sigma$ is a face of each.
\end{enumerate}

Given a fan with edges $v_1,\cdots,v_n$ in $N$, we can construct the toric geometry as follows. First we associate a coordinate system $(x_1,\cdots, x_n)$ to every edge. This defines a $\mathbb{C}^n$. Define
\begin{align}
Z(\Sigma) = \cup_I \left\{ (x_1, \cdots , x_n )| x_i=0 ~\forall i\in I\right\}
\end{align}
where $I\subset \{1,\cdots,n\}$ for which $\{x_i | i\in I\}$ does not belong to a cone in $\Sigma$.

Next, we identify all the integers $Q^i_a$ such that
\begin{align}
\sum_{i=1}^n Q^i_a v_i =0
\end{align}
Note that since we have $n$ lattice vectors on a $r$-dimensional lattice $N$, there are going to be $(n-r)$ $Q_{a}$'s.

The toric geometry is then given by
\begin{align}
X_\Sigma = {\mathbb{C}^n -Z(\Sigma) \over G  }
\end{align}
where $G$ is $(n-r)$-dimensional action on $(x_1,\cdots,x_n)$ defined by
\begin{align}
G: (x_1,\cdots, x_n ) \mapsto (\lambda^{Q_a^1}x_1, \cdots,\lambda^{ Q_a^n}x_n).
\end{align}
The set $Z(\Sigma)$ is going to be identified as the SR ideal of the resolution.

\subsubsection{Toric flops between Type I, II, III}

After the review above, we are now ready to give the toric descriptions for the ambient spaces for resolutions of type I, II, III along the normal direction of the divisor $e_0$. We will illustrate the case in details for type II while the other two can be straightforwardly reproduced. 

Starting with the $\mathbb{C}^3:(x,y,e_0)$ for the original Weierstrass model $\mathscr{E}_0$, the fan diagram consists of three linearly independent three-dimensional lattice vectors, say\footnote{We will denote the lattice vectors in the fan diagram by adding vector symbols $\vec{~}$ to the original symbols $x,y,e_0$ etc.},
\begin{align}
\vec{x}=(1,0,0),~\vec{y}=(0,1,0),~\vec{e_0} = (0,0,1).
\end{align}
The only cone in the fan diagram is one generated by the above three vectors. 
Since there is no linear relation between the three vectors, there is no scaling $G$ to mod out.

When we blow up $\mathscr{E}_0$ along the ideal $(x,y,e_0)$ to obtain $\mathscr{E}_1$, we perform the following replacements
\begin{align}
x\rightarrow e_1 x,~~y\rightarrow e_1 y,~~e_0\rightarrow e_1 e_0 ,
\end{align}
with the new $x,y,e_0$ being the homogeneous coordinates of a $\mathbb{P}^2:[x:y:e_0]$ while the exceptional divisor $e_1$ being a section of the $\mathscr{O}_{\mathbb{P}^2}(-1)$ bundle. The scalings can be summarized as:
\begin{align}
\left.\begin{tabular}{|c|c|c|c|c|}
\cline{2-5}
\multicolumn{1}{c|}{} &$x$ &$y$& $e_0$ & $ e_1$  \\\hline  $\ell_1$& 1 & 1 & 1 & -1 \\\hline   \end{tabular}\right.
\end{align}
The SR ideal is $xye_0$, namely, those three coordinates can not vanish at the same time because they are coordinates of a $\mathbb{P}^2$.

From the above scaling, the fan diagram for the blow up $\mathscr{O}_{\mathbb{P}^2}(-1)$ of $\mathbb{C}^3$ can be obtained by adding an additional lattice vector $\vec{e_1}$
\begin{align}
\vec{e_1} = \vec{x}+ \vec{y}+\vec{e_0}.
\end{align}
With this extra lattice vector, there are now three cones  generated by $(\vec{e_1},\vec{x},\vec{y})$, $(\vec{e_1},\vec{x},\vec{e_0})$, and $(\vec{e_1},\vec{e_0},\vec{y})$, respectively. In particular, since $\vec{x},\vec{y},\vec{e_0}$ are not the faces of a single cone, they cannot vanish at the same time. That is, the SR ideal includes $xye_0$.

Now it is clear that each step of blow up will introduce a new lattice vector $\vec{e_i}$ corresponding to the exceptional divisor of the blow up to the fan diagram. The fan diagrams for the type I are presented following three paths $\mathscr{B}_{1,3}$, $\mathscr{T}^+_{2+}$, $\mathscr{T}^+_{1+}$ in Figure \ref{network} are presented in Figure \ref{typeI:fan1}, \ref{typeI:fan2}, and \ref{typeI:fan3}, respectively. From the toric description it is now obvious why the three resolutions $\mathscr{B}_{1,3}$, $\mathscr{T}^+_{2+}$, $\mathscr{T}^+_{1+}$ are identified. The type II resolution and its partial resolutions are shown in 
figure \ref{typeII:fan}. 

For readers' convenience, we list the scalings and the relevant part of the SR ideals\footnote{There is a common SR ideal \eqref{SRIdeal} shared by all three types of resolutions that is not listed here.} of the type I and type II resolutions here\footnote{There are three equivalent ways $\mathscr{T}^+_{1+}$, $\mathscr{T}^+_{2+}$, $\mathscr{B}_{1,3}$ to obtain the type I resolution as discussed in Section \ref{section:toric2}. Here we present the sequence of resolution for $\mathscr{B}_{1,3}$. We have exchanged $e_3$ with $e_4$ compared with the previous text for convenience  of comparison here.} 
\begin{align}
\text{Type I}:~~ &\mathscr{E}_0 \xleftarrow{(x,y,e_0|e_1)}\mathscr{E}_1 \xleftarrow{(x,y,e_1|e_2)} \mathscr{B}\xleftarrow{(y,e_1|e_4)}\mathscr{B}_{1,\bullet}\xleftarrow{(y,e_2|e_3)}\mathscr{B}_{1,3}
  \end{align}
\begin{equation}
\mathscr{B}_{1,3}:~\begin{tabular}{|c|c|c|c|c|c|c|c|}
\cline{2-8}
\multicolumn{1}{c|}{}& $x$ & $y$ & $e_0$ & $e_1$ & $e_2$ & $e_4$ & $e_3$\\
\hline 
$\ell_1$ & 1 & 1 & 1& -1 &  0&0  &0  \\
\hline
$\ell_2$ & 1 & 1 & 0 &  1 & -1 &0  &0  \\
\hline
$\ell_3$ & 0 & 1 &0 &1  & 0& -1 &0  \\
\hline
$\ell_4$ & 0 & 1 &0 & 0&  1& 0 & -1  \\
\hline
\end{tabular}\quad\text{SR:~$e_0e_3,~e_1e_3$}
\end{equation}

\begin{align}
\text{Type II}:~~& \mathscr{E}_0 \xleftarrow{(x,y,e_0|e_1)}\mathscr{E}_1 \xleftarrow{(x,y,e_1|e_2)} \mathscr{B}\xleftarrow{(y,e_2|e_3)}\mathscr{B}_{2,\bullet} \xleftarrow{(y,e_1|e_4)}\mathscr{B}_{2,3}
\end{align}
\begin{equation}
\mathscr{B}_{2,3}:~\begin{tabular}{|c|c|c|c|c|c|c|c|}
\cline{2-8}
\multicolumn{1}{c|}{}& $x$ & $y$ & $e_0$ & $e_1$ & $e_2$ & $e_3$ & $e_4$ \\
\hline 
$\ell_1$ & 1 & 1 & 1& -1 &  0&0  &0   \\
\hline
$\ell_2$ & 1 & 1 & 0 &  1 & -1 &0    &0  \\
\hline
$\ell_3$ & 0 & 1 &0 &0  & 1 & -1  & 0 \\
\hline
$\ell_4$ & 0 & 1 &0 & 1 &  0& 0 & -1   \\
\hline
\end{tabular}
\quad \text{SR:~$e_0e_3,~e_2e_4$}
\end{equation}
For the type III resolution, it is obtained by blowing up type II first and then blowing down. Its scalings and the relevant SR ideal are\footnote{We have removed the tildes on $e_0$ and $e_3$ for convenience of comparison here.}
\begin{equation}
\text{Type III}:~\begin{tabular}{|c|c|c|c|c|c|c|c|}
\cline{2-8}
\multicolumn{1}{c|}{}& $x$ & $y$ & $e_0$ & $e_1$ & $e_2$ & $e_3$ & $e_4$ \\
\hline 
$\ell_2'$ & 1 & 0 & 0& -1 &  0&-1 &2   \\
\hline
$\ell_3'$ & 0 & 0 & 0 &  -1 & 1 &-1   &1  \\
\hline
$\ell_4'$ & 0 & 1 &0 &1  & 0 & 0  & -1 \\
\hline
$\ell_5'$ & 0 & 0 &-1 & 1 &  0& -1 & 1   \\
\hline
\end{tabular}\quad
\text{SR:~$e_1e_4,~e_2e_4$  }
\end{equation}

\begin{figure}[htb]
\begin{center}
\includegraphics[scale=.9]{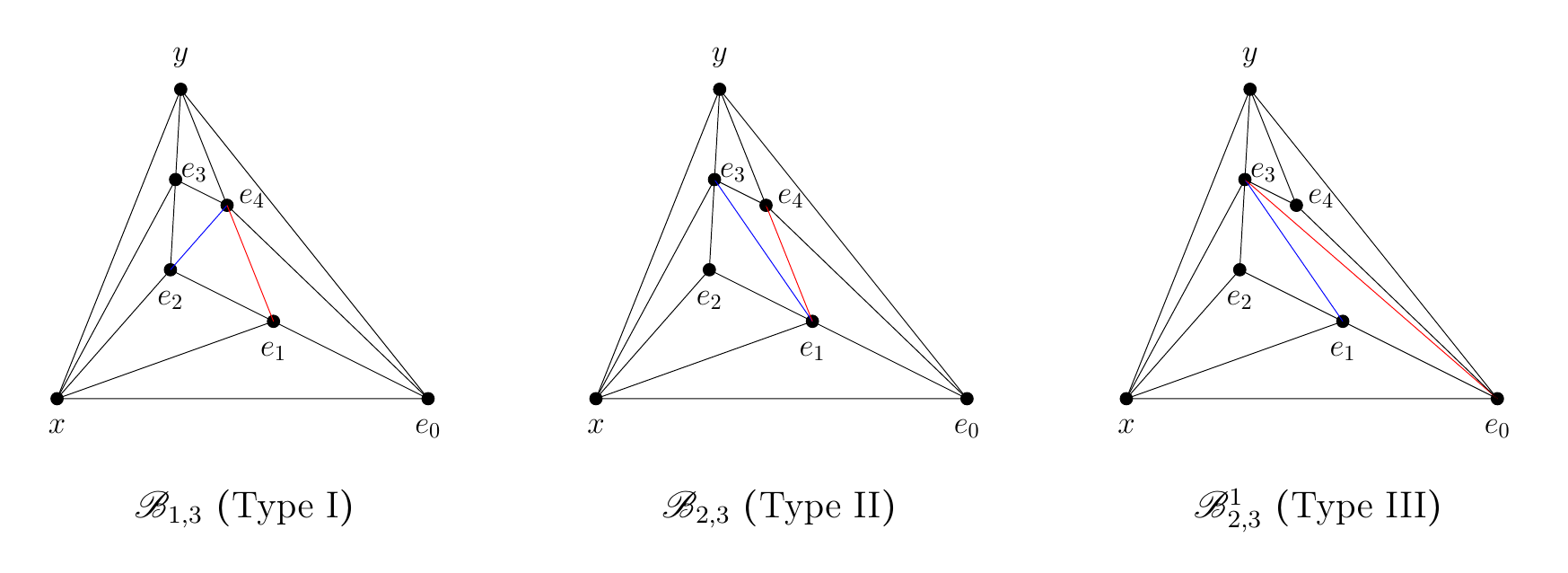}
\end{center}
\caption{Toric diagrams for the ambient spaces along the $x,y,e_0$ directions of the type I, II, and III models. \label{AllTorics}}
\end{figure}

From the scalings and the SR ideal, one can construct the fan diagrams as shown in  figure \ref{AllTorics}. From the fan diagrams, it is then clear that type II resolution is related to type I and III by Atiyah flops.

\begin{figure}[htb]
\begin{center}
\includegraphics[scale=.9]{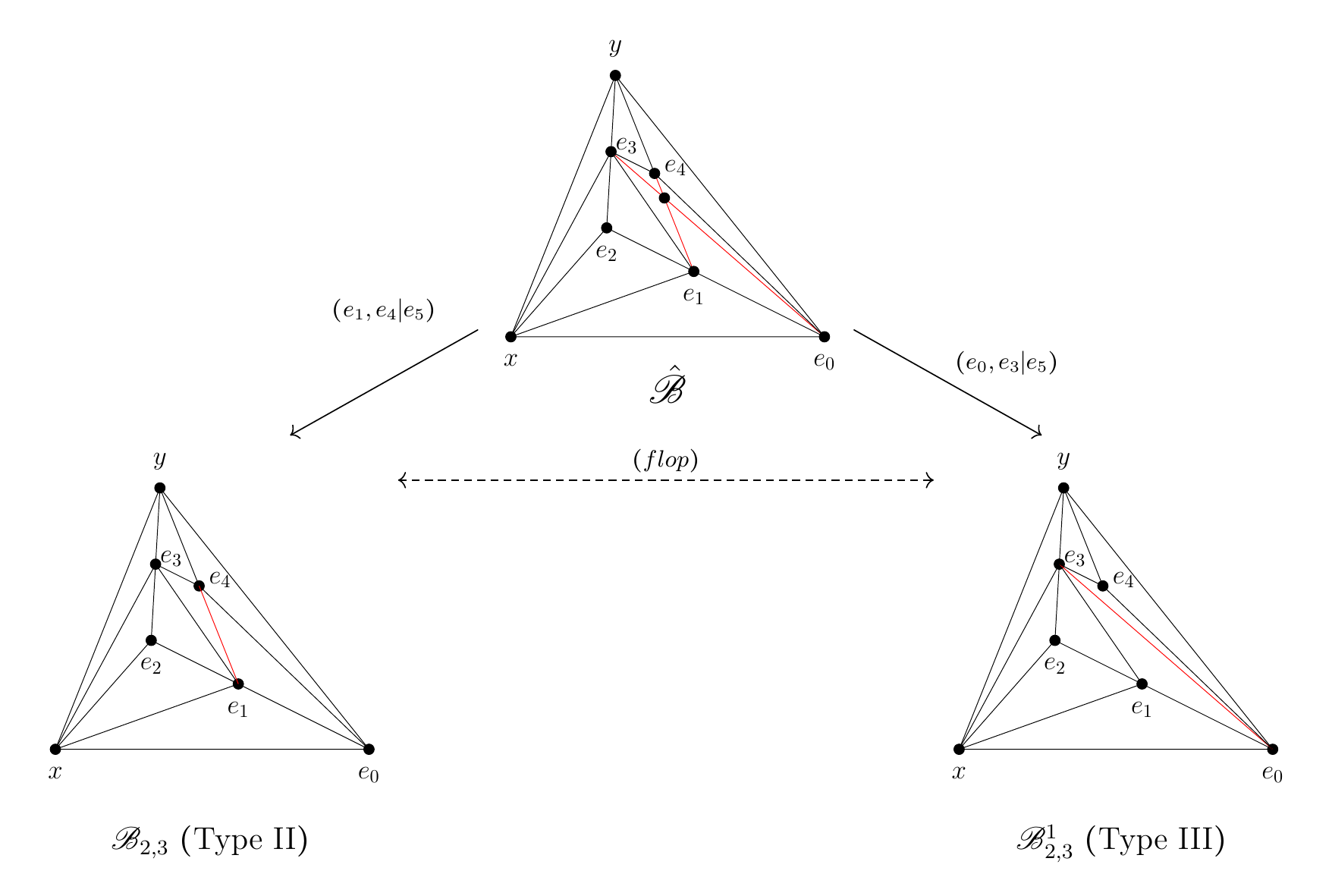}
\end{center}
\caption{Type III is obtained by blowing up $\mathscr{B}_{2,3}$ along the node $\mathbb{P}^1:[e_1:e_4]$ and then blowing down along the $\mathbb{P}^1:[e_0:e_3]$.}
\end{figure}

\begin{center}

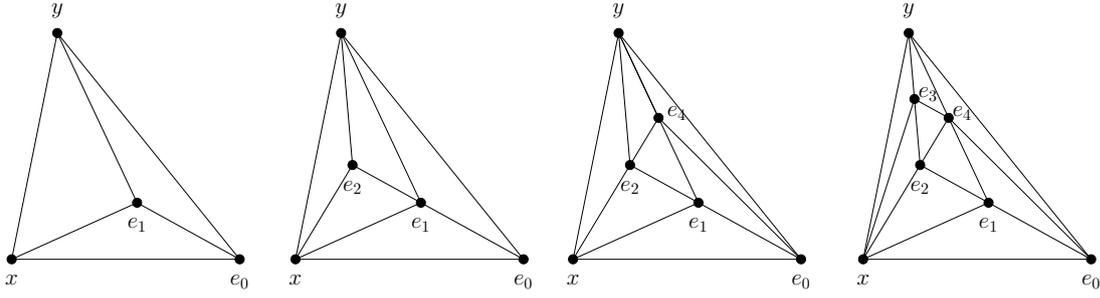
\begin{figure}[htb]
\scalebox{.6}
{\unitlength = .2 cm
\begin{tikzpicture}
\coordinate (x) at (-5,-5);
\coordinate (y) at (-4,0);
\coordinate (e0) at (0,-5);
\coordinate (v) at (0,.2);
\coordinate (w) at (.2,0);
\coordinate (e1) at (barycentric cs:x=1,y=1,e0=2) ;
\draw ($(x)-(v)$) node[below]{\large $x$};
\draw ($(y)+(v)$) node[above]{\large $y$};
\draw ($(e0)-(v)$) node[below]{\large $e_0$};
\draw ($(e1)-(v)$) node[below]{ \large $e_1$};
\draw[fill,color=black]  (x) circle(.1);
\draw[fill,color=black]  (y) circle(.1);
\draw[fill,color=black]  (e0) circle(.1);
\draw[fill,color=black]  (e1) circle(.1);
\draw (x)--(y)--(e0)--(x);
\draw (x)--(e1);\draw (y)--(e1);\draw (e0)--(e1);
\end{tikzpicture}} \    
\scalebox{.6}
{
\unitlength = .2 cm
\begin{tikzpicture}
\coordinate (x) at (-5,-5);
\coordinate (y) at (-4,0);
\coordinate (e0) at (0,-5);
\coordinate (v) at (0,.2);
\coordinate (w) at (.2,0);
\coordinate (e1) at (barycentric cs:x=1,y=1,e0=2) ;
\coordinate (e2) at (barycentric cs:x=1,y=1,e1=1) ;
\draw ($(x)-(v)$) node[below]{\large $x$};
\draw ($(y)+(v)$) node[above]{\large $y$};
\draw ($(e0)-(v)$) node[below]{\large $e_0$};
\draw ($(e1)-(v)$) node[below]{ \large $e_1$};
\draw ($(e2)-(v)$) node[below]{\large $e_2$};
\draw[fill,color=black]  (x) circle(.1);
\draw[fill,color=black]  (y) circle(.1);
\draw[fill,color=black]  (e0) circle(.1);
\draw[fill,color=black]  (e1) circle(.1);
\draw[fill,color=black]  (e2) circle(.1);
\draw (x)--(y)--(e0)--(x);
\draw (x)--(e1);\draw (y)--(e1);\draw (e0)--(e1);
\draw (x)--(e2);\draw (y)--(e2);\draw (e1)--(e2);
\end{tikzpicture}}
\  
\scalebox{.6}{\unitlength = .2 cm
\begin{tikzpicture}
\coordinate (x) at (-5,-5);
\coordinate (y) at (-4,0);
\coordinate (e0) at (0,-5);
\coordinate (v) at (0,.2);
\coordinate (w) at (.2,0);
\coordinate (e1) at (barycentric cs:x=1,y=1,e0=2) ;
\coordinate (e2) at (barycentric cs:x=1,y=1,e1=1) ;
\coordinate (e4) at (barycentric cs:y=1,e1=1) ;
\coordinate (e3) at (barycentric cs:y=1,e2=1) ;
\draw ($(x)-(v)$) node[below]{\large $x$};
\draw ($(y)+(v)$) node[above]{\large $y$};
\draw ($(e0)-(v)$) node[below]{\large $e_0$};
\draw ($(e1)-(v)$) node[below]{ \large $e_1$};
\draw ($(e4)+2*(w)+2*(v)$) node[below]{\large $e_4$};
\draw ($(e2)-(v)$) node[below]{\large $e_2$};
\draw[fill,color=black]  (x) circle(.1);
\draw[fill,color=black]  (y) circle(.1);
\draw[fill,color=black]  (e0) circle(.1);
\draw[fill,color=black]  (e1) circle(.1);
\draw[fill,color=black]  (e2) circle(.1);
\draw[fill,color=black]  (e4) circle(.1);
\draw (x)--(y)--(e0)--(x);
\draw (x)--(e1);
\draw (y)--(e4);
\draw (e0)--(e1);
\draw (x)--(e2);\draw (y)--(e2);\draw (e1)--(e2);
\draw (e1)--(y);
\draw (e0)--(e4);
\draw (e2)--(e4);
\end{tikzpicture}
}
\ 
\scalebox{.6}
{
\unitlength = .2 cm
\begin{tikzpicture}
\coordinate (x) at (-5,-5);
\coordinate (y) at (-4,0);
\coordinate (e0) at (0,-5);
\coordinate (v) at (0,.2);
\coordinate (w) at (.2,0);
\coordinate (e1) at (barycentric cs:x=1,y=1,e0=2) ;
\coordinate (e2) at (barycentric cs:x=1,y=1,e1=1) ;
\coordinate (e4) at (barycentric cs:y=1,e1=1) ;
\coordinate (e3) at (barycentric cs:y=1,e2=1) ;
\draw ($(x)-(v)$) node[below]{\large $x$};
\draw ($(y)+(v)$) node[above]{\large $y$};
\draw ($(e0)-(v)$) node[below]{\large $e_0$};
\draw ($(e1)-(v)$) node[below]{ \large $e_1$};
\draw ($(e4)+1.5*(w)+2*(v)$) node[below]{\large $e_4$};
\draw ($(e3)+1.5*(w)+2*(v)$) node[below]{\large $e_3$};
\draw ($(e2)-(v)$) node[below]{\large $e_2$};
\draw[fill,color=black]  (x) circle(.1);
\draw[fill,color=black]  (y) circle(.1);
\draw[fill,color=black]  (e0) circle(.1);
\draw[fill,color=black]  (e1) circle(.1);
\draw[fill,color=black]  (e2) circle(.1);
\draw[fill,color=black]  (e4) circle(.1);
\draw[fill,color=black]  (e3) circle(.1);
\draw (x)--(y)--(e0)--(x);
\draw (x)--(e1);\draw (y)--(e1);\draw (e0)--(e1);
\draw (x)--(e2);\draw (y)--(e2);\draw (e1)--(e2);
\draw (e2)--(e4);
\draw (e0)--(e4);
\draw (e3)--(e4);
\draw (x)--(e3);
\end{tikzpicture}}
\caption{$
\mathscr{E}_0 \xleftarrow{(x,y,e_0|e_1)}\mathscr{E}_1 \xleftarrow{(x,y,e_1|e_2)} \mathscr{B}\xleftarrow{(y,e_1|e_4)}\mathscr{B}_{1,\bullet } \xleftarrow{(y,e_2|e_3)}\mathscr{B}_{1,3} \   (\text{Type I})$
\label{typeI:fan1}}
\end{figure}

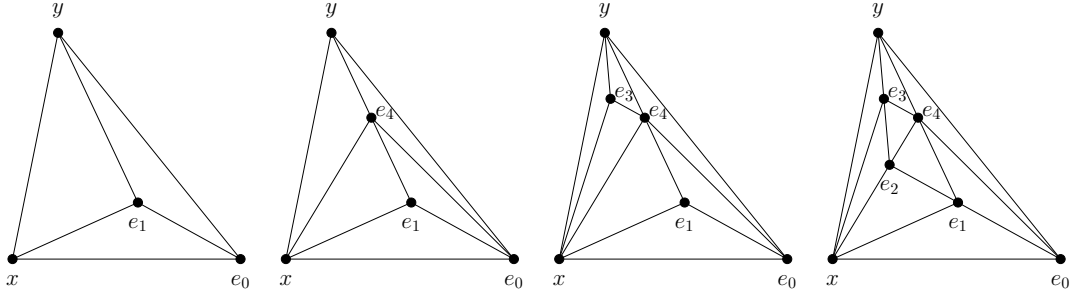
\begin{figure}[htb]
\scalebox{.6}
{
\unitlength = .2 cm
\begin{tikzpicture}
\coordinate (x) at (-5,-5);
\coordinate (y) at (-4,0);
\coordinate (e0) at (0,-5);
\coordinate (v) at (0,.2);
\coordinate (w) at (.2,0);
\coordinate (e1) at (barycentric cs:x=1,y=1,e0=2) ;
\coordinate (e2) at (barycentric cs:x=1,y=1,e1=1) ;
\coordinate (e4) at (barycentric cs:y=1,e1=1) ;
\coordinate (e3) at (barycentric cs:y=1,e2=1) ;
\draw ($(x)-(v)$) node[below]{\large $x$};
\draw ($(y)+(v)$) node[above]{\large $y$};
\draw ($(e0)-(v)$) node[below]{\large $e_0$};
\draw ($(e1)-(v)$) node[below]{ \large $e_1$};
\draw[fill,color=black]  (x) circle(.1);
\draw[fill,color=black]  (y) circle(.1);
\draw[fill,color=black]  (e0) circle(.1);
\draw[fill,color=black]  (e1) circle(.1);
\draw (x)--(y)--(e0)--(x);
\draw (x)--(e1);\draw (y)--(e1);\draw (e0)--(e1);
\end{tikzpicture}}\  
\scalebox{.6}
{
\unitlength = .2 cm
\begin{tikzpicture}
\coordinate (x) at (-5,-5);
\coordinate (y) at (-4,0);
\coordinate (e0) at (0,-5);
\coordinate (v) at (0,.2);
\coordinate (w) at (.2,0);
\coordinate (e1) at (barycentric cs:x=1,y=1,e0=2) ;
\coordinate (e2) at (barycentric cs:x=1,y=1,e1=1) ;
\coordinate (e4) at (barycentric cs:y=1,e1=1) ;
\coordinate (e3) at (barycentric cs:y=1,e2=1) ;
\draw ($(x)-(v)$) node[below]{\large $x$};
\draw ($(y)+(v)$) node[above]{\large $y$};
\draw ($(e0)-(v)$) node[below]{\large $e_0$};
\draw ($(e1)-(v)$) node[below]{ \large $e_1$};
\draw ($(e4)+1.5*(w)+2*(v)$) node[below]{\large $e_4$};
\draw[fill,color=black]  (x) circle(.1);
\draw[fill,color=black]  (y) circle(.1);
\draw[fill,color=black]  (e0) circle(.1);
\draw[fill,color=black]  (e1) circle(.1);
\draw[fill,color=black]  (e4) circle(.1);
\draw (x)--(y)--(e0)--(x);
\draw (x)--(e1);\draw (y)--(e1);\draw (e0)--(e1);
\draw (e0)--(e4);
\draw (x)--(e4);
\end{tikzpicture}}\  
\scalebox{.6}
{
\unitlength = .2 cm
\begin{tikzpicture}
\coordinate (x) at (-5,-5);
\coordinate (y) at (-4,0);
\coordinate (e0) at (0,-5);
\coordinate (v) at (0,.2);
\coordinate (w) at (.2,0);
\coordinate (e1) at (barycentric cs:x=1,y=1,e0=2) ;
\coordinate (e4) at (barycentric cs:y=1,e1=1) ;
\coordinate (e3) at (barycentric cs:y=1,e2=1) ;
\draw ($(x)-(v)$) node[below]{\large $x$};
\draw ($(y)+(v)$) node[above]{\large $y$};
\draw ($(e0)-(v)$) node[below]{\large $e_0$};
\draw ($(e1)-(v)$) node[below]{ \large $e_1$};
\draw ($(e4)+1.5*(w)+2*(v)$) node[below]{\large $e_4$};
\draw ($(e3)+1.5*(w)+2*(v)$) node[below]{\large $e_3$};
\draw[fill,color=black]  (x) circle(.1);
\draw[fill,color=black]  (y) circle(.1);
\draw[fill,color=black]  (e0) circle(.1);
\draw[fill,color=black]  (e1) circle(.1);
\draw[fill,color=black]  (e4) circle(.1);
\draw[fill,color=black]  (e3) circle(.1);
\draw (x)--(y)--(e0)--(x);
\draw (x)--(e1);\draw (y)--(e1);\draw (e0)--(e1);
\draw (y)--(e3);
\draw (x)--(e4);
\draw (e0)--(e4);
\draw (e3)--(e4);\draw (x)--(e3);
\end{tikzpicture}}\  
\scalebox{.6}
{
\unitlength = .2 cm
\begin{tikzpicture}
\coordinate (x) at (-5,-5);
\coordinate (y) at (-4,0);
\coordinate (e0) at (0,-5);
\coordinate (v) at (0,.2);
\coordinate (w) at (.2,0);
\coordinate (e1) at (barycentric cs:x=1,y=1,e0=2) ;
\coordinate (e2) at (barycentric cs:x=1,y=1,e1=1) ;
\coordinate (e4) at (barycentric cs:y=1,e1=1) ;
\coordinate (e3) at (barycentric cs:y=1,e2=1) ;
\draw ($(x)-(v)$) node[below]{\large $x$};
\draw ($(y)+(v)$) node[above]{\large $y$};
\draw ($(e0)-(v)$) node[below]{\large $e_0$};
\draw ($(e1)-(v)$) node[below]{ \large $e_1$};
\draw ($(e4)+1.5*(w)+2*(v)$) node[below]{\large $e_4$};
\draw ($(e3)+1.5*(w)+2*(v)$) node[below]{\large $e_3$};
\draw ($(e2)-(v)$) node[below]{\large $e_2$};
\draw[fill,color=black]  (x) circle(.1);
\draw[fill,color=black]  (y) circle(.1);
\draw[fill,color=black]  (e0) circle(.1);
\draw[fill,color=black]  (e1) circle(.1);
\draw[fill,color=black]  (e2) circle(.1);
\draw[fill,color=black]  (e4) circle(.1);
\draw[fill,color=black]  (e3) circle(.1);
\draw (x)--(y)--(e0)--(x);
\draw (x)--(e1);\draw (y)--(e1);\draw (e0)--(e1);
\draw (x)--(e2);\draw (y)--(e2);\draw (e1)--(e2);
\draw (e2)--(e4);
\draw (e0)--(e4);
\draw (e3)--(e4);\draw (x)--(e3);
\end{tikzpicture}}
\caption{$\mathscr{E}_0 \xleftarrow{(x,y,e_0|e_1)}\mathscr{E}_1 \xleftarrow{(y,e_1|e_4)} \mathscr{T}^+\xleftarrow{(y,x,e_4|e_3)}\mathscr{T}^+_2 \xleftarrow{(x,e_4|e_2)}\mathscr{T}^+_{2+}    (\text{Type I})$
}\label{typeI:fan2}
\end{figure}

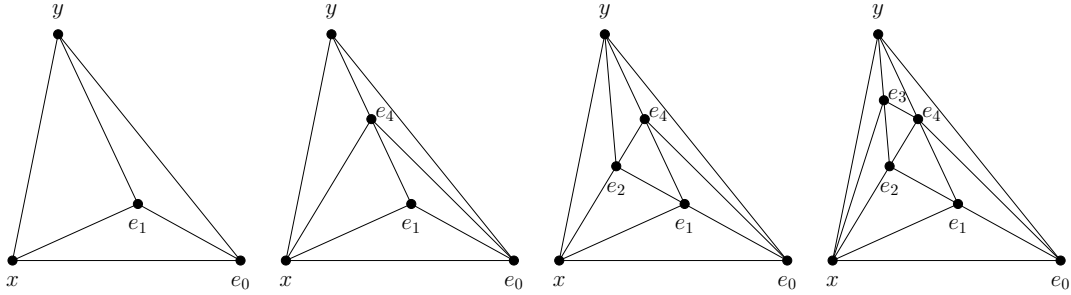
\begin{figure}
\scalebox{.6}
{
\unitlength = .2 cm
\begin{tikzpicture}
\coordinate (x) at (-5,-5);
\coordinate (y) at (-4,0);
\coordinate (e0) at (0,-5);
\coordinate (v) at (0,.2);
\coordinate (w) at (.2,0);
\coordinate (e1) at (barycentric cs:x=1,y=1,e0=2) ;
\coordinate (e2) at (barycentric cs:x=1,y=1,e1=1) ;
\coordinate (e4) at (barycentric cs:y=1,e1=1) ;
\coordinate (e3) at (barycentric cs:y=1,e2=1) ;
\draw ($(x)-(v)$) node[below]{\large $x$};
\draw ($(y)+(v)$) node[above]{\large $y$};
\draw ($(e0)-(v)$) node[below]{\large $e_0$};
\draw ($(e1)-(v)$) node[below]{ \large $e_1$};
\draw[fill,color=black]  (x) circle(.1);
\draw[fill,color=black]  (y) circle(.1);
\draw[fill,color=black]  (e0) circle(.1);
\draw[fill,color=black]  (e1) circle(.1);
\draw (x)--(y)--(e0)--(x);
\draw (x)--(e1);\draw (y)--(e1);\draw (e0)--(e1);
\end{tikzpicture}}\  
\scalebox{.6}
{
\unitlength = .2 cm
\begin{tikzpicture}
\coordinate (x) at (-5,-5);
\coordinate (y) at (-4,0);
\coordinate (e0) at (0,-5);
\coordinate (v) at (0,.2);
\coordinate (w) at (.2,0);
\coordinate (e1) at (barycentric cs:x=1,y=1,e0=2) ;
\coordinate (e2) at (barycentric cs:x=1,y=1,e1=1) ;
\coordinate (e4) at (barycentric cs:y=1,e1=1) ;
\coordinate (e3) at (barycentric cs:y=1,e2=1) ;
\draw ($(x)-(v)$) node[below]{\large $x$};
\draw ($(y)+(v)$) node[above]{\large $y$};
\draw ($(e0)-(v)$) node[below]{\large $e_0$};
\draw ($(e1)-(v)$) node[below]{ \large $e_1$};
\draw ($(e4)+1.5*(w)+2*(v)$) node[below]{\large $e_4$};
\draw[fill,color=black]  (x) circle(.1);
\draw[fill,color=black]  (y) circle(.1);
\draw[fill,color=black]  (e0) circle(.1);
\draw[fill,color=black]  (e1) circle(.1);
\draw[fill,color=black]  (e4) circle(.1);
\draw (x)--(y)--(e0)--(x);
\draw (x)--(e1);\draw (y)--(e1);\draw (e0)--(e1);
\draw (e0)--(e4);
\draw (x)--(e4);
\end{tikzpicture}}\  
\scalebox{.6}
{
\unitlength = .2 cm
\begin{tikzpicture}
\coordinate (x) at (-5,-5);
\coordinate (y) at (-4,0);
\coordinate (e0) at (0,-5);
\coordinate (v) at (0,.2);
\coordinate (w) at (.2,0);
\coordinate (e1) at (barycentric cs:x=1,y=1,e0=2) ;
\coordinate (e2) at (barycentric cs:x=1,y=1,e1=1) ;
\coordinate (e4) at (barycentric cs:y=1,e1=1) ;
\coordinate (e3) at (barycentric cs:y=1,e2=1) ;
\draw ($(x)-(v)$) node[below]{\large $x$};
\draw ($(y)+(v)$) node[above]{\large $y$};
\draw ($(e0)-(v)$) node[below]{\large $e_0$};
\draw ($(e1)-(v)$) node[below]{ \large $e_1$};
\draw ($(e4)+1.5*(w)+2*(v)$) node[below]{\large $e_4$};
\draw ($(e2)-(v)$) node[below]{\large $e_2$};
\draw[fill,color=black]  (x) circle(.1);
\draw[fill,color=black]  (y) circle(.1);
\draw[fill,color=black]  (e0) circle(.1);
\draw[fill,color=black]  (e1) circle(.1);
\draw[fill,color=black]  (e2) circle(.1);
\draw[fill,color=black]  (e4) circle(.1);
\draw (x)--(y)--(e0)--(x);
\draw (x)--(e1);\draw (y)--(e1);\draw (e0)--(e1);
\draw (x)--(e2);\draw (y)--(e2);\draw (e1)--(e2);
\draw (e2)--(e4);
\draw (e0)--(e4);
\end{tikzpicture}}\  
\scalebox{.6}
{
\unitlength = .2 cm
\begin{tikzpicture}
\coordinate (x) at (-5,-5);
\coordinate (y) at (-4,0);
\coordinate (e0) at (0,-5);
\coordinate (v) at (0,.2);
\coordinate (w) at (.2,0);
\coordinate (e1) at (barycentric cs:x=1,y=1,e0=2) ;
\coordinate (e2) at (barycentric cs:x=1,y=1,e1=1) ;
\coordinate (e4) at (barycentric cs:y=1,e1=1) ;
\coordinate (e3) at (barycentric cs:y=1,e2=1) ;
\draw ($(x)-(v)$) node[below]{\large $x$};
\draw ($(y)+(v)$) node[above]{\large $y$};
\draw ($(e0)-(v)$) node[below]{\large $e_0$};
\draw ($(e1)-(v)$) node[below]{ \large $e_1$};
\draw ($(e4)+1.5*(w)+2*(v)$) node[below]{\large $e_4$};
\draw ($(e3)+1.5*(w)+2*(v)$) node[below]{\large $e_3$};
\draw ($(e2)-(v)$) node[below]{\large $e_2$};
\draw[fill,color=black]  (x) circle(.1);
\draw[fill,color=black]  (y) circle(.1);
\draw[fill,color=black]  (e0) circle(.1);
\draw[fill,color=black]  (e1) circle(.1);
\draw[fill,color=black]  (e2) circle(.1);
\draw[fill,color=black]  (e4) circle(.1);
\draw[fill,color=black]  (e3) circle(.1);
\draw (x)--(y)--(e0)--(x);
\draw (x)--(e1);\draw (y)--(e1);\draw (e0)--(e1);
\draw (x)--(e2);\draw (y)--(e2);\draw (e1)--(e2);
\draw (e2)--(e4);
\draw (e0)--(e4);
\draw (e3)--(e4);\draw (x)--(e3);
\end{tikzpicture}}\\
\caption{  $\mathscr{E}_0 \xleftarrow{(x,y,e_0|e_1)}\mathscr{E}_1 \xleftarrow{(y,e_1|e_4)} \mathscr{T}^+\xleftarrow{(x,e_4|e_2)}\mathscr{T}^+_1 \xleftarrow{(y,e_2|e_3)}\mathscr{T}^+_{1+}    (\text{Type I})$
}\label{typeI:fan3}
\end{figure}

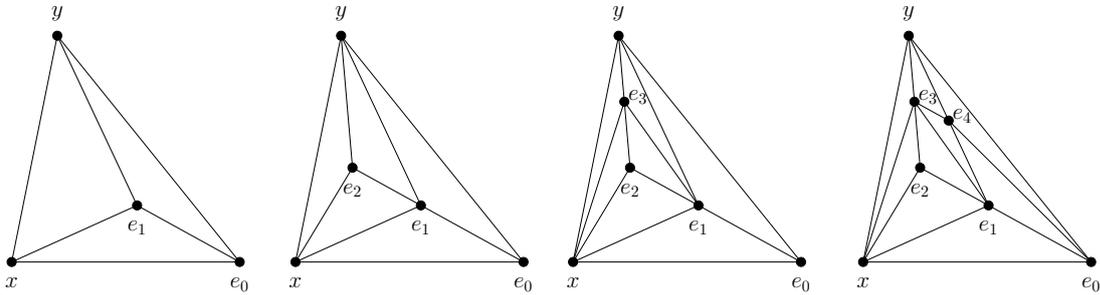
\begin{figure}[htb]
\scalebox{.6}
{\unitlength = .2 cm
\begin{tikzpicture}
\coordinate (x) at (-5,-5);
\coordinate (y) at (-4,0);
\coordinate (e0) at (0,-5);
\coordinate (v) at (0,.2);
\coordinate (w) at (.2,0);
\coordinate (e1) at (barycentric cs:x=1,y=1,e0=2) ;
\draw ($(x)-(v)$) node[below]{\large $x$};
\draw ($(y)+(v)$) node[above]{\large $y$};
\draw ($(e0)-(v)$) node[below]{\large $e_0$};
\draw ($(e1)-(v)$) node[below]{ \large $e_1$};
\draw[fill,color=black]  (x) circle(.1);
\draw[fill,color=black]  (y) circle(.1);
\draw[fill,color=black]  (e0) circle(.1);
\draw[fill,color=black]  (e1) circle(.1);
\draw (x)--(y)--(e0)--(x);
\draw (x)--(e1);\draw (y)--(e1);\draw (e0)--(e1);
\end{tikzpicture}} \    
\scalebox{.6}
{
\unitlength = .2 cm
\begin{tikzpicture}
\coordinate (x) at (-5,-5);
\coordinate (y) at (-4,0);
\coordinate (e0) at (0,-5);
\coordinate (v) at (0,.2);
\coordinate (w) at (.2,0);
\coordinate (e1) at (barycentric cs:x=1,y=1,e0=2) ;
\coordinate (e2) at (barycentric cs:x=1,y=1,e1=1) ;
\draw ($(x)-(v)$) node[below]{\large $x$};
\draw ($(y)+(v)$) node[above]{\large $y$};
\draw ($(e0)-(v)$) node[below]{\large $e_0$};
\draw ($(e1)-(v)$) node[below]{ \large $e_1$};
\draw ($(e2)-(v)$) node[below]{\large $e_2$};
\draw[fill,color=black]  (x) circle(.1);
\draw[fill,color=black]  (y) circle(.1);
\draw[fill,color=black]  (e0) circle(.1);
\draw[fill,color=black]  (e1) circle(.1);
\draw[fill,color=black]  (e2) circle(.1);
\draw (x)--(y)--(e0)--(x);
\draw (x)--(e1);\draw (y)--(e1);\draw (e0)--(e1);
\draw (x)--(e2);\draw (y)--(e2);\draw (e1)--(e2);
\end{tikzpicture}}
\  
\scalebox{.6}{\unitlength = .2 cm
\begin{tikzpicture}
\coordinate (x) at (-5,-5);
\coordinate (y) at (-4,0);
\coordinate (e0) at (0,-5);
\coordinate (v) at (0,.2);
\coordinate (w) at (.2,0);
\coordinate (e1) at (barycentric cs:x=1,y=1,e0=2) ;
\coordinate (e2) at (barycentric cs:x=1,y=1,e1=1) ;
\coordinate (e4) at (barycentric cs:y=1,e1=1) ;
\coordinate (e3) at (barycentric cs:y=1,e2=1) ;
\draw ($(x)-(v)$) node[below]{\large $x$};
\draw ($(y)+(v)$) node[above]{\large $y$};
\draw ($(e0)-(v)$) node[below]{\large $e_0$};
\draw ($(e1)-(v)$) node[below]{ \large $e_1$};
\draw ($(e3)+1.5*(w)+2*(v)$) node[below]{\large $e_3$};
\draw ($(e2)-(v)$) node[below]{\large $e_2$};
\draw[fill,color=black]  (x) circle(.1);
\draw[fill,color=black]  (y) circle(.1);
\draw[fill,color=black]  (e0) circle(.1);
\draw[fill,color=black]  (e1) circle(.1);
\draw[fill,color=black]  (e2) circle(.1);
\draw[fill,color=black]  (e3) circle(.1);
\draw (x)--(y)--(e0)--(x);
\draw (x)--(e1);
\draw (e0)--(e1);
\draw (x)--(e2);\draw (y)--(e2);\draw (e1)--(e2);
\draw (e1)--(e3);\draw (e1)--(y);
\draw (x)--(e3);
\end{tikzpicture}
}
\ 
\scalebox{.6}
{
\unitlength = .2 cm
\begin{tikzpicture}
\coordinate (x) at (-5,-5);
\coordinate (y) at (-4,0);
\coordinate (e0) at (0,-5);
\coordinate (v) at (0,.2);
\coordinate (w) at (.2,0);
\coordinate (e1) at (barycentric cs:x=1,y=1,e0=2) ;
\coordinate (e2) at (barycentric cs:x=1,y=1,e1=1) ;
\coordinate (e4) at (barycentric cs:y=1,e1=1) ;
\coordinate (e3) at (barycentric cs:y=1,e2=1) ;
\draw ($(x)-(v)$) node[below]{\large $x$};
\draw ($(y)+(v)$) node[above]{\large $y$};
\draw ($(e0)-(v)$) node[below]{\large $e_0$};
\draw ($(e1)-(v)$) node[below]{ \large $e_1$};
\draw ($(e4)+1.5*(w)+2*(v)$) node[below]{\large $e_4$};
\draw ($(e3)+1.5*(w)+2*(v)$) node[below]{\large $e_3$};
\draw ($(e2)-(v)$) node[below]{\large $e_2$};
\draw[fill,color=black]  (x) circle(.1);
\draw[fill,color=black]  (y) circle(.1);
\draw[fill,color=black]  (e0) circle(.1);
\draw[fill,color=black]  (e1) circle(.1);
\draw[fill,color=black]  (e2) circle(.1);
\draw[fill,color=black]  (e4) circle(.1);
\draw[fill,color=black]  (e3) circle(.1);
\draw (x)--(y)--(e0)--(x);
\draw (x)--(e1);\draw (y)--(e1);\draw (e0)--(e1);
\draw (x)--(e2);\draw (y)--(e2);\draw (e1)--(e2);
\draw (e1)--(e3);
\draw (e0)--(e4);
\draw (e3)--(e4);\draw (x)--(e3);
\end{tikzpicture}}
\caption{$
 \mathscr{E}_0 \xleftarrow{(x,y,e_0|e_1)}\mathscr{E}_1 \xleftarrow{(x,y,e_1|e_2)} \mathscr{B}	\xleftarrow{(y,e_2|e_3)}\mathscr{B}_{2,\bullet} \xleftarrow{(y,e_1|e_4)}\mathscr{B}_{2,3}\   (\text{Type II})$
}\label{typeII:fan}
\end{figure}

\end{center}

\clearpage
\newpage

\subsection{Weighted blow ups for type III}\label{section:toric5}
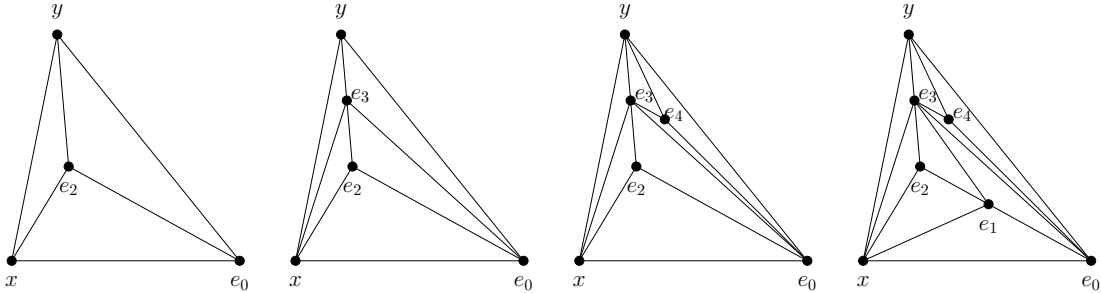
\begin{figure}[htb]
\scalebox{.6}
{\unitlength = .2 cm
\begin{tikzpicture}
\coordinate (x) at (-5,-5);
\coordinate (y) at (-4,0);
\coordinate (e0) at (0,-5);
\coordinate (v) at (0,.2);
\coordinate (w) at (.2,0);
\coordinate (e2) at (barycentric cs:x=1,y=1,e1=1) ;
\draw ($(x)-(v)$) node[below]{\large $x$};
\draw ($(y)+(v)$) node[above]{\large $y$};
\draw ($(e0)-(v)$) node[below]{\large $e_0$};
\draw ($(e2)-(v)$) node[below]{\large $e_2$};
\draw[fill,color=black]  (x) circle(.1);
\draw[fill,color=black]  (y) circle(.1);
\draw[fill,color=black]  (e0) circle(.1);
\draw[fill,color=black]  (e2) circle(.1);
\draw (x)--(y)--(e0)--(x);
\draw (x)--(e2);\draw (y)--(e2); \draw (e2)--(e0);
\end{tikzpicture}} \    
\scalebox{.6}
{
\unitlength = .2 cm
\begin{tikzpicture}
\coordinate (x) at (-5,-5);
\coordinate (y) at (-4,0);
\coordinate (e0) at (0,-5);
\coordinate (v) at (0,.2);
\coordinate (w) at (.2,0);
\coordinate (e2) at (barycentric cs:x=1,y=1,e1=1) ;
\coordinate (e3) at (barycentric cs:y=1,e2=1) ;
\draw ($(x)-(v)$) node[below]{\large $x$};
\draw ($(y)+(v)$) node[above]{\large $y$};
\draw ($(e0)-(v)$) node[below]{\large $e_0$};
\draw ($(e3)+1.5*(w)+2*(v)$) node[below]{\large $e_3$};
\draw ($(e2)-(v)$) node[below]{\large $e_2$};
\draw[fill,color=black]  (x) circle(.1);
\draw[fill,color=black]  (y) circle(.1);
\draw[fill,color=black]  (e0) circle(.1);
\draw[fill,color=black]  (e2) circle(.1);
\draw[fill,color=black]  (e3) circle(.1);
\draw (x)--(y)--(e0)--(x);
\draw (x)--(e2);\draw (y)--(e2);\draw (e0)--(e2);
\draw (e3)--(x);\draw (e0)--(e3);
\end{tikzpicture}}
\  
\scalebox{.6}
{
\unitlength = .2 cm
\begin{tikzpicture}
\coordinate (x) at (-5,-5);
\coordinate (y) at (-4,0);
\coordinate (e0) at (0,-5);
\coordinate (v) at (0,.2);
\coordinate (w) at (.2,0);
\coordinate (e2) at (barycentric cs:x=1,y=1,e1=1) ;
\coordinate (e4) at (barycentric cs:y=1,e1=1) ;
\coordinate (e3) at (barycentric cs:y=1,e2=1) ;
\draw ($(x)-(v)$) node[below]{\large $x$};
\draw ($(y)+(v)$) node[above]{\large $y$};
\draw ($(e0)-(v)$) node[below]{\large $e_0$};
\draw ($(e4)+(w)+2*(v)$) node[below]{\large $e_4$};
\draw ($(e3)+1.5*(w)+2*(v)$) node[below]{\large $e_3$};
\draw ($(e2)-(v)$) node[below]{\large $e_2$};
\draw[fill,color=black]  (x) circle(.1);
\draw[fill,color=black]  (y) circle(.1);
\draw[fill,color=black]  (e0) circle(.1);
\draw[fill,color=black]  (e2) circle(.1);
\draw[fill,color=black]  (e4) circle(.1);
\draw[fill,color=black]  (e3) circle(.1);
\draw (x)--(y)--(e0)--(x);
\draw (y)--(e4);\draw (e3)--(e4);\draw (e0)--(e4);
\draw (x)--(e2);\draw (y)--(e2);\draw (e0)--(e2);
\draw (e3)--(x);\draw (e0)--(e3);
\end{tikzpicture}}
\ 
\scalebox{.6}
{
\unitlength = .2 cm
\begin{tikzpicture}
\coordinate (x) at (-5,-5);
\coordinate (y) at (-4,0);
\coordinate (e0) at (0,-5);
\coordinate (v) at (0,.2);
\coordinate (w) at (.2,0);
\coordinate (e1) at (barycentric cs:x=1,y=1,e0=2) ;
\coordinate (e2) at (barycentric cs:x=1,y=1,e1=1) ;
\coordinate (e4) at (barycentric cs:y=1,e1=1) ;
\coordinate (e3) at (barycentric cs:y=1,e2=1) ;
\draw ($(x)-(v)$) node[below]{\large $x$};
\draw ($(y)+(v)$) node[above]{\large $y$};
\draw ($(e0)-(v)$) node[below]{\large $e_0$};
\draw ($(e1)-(v)$) node[below]{ \large $e_1$};
\draw ($(e4)+1.5*(w)+2*(v)$) node[below]{\large $e_4$};
\draw ($(e3)+1.5*(w)+2*(v)$) node[below]{\large $e_3$};
\draw ($(e2)-(v)$) node[below]{\large $e_2$};
\draw[fill,color=black]  (x) circle(.1);
\draw[fill,color=black]  (y) circle(.1);
\draw[fill,color=black]  (e0) circle(.1);
\draw[fill,color=black]  (e1) circle(.1);
\draw[fill,color=black]  (e2) circle(.1);
\draw[fill,color=black]  (e4) circle(.1);
\draw[fill,color=black]  (e3) circle(.1);
\draw (x)--(y)--(e0)--(x);
\draw (x)--(e1);\draw (y)--(e4);\draw (e0)--(e1);
\draw (x)--(e2);\draw (y)--(e2);\draw (e1)--(e2);
\draw (e1)--(e3);
\draw (e0)--(e4);
\draw (e3)--(e4);\draw (x)--(e3);\draw (e0)--(e3);
\end{tikzpicture}}
\caption{
{$\mathscr{E}_0 \xleftarrow{( x,y,e_0 |e_2 )_{(2,2,1)} }\mathscr{E}_1' \xleftarrow{( y,e_2 |e_3)}\mathscr{E}_2' \xleftarrow{( y,e_0,e_3|e_4 )_{\frac{1}{2},\frac{1}{2},\frac{1}{2} }}\mathscr{E}_3'\xleftarrow{( e_0,e_2 |e_1 )_{\frac{1}{2},\frac{1}{2}}} \mathscr{B}^1_{2,3}$ 
} 
{The toric description for the weighted blow up for type III $(\mathscr{B}_{2,3}^1)$.} \label{fig.toric.typeIII}}
\end{figure}

Using the previous construction of  the resolution $\mathscr{B}^1_{2,3}$, it is easy to express it as a sequence of toric blow ups after the following  linear transformation from the scaling of  equation \eqref{typeIII.eq1}:
\begin{equation}
\begin{pmatrix}
\ell_1'' \\
\ell_2'' \\
\ell_3''\\
\ell_4''
\end{pmatrix}=
\begin{pmatrix}
2 & -1 & 2 & -1 \\
0 & 0 & 1 & -1 \\
0 & 1 & 1 & 0 \\
0 & 1 & 0 & -1 
\end{pmatrix}
\begin{pmatrix}
\ell_2' \\
\ell_3' \\
\ell_4'\\
\ell_5'
\end{pmatrix}
\end{equation}
 which gives:
\begin{equation}\label{typeIII.scaling}
\text{Type III}:~\begin{tabular}{|c|c|c|c|c|c|c|c|}
\cline{2-8}
\multicolumn{1}{c|}{}& $x$ & $y$ & $e_0$ & $e_1$ & $e_2$ & $e_3$ & $e_4$ \\
\hline 
$\ell_1''$ & 2 & 2 & 1 & 0 &  -1&0  & 0  \\
\hline
$\ell_2''$ & 0 & 1 & 0 &  0 & 1 &-1   &0  \\
\hline
$\ell_3''$ & 0 & 1 & 1 & 0  & 0 & 1  & -2 \\
\hline
$\ell_4''$ & 0 & 0 & 1 & -2 &  1& 0  & 0   \\
\hline
\end{tabular}
\end{equation}
From these scalings, we derive the following sequence of weighted blow ups for  $\mathscr{B}^1_{2,3}$:
\begin{equation}
\begin{tikzcd}[column sep=2.5cm]
\mathscr{E}_0 \arrow[leftarrow]{r}{ \textstyle{( x,y,e_0 |e_2 )}_{(2,2,1)} } & 
\mathscr{E}_1' \arrow[leftarrow]{r}{ \textstyle{( y,e_2 |e_3)}} & 
\mathscr{E}_2 '\arrow[leftarrow]{r}{ \textstyle{( y,e_0,e_3|e_4 )}_{\frac{1}{2},\frac{1}{2},\frac{1}{2}} } & 
\mathscr{E}_3 '\arrow[leftarrow]{r}{ \textstyle{( e_0,e_2 |e_1 )}_{\frac{1}{2},\frac{1}{2}}  } & \mathscr{B}^1_{2,3}.
\end{tikzcd}
\end{equation}
Here the subscript stands for the weight for the blow up. For example, $(x,y,e_0|e_2)_{(2,2,1)}$ means we do the replacements $x\rightarrow e_2^2 x,~y\rightarrow e_2^2 y,~e_0\rightarrow e_2e_0$ for the blow up. 
 The composition of these weighted blow ups reproduces the defining equation of $\mathscr{B}^1_{2,3}$ with the correct SR ideal. 
 See figure \ref{fig.toric.typeIII} for a toric illustration of this sequence of blow ups. The SR ideal can be read off from the ambient space from the above weighted blow up:
\begin{equation}
[x e_2^2 e_3^2 e_4 e_1:ye_2^2 e_3^3 e_4^2 e_1:z][x:e_4e_3y: e_0^2 e_4 e_1 ] [e_4y^2: e_2^2 e_1][y^2:e_0^2 e_1 :e_3^2][e_0^2:e_2^2]
\end{equation}
where we have used the isomorphism $\mathbb{P}^n_{ k,\cdots ,k}\cong \mathbb{P}^n$ and $\mathbb{P}^2\cong \mathbb{P}^2_{2,2,1}$ (see, for example, Lecture 10 of  \cite{Harris}).

\newpage

\section{The $\mathscr{T}^+_{2}$ and $\mathscr{T}^+_{3}$ Branches}\label{section:T23}

Let us summarize our construction for the network in Figure \ref{network} so far. In Section \ref{EYdescription}, we have explored the $\mathscr{B}$ branch and obtained the six resolutions $\mathscr{B}_{i,j}$ in \cite{EY}. In Section \ref{section:toric}, we have identified the type I resolution as $\mathscr{T}^+_{1+}\cong\mathscr{T}^+_{2+}\cong\mathscr{B}_{1,3}$, type II as $\mathscr{B}_{2,3}$, and type III as $\mathscr{B}_{2,3}^1$. Furthermore, as noted before and  will be discussed in Section \ref{section:iso}, $\mathscr{T}^+_{1-}\cong\mathscr{B}_{1,2}$. 

Among the resolutions in Figure \ref{network}, we are then left with
\begin{align}
\mathscr{T}^+_{2-},~~\mathscr{T}^+_{3\pm},
\end{align}
while the $\mathscr{T}^-$ branch and the other three of $\mathscr{B}_{i,j}$ are trivially obtained by the Mordell-Weil involutions.  In this section we will study the $\mathscr{T}^+_2$ and $\mathscr{T}^+_3$ branches. In the end, we find 
\begin{align}
\begin{split}
&\mathscr{T}^+_{2+} \cong\mathscr{B}_{1,3} \cong\mathscr{T}^+_{1+}~\text{(type I)},\\
&\mathscr{T}^+_{2-}\cong\mathscr{T}^+_{3+} = : \mathscr{B}_{1,3}^1,\\
&\mathscr{T}^+_{3-}=:\mathscr{B}^2_{1,3}.
\end{split}
\end{align}

\subsection{Resolutions $\mathscr{T}^+_{2\pm}$}

The partial resolution $\mathscr{T}^+_2$ can be written as
\begin{equation}\label{r}
\mathscr{T}^+_2:\quad e_2 (e_3 y^2+e_1  e_0^2 a_{3,2} y-e_1^3 e_2 e_0^5 a_{6,5})=x\   \   \overbrace{(e_1 e_3 x^2+e_0 e_1 a_{2,1}x+e_1^2 e_2 e_0^3  a_{4,3}-a_1  y)}^{\displaystyle{r}}
\end{equation}
with the projective coordinates of the ambient space:
\begin{equation}
[e_1e_2e_3^2 x:e_1e_2^2 e_3^3 y:z] \quad [e_3x:e_2e_3^2 y:e_0] \quad [e_3y:e_1] \quad[x:y:e_2]
\end{equation}
Over the codimension one locus $e_0e_1 e_2e_3=0$, we have the following five nodes in the singular fiber:
\begin{align}
\begin{split}
C_0&: \  e_0=e_2 e_3 y^2+a_1 x y -e_1 e_3 x^2=0,\\
C_1&:\   e_1= e_2 e_3 y + a_1 x =0,\\
C_{2+}&:\   e_2=x=0, \quad C_{2-}: e_2=r=0, \\
C_3&:\    e_3= e_2 e_0^2e_1(  a_{3,2} y-e_1^2 e_2 e_0^3 a_{6,5})-x  (e_0 e_1 a_{2,1}x+e_1^2 e_2 e_0^3  a_{4,3}-a_1  y) =0  .
\end{split}
\end{align}

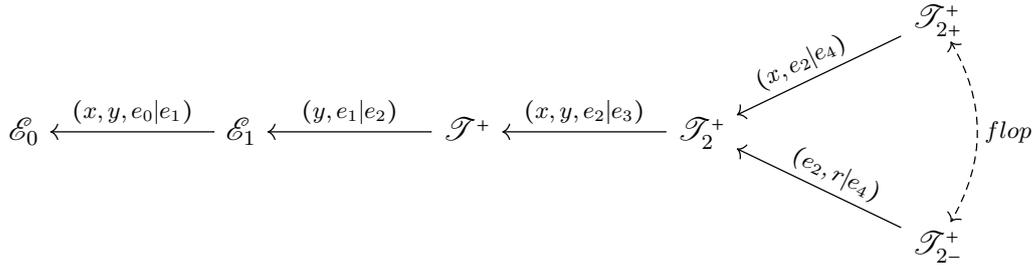
\begin{figure}
\begin{tikzcd}[column sep=huge]
& & & & \mathscr{T}^+_{2+} \arrow[leftrightarrow, bend left, dashed]{dd}{\textstyle{flop}}\\
\mathscr{E}_0 \arrow[leftarrow]{r}{ \textstyle{( x,y,e_0 |e_1 )}} & 
\mathscr{E}_1 \arrow[leftarrow]{r}{ \textstyle{( y,e_1 |e_2)}} & 
\mathscr{T}^+ \arrow[leftarrow]{r}{ \textstyle{( x,y,e_2 |e_3)}} & 
\mathscr{T}^+_2 \arrow[leftarrow]{ru}[sloped, near end]{ \textstyle{( x,e_2 |e_4)}}  \arrow[leftarrow]{rd}[sloped, near start]{ \textstyle{( e_2, r |e_4)}}  & \\
& & & & \mathscr{T}^+_{2-}
\end{tikzcd}
\caption{The sequences of blow ups for $\mathscr{T}^+_{2\pm}$.}
\end{figure} 

We would like to blow up the non-Cartier divisor $(x,e_2)$ or $(r,e_2)$ corresponding respectively to $C_{2+}$ and $C_{2-}$. The blow ups defined in this way correspond respectively to  $\mathscr{T}^+_{2+}$ and  $\mathscr{T}^+_{2-}$  related by a flop. Since $\mathscr{T}^+_{2+}$ is a model of type I, we know its fiber structure already. We will therefore focus on $\mathscr{T}^+_{2-}$:
  \begin{equation}\label{eqT+2-}
\mathscr{T}^+_{2-}
\begin{cases}
e_2 (e_3 y^2+e_1  e_0^2 a_{3,2} y-e_1^3 e_2 e_4 e_0^5 a_{6,5})=r x,\\
e_1 e_3 x^2+e_0 e_1 a_{2,1}x+e_1^2 e_2 e_4 e_0^3  a_{4,3}-a_1  y=e_4 r,
\end{cases}
\end{equation}
with the projective coordinates:
\begin{equation}
[e_1e_2e_4e_3^2 x:e_1e_2^2 e_4^2e_3^3 y:z]\quad [e_3x:e_2e_3^2 e_4y:e_0]\quad [e_3y:e_1]\quad[x:y:e_2e_4]\quad[e_2:r].
\end{equation}
The fiber enhancements are obtained in Table \ref{Table.Fplus2minus} by a straightforward calculation.

\begin{table}[htb]
\begin{center}
\scalebox{.8}{\begin{tabular}{|c|c|c|c|c|}
\hline
$w=0$& $w=P=0$ &  $w=a_1=0$ & $w=a_1=a_{3,2}=0$ & $w=a_1=a_{2,1}=0$ \\ 
\hline
& 
{\footnotesize $C_3\to C_{3+}+C_{3-}$ }
&
\scalebox{.8}{ \begin{tabular}{l}
$C_1\to C_{14}$\\
$C_2\to C_{24}$\\
$C_4\to C_{14}+C_{24}+C_4'+C_4''$
\end{tabular} }
 & 
\scalebox{.8}{ \begin{tabular}{l}
$C_1\to C_{14}$\\
$C_2\to C_{24}$\\
$C_3\to C_{34}+C_{3+}'+C_{3-}'$\\
$C_4\to C_{14}+C_{24}+C_{34}+C_4''$
\end{tabular} }
 & 
\scalebox{.8}{\begin{tabular}{l}
$C_1\to C_{14}$\\
$C_2\to C_{24}$\\
$C_3\to C_{34}+C_3'$\\
$C_4\to C_{14}+2C_{24}+C_{34}+2C_4'$
\end{tabular} }
 \\
\hline
\scalebox{.9}{\begin{tikzpicture}[every node/.style={circle,draw, minimum size= 10 mm}]
\node (C2)  at (18+72*5:1.4cm)  {\tiny $C_{4}$};
\node (C4) at (18+72*4:1.4cm) {\tiny $C_{3}$};
\node (C3) at (18+72*3:1.4cm) {\tiny $C_{2-}$};
\node (C1) at (18+72*2:1.4cm) {\tiny $C_{1}$};
\node (C0) at (18+72*1:1.4cm) {\tiny $C_0$};
\draw (C1)--(C3)--(C4)--(C2)--(C0)--(C1);
\end{tikzpicture}}& 
\scalebox{.9}{\begin{tikzpicture}[every node/.style={circle,draw, minimum size= 5 mm}]
\node (C2b)  at (30+60*6:1.4cm)  {\tiny $C_{4}$};
\node (C2)  at (30+60*5:1.4cm)  {\tiny $C_{3-}$};
\node (C4) at (30+60*4:1.4cm) {\scalebox{.7}{ $C_{3+}$}};
\node (C3) at (30+60*3:1.4cm) {\scalebox{.7}{$C_{2-}$}};
\node (C1) at (30+60*2:1.4cm) {\tiny $C_{1}$};
\node (C0) at (30+60*1:1.4cm) {\tiny $C_0$};
\draw (C1)--(C3)--(C4)--(C2)--(C2b)--(C0)--(C1);
\end{tikzpicture}}
& 
\scalebox{1}{\begin{tikzpicture}[every node/.style={circle,draw, minimum size= 8 mm}, scale=.8]
\node (C2) at (0,0) { \scalebox{.8}{$2C_{14}$}};
\node (C5) at (90:-1.8cm) { \scalebox{.8}{$2C_{24}$}};
\node (C0) at (90-35:1.8cm) { \scalebox{.8}{$C_4''$}} ;
\node (C1) at (90+35:1.8cm) { $C_0$};
\node [yshift=-1.8cm] (C3) at  (-90-35:1.8cm){ \scalebox{.8}{ $C_4'$}};
\node [yshift=-1.8cm] (C4) at (-90+35:1.8cm) {  $C_3$};
\draw (C2)--(C0);
\draw (C2)--(C1);
\draw (C2)--(C5);
\draw (C5)--(C3);
\draw (C5)--(C4);
\end{tikzpicture}}
& 
\scalebox{1} {\begin{tikzpicture}[every node/.style={circle,draw, minimum size= 4 mm}, scale=.8]
\node (C2) at (0,0) {\tiny $2C_{14}$};
\node (C5) at (90:-1.8cm) {\tiny $2C_{24}$};
\node (C0) at (90-35:1.8cm) {\tiny $C_4''$} ;
\node (C1) at (90+35:1.8cm) {\small $C_0$};
\node (C6) at (90:-2*1.8cm) { \tiny {$2C_{34}$}};
\node [yshift=-1.8*2cm] (C3) at  (-90-35:1.4cm){\tiny $C_{3+}'$};
\node [yshift=-1.8*2cm] (C4) at (-90+35:1.4cm) {\tiny $C_{3-}'$};
\draw (C2)--(C0);
\draw (C2)--(C1);
\draw (C2)--(C5);
\draw (C6)--(C3);
\draw (C6)--(C4);
\draw (C5)--(C6);
\end{tikzpicture}}
& 
\scalebox{1}{\begin{tikzpicture}[every node/.style={circle,draw, minimum size= 4 mm}, scale=.8]
\node (C2) at (0,0) {\scalebox{.65}{$2C_{14}$}};
\node (C5) at (90:-1.8cm) {\scalebox{.65}{$3C_{24}$}};
\node (C0) at (90:1.8cm) {\tiny $C_0$} ;
\node [xshift=-1.4cm] (C3) at  (90:-1.8cm){\tiny $2C_4'$};
\node  (C6) at (90:-2*1.8cm) {\scalebox{.65}{$2C_{34}$}};
\node  (C4) at (90:-3*1.8cm) {\tiny $C_{3}'$};
\draw (C2)--(C0);
\draw (C2)--(C5)--(C6)--(C4);
\draw (C5)--(C3);
\end{tikzpicture}}
\\ 
\hline
\end{tabular}}
\end{center}
\caption{Fibers of the resolution $\mathscr{B}_{1,3}^1:=\mathscr{T}^+_{2-}\cong\mathscr{T}^+_{3+}$.  Here $w=e_0 e_1 e_2 e_3 e_4$ and $P=
a_{2,1} a_{3,2}^2 - a_{4,3} a_1 a_{3,2}  +a_{6,5}a_1^2=0$.\label{Table.Fplus2minus}}
\end{table}

\subsection{Resolutions $\mathscr{T}^+_{3\pm}$}

The partial resolution $\mathscr{T}^+_3$ can be written as follows:
\begin{equation}
\mathscr{T}^+_3 :
\begin{cases}
& e_2  (y^2 +a_{3,2}e_0^2e_1 y -a_{6,5}e_0^5 e_1^3 e_2e_3) = x r,\\
& e_1x^2 + a_{2,1}e_0e_1 x +a_{4,3}e_0^3 e_1^2 e_2e_3- a_1 y = e_3r,
\end{cases}
\end{equation}
with projective coordinates
\begin{align}
\,[e_3e_2e_1 x :e_3^2 e_2^2 e_1 y :z]\quad[x:e_3e_2 y :e_0] \quad [y:e_1] \quad[r:e_2].
\end{align}

The codimension one fiber for $\mathscr{T}^+_3$ is
\begin{align}
\begin{split}
C_0:~&e_0=e_2y^2 -xr = e_1x^2 -a_1 y-e_3r=0,\\
C_1:~&e_1 =  e_2 y^2 - r x=  a_1 y +e_3r=0,\\
C_r:~& e_3 = e_1x( x+a_{2,1} e_0)  - a_1 y = 
e_2 y(  y  + a_{3,2} e_0^2 e_1 )  - xr=0 , \\ 
C_2:~& e_2 = x = a_1 y+e_3r=0.\\
\end{split}
\end{align}
$C_r$ in fact splits into two. First notice that over $r\neq 0$, we can write 
\begin{align}
x = {1\over r  } \, e_2 y(y +a_{3,2} e_0^2e_1).
\end{align}
Substitute this into the other equation in $C_r$ we have
\begin{align}
\begin{split}
 e_1 x^2 + a_{2,1} e_0e_1 x -a_1y 
&= {1\over r^2} y\left[
e_1 e_2^2 y (y+a_{3,2} e_0^2e_1)^2 + a_{2,1} e_0 e_1 e_2 r (y+a_{3,2} e_0^2e_1) - a_1r^2
\right]\\
&=: {1\over r^2 } \, y u,
\end{split}
\end{align}
where we have defined
\begin{align}\label{u}
u:= 
e_1 e_2^2 y (y+a_{3,2} e_0^2e_1)^2 + a_{2,1} e_0 e_1 e_2 r (y+a_{3,2} e_0^2e_1) - a_1r^2.
\end{align}
It follows that over $r\neq0$, $C_r$ splits into two components:
\begin{align}
\begin{split}
C_r\rightarrow \,&C_3:~e_3= y= x=0,\\
&C_4 :~e_3=  u =v= e_1( x^2 +a_{2,1}e_0 x) -a_1y=0,\\
\end{split}
\end{align}
where we have defined
\begin{align}\label{v}
v:= e_2 y(  y  + a_{3,2} e_0^2 e_1 )  - xr.
\end{align}

The partial resolution is singular at the intersection between $C_3$ and $C_4$. It is then clear that we should blow up $C_3$ and $C_4$ to obtain the final resolutions $\mathscr{T}^+_{3+}$ and $\mathscr{T}^+_{3-}$:

\begin{tikzcd}[column sep=2.5cm]
& & & & \mathscr{T}^+_{3+} 
\\
\mathscr{E}_0 \arrow[leftarrow]{r}{ \textstyle{( x,y,e_0 |e_1 )}} & 
\mathscr{E}_1 \arrow[leftarrow]{r}{ \textstyle{( y,e_1 |e_2)}} & 
\mathscr{T}^+ \arrow[leftarrow]{r}[above]{ \textstyle{( e_2,r |e_3)}}  & 
\mathscr{T}^+_3 \arrow[leftarrow]{ru}[sloped,midway, above= .1 cm]{ \textstyle{(x,y,e_3 |e_4)}}  \arrow[leftarrow]{rd}[sloped, near start]{ \textstyle{(u,v, e_3 |e_4)}}  & \\
& & & & \mathscr{T}^+_{3-}
\end{tikzcd}

Let us start with the easier one, $\mathscr{T}^+_{3+}$:
\begin{equation}
\mathscr{T}^+_{3+}: \   
\begin{cases}
e_2 (e_4  y^2 + a_{3,2}  e_1 e_0^2  y  - a_{6,5} e_0^5 e_1^3e_2 e_3  )=x r,\\
e_1 e_4 x^2 + a_{2,1} e_1 e_0  x + a_{4,3} e_0^3 e_1^2 e_2  e_3 -a_1 y =e_3 r ,
\end{cases}
\end{equation}
with projective coordinates 
\begin{equation}
[e_1 e_2 e_3 e_4^2 x:e_1 e_2^2 e_3^2 e_4^2 y:z]  \  [e_4 x:e_2 e_3 e_4^2 y:e_0]\   [e_4y:e_1]\  [ r:e_2]\ [x:y: e_3].
\end{equation}
We immediately notice that $\mathscr{T}^+_{3+}$ shares the same defining equation as that for $\mathscr{T}^+_{2-}$ \eqref{eqT+2-}. The ambient spaces and SR ideals can also be shown to be the same following a similar calculation in Appendix B.4.1 of \cite{ESY}. Since $\mathscr{T}^+_{2-}$ is related to $\mathscr{T}^+_{2+}$, which is identified as $\mathscr{B}_{1,3}$ as the type I resolution, by a flop, we will denote it by $\mathscr{B}_{1,3}^1$. We therefore conclude
\begin{align}
\mathscr{B}^1_{1,3}:=\mathscr{T}^+_{3+}\cong\mathscr{T}^+_{2-}.
\end{align}

Moving on to the other resolution $\mathscr{T}^+_{3-}$, we are going to do the blow up in the $r\neq 0$ patch. First let us rewrite $\mathscr{T}^+_3$ in this patch as
\begin{align}
\mathscr{T}^+_3:
\begin{cases}
& v -a_{6,5}e_0^5 e_1^3e_2e_3=0,\\
& e_1x^2 + a_{2,1}e_0e_1 x +a_{4,3}e_0^3 e_1^2 e_2e_3- a_1 y = e_3r,\\
& uy =\\
&
e_3\left[  {2} a_{6,5} e_0^5 e_1^4 e_2^3  y(y+a_{3,2}e_0^2e_1) 
- a_{6,5}^2 e_0^{10} e_1^7 e_2^4 e_3
+{ r} a_{2,1}a_{6,5} e_0^6 e_1^4e_2^2 
-a_{4,3} e_0^3 e_1^2e_2r^2 +r^3\right]=0,
\end{cases}
\end{align}
where we have used $x = {1\over r} e_2 y(y +a_{3,2} e_0^2 e_1) - {1\over r} a_{6,5} e_0^5 e_1^3 e_2^2 e_3$. Next, we blow up the ideal $(u,v,e_3)$ to obtain $\mathscr{T}^+_{3-}$:
\begin{align}
\mathscr{T}^+_{3-} :~
\begin{cases}
 & v -a_{6,5}e_0^5 e_1^3e_2e_3=0.\\
& e_1x^2 + a_{2,1}e_0e_1 x +a_{4,3}e_0^3 e_1^2 e_2e_3e_4- a_1 y = e_3e_4r,\\
& uy =\\
&~
e_3\left[  {2} a_{6,5} e_0^5 e_1^4 e_2^3  y(y+a_{3,2}e_0^2e_1) 
- a_{6,5}^2 e_0^{10} e_1^7 e_2^4 e_3e_4
+{ r} a_{2,1}a_{6,5} e_0^6 e_1^4e_2^2 
-a_{4,3} e_0^3 e_1^2e_2r^2 +r^3\right]=0,\\
& e_4 u =e_1 e_2^2 y (y+a_{3,2} e_0^2e_1)^2 + a_{2,1} e_0 e_1 e_2 r (y+a_{3,2} e_0^2e_1) - a_1r^2,\\
&e_4 v = e_2 y(  y  + a_{3,2} e_0^2 e_1 )  - xr,
\end{cases}
\end{align}
with the ambient space parametrized by
\begin{align}
\, [ e_4 e_3e_2e_1 x : e_4^2 e_3^2 e_2^2 e_1 y:1]\quad [x:e_4 e_3 e_2 y: e_0] \quad [ y:e_1 ]\quad [r:e_2]\quad[u:v:e_3].
\end{align}
After a straightforward but rather tedious calculation, one obtains the fiber enhancements in Table \ref{Table.B132}. Since $\mathscr{T}^+_{3-}$ is related to $\mathscr{B}_{1,3}^1\cong\mathscr{T}^+_{3+}$ by a flop, we will give it a new name to fit in the network in Figure \ref{network}:
\begin{align}
\mathscr{B}^2_{1,3} := \mathscr{T}^+_{3-}.
\end{align} 

This completes our constructions for the twelve small resolutions of the $SU(5)$ model.

\begin{table}[htb]
\begin{center}
\scalebox{.8}{\begin{tabular}{|c|c|c|c|c|}
\hline
$w=0$& $w=P=0$ &  $w=a_1=0$ & $w=a_1=a_{3,2}=0$ & $w=a_1=a_{2,1}=0$ \\ 
\hline
& 
{\footnotesize $C_{4}\to C_{4a}+C_{4b}$ }
&
\scalebox{.8}{ \begin{tabular}{l}
$C_1\to C_{14}$\\
$C_2\to C_{24}$\\
$C_{4}\to C_{14}+C_{24}+C_{4}'+C_4''$\\
\end{tabular} }
 & 
\scalebox{.8}{ \begin{tabular}{l}
$C_1\to C_{14}$\\
$C_2\to C_{24}$\\
$C_{3}\to C_{34}$\\
$C_{4} \to C_{14}+C_{24}+C_{34}$\\
$\qquad+C_4' +C_{4+}+C_{4-}$
\end{tabular} }
& 
\scalebox{.7}{ \begin{tabular}{l}
$C_1\to C_{14}$\\
$C_2\to C_{24}$\\
$C_{3}\to C_{34}$\\
$C_{4} \to C_{14}+2C_{24}$\\
$\qquad\qquad + C_{34}+2C_4'+C_4''$\\
\end{tabular} }
 \\
\hline
\scalebox{.7}{\begin{tikzpicture}[every node/.style={circle,draw, minimum size= 10 mm}]
\node (C2)  at (18+72*5:1.4cm)  {\tiny $C_{4}$};
\node (C4) at (18+72*4:1.4cm) {\tiny $C_{3}$};
\node (C3) at (18+72*3:1.4cm) {\tiny $C_{2}$};
\node (C1) at (18+72*2:1.4cm) {\tiny $C_{1}$};
\node (C0) at (18+72*1:1.4cm) {\tiny $C_0$};
\draw (C1)--(C3)--(C4)--(C2)--(C0)--(C1);
\end{tikzpicture}}& 
\scalebox{.9}{\begin{tikzpicture}[every node/.style={circle,draw, minimum size= 5 mm}]
\node (C2b)  at (30+60*6:1.4cm)  {\tiny $C_{4b}$};
\node (C2)  at (30+60*5:1.4cm)  {\tiny $C_{4a}$};
\node (C4) at (30+60*4:1.4cm) {\scalebox{.7}{ $C_{3}$}};
\node (C3) at (30+60*3:1.4cm) {\scalebox{.7}{$C_{2}$}};
\node (C1) at (30+60*2:1.4cm) {\tiny $C_{1}$};
\node (C0) at (30+60*1:1.4cm) {\tiny $C_0$};
\draw (C1)--(C3)--(C4)--(C2)--(C2b)--(C0)--(C1);
\end{tikzpicture}}
& 
\scalebox{1}{\begin{tikzpicture}[every node/.style={circle,draw, minimum size= 8 mm}, scale=.8]
\node (C2) at (0,0) { \scalebox{.8}{$2C_{14}$}};
\node (C5) at (90:-1.8cm) { \scalebox{.8}{$2C_{24}$}};
\node (C0) at (90-35:1.8cm) { \scalebox{.8}{$C_{4}'$}} ;
\node (C1) at (90+35:1.8cm) { $C_0$};
\node [yshift=-1.8cm] (C3) at  (-90-35:1.8cm){ \scalebox{.8}{ $C_{3}$}};
\node [yshift=-1.8cm] (C4) at (-90+35:1.8cm) {  $C_{4}''$};
\draw (C2)--(C0);
\draw (C2)--(C1);
\draw (C2)--(C5);
\draw (C5)--(C3);
\draw (C5)--(C4);
\end{tikzpicture}}
& 
\scalebox{1} {\begin{tikzpicture}[every node/.style={circle,draw, minimum size= 4 mm}, scale=.8]
\node (C2) at (0,0) {\tiny $2C_{14}$};
\node (C5) at (90:-1.8cm) {\tiny $2C_{24}$};
\node (C0) at (90-35:1.8cm) {\tiny $C_4'$} ;
\node (C1) at (90+35:1.8cm) {\small $C_0$};
\node (C6) at (90:-2*1.8cm) { \tiny {$2C_{34}$}};
\node [yshift=-1.8*2cm] (C3) at  (-90-35:1.4cm){\tiny $C_{4+}$};
\node [yshift=-1.8*2cm] (C4) at (-90+35:1.4cm) {\tiny $C_{4-}$};
\draw (C2)--(C0);
\draw (C2)--(C1);
\draw (C2)--(C5);
\draw (C6)--(C3);
\draw (C6)--(C4);
\draw (C5)--(C6);
\end{tikzpicture}}
& 
\scalebox{1}{\begin{tikzpicture}[every node/.style={circle,draw, minimum size= 4 mm}, scale=.8]
\node (C2) at (0,0) {\scalebox{.65}{$2C_{14}$}};
\node (C5) at (90:-1.8cm) {\scalebox{.65}{$3C_{24}$}};
\node (C0) at (90:1.8cm) {\tiny $C_0$} ;
\node  (C4) at (90:-3*1.8cm) {\tiny $C_{3}'$};
\node [xshift=-1.4cm] (C3) at  (90:-1.8cm){\tiny $2C_4'$};
\node  (C6) at (90:-2*1.8cm) {\scalebox{.65}{\color{black}  \bf $2C_{34}$}};
\draw (C2)--(C0);
\draw (C2)--(C5)--(C6)--(C4);
\draw (C5)--(C3);
\end{tikzpicture}}
\\ 
\hline
\end{tabular}}
\end{center}
\caption{Fibers of the partial resolution $\mathscr{B}_{1,3}^2:=\mathscr{T}^+_{3-}$.  Here $w=e_0 e_1 e_2 e_3 $ and  $P=
a_{2,1} a_{3,2}^2 - a_{4,3} a_1 a_{3,2}  +a_{6,5}a_1^2=0$. }\label{Table.B132}
\end{table}

\section{Isomorphisms}\label{section:iso}

In the network of resolutions in Figure \ref{network}, seemingly different sequences of blow ups can result in the same resolution. By that we mean the composition of the sequences of blow ups are the same, even if each individual step might be different. That is to say, a given resolution can admit more than one ``history" of blow ups when we trace back along the network in Figure \ref{network} to the original Weierstrass model $\mathscr{E}_0$.
 It is only after identifying isomorphic resolutions that the network can match with the Coulomb branch.

In this section we summarize all the isomorphisms in the network. The explicit calculation is similar to the $SU(4)$ model and we refer the readers to Appendix B.4.1 of \cite{ESY} for more details. As described in \cite{ESY}, we identify two resolutions if they share the same defining equation, the same scalings for each variables, and the same SR ideal.

The  partial resolutions $\mathscr{T}^+_1$ and $\mathscr{B}_1$ are isomorphic to each other,
\begin{align}\label{T=B}
\mathscr{T}^+_1\cong \mathscr{B}^+,
\end{align}
after relabeling $e_2$ with $e_3$ and vice versa. 
 This is similar to the isomorphism between $\mathscr{B}^+$ and $\mathscr{T}^+$ in the $SU(4)$ model \cite{ESY}.

We immediately have the more isomorphisms between resolutions inherited from \eqref{T=B}. 
Recall that $\mathscr{T}^+_{1-}$ is obtained  from $\mathscr{T}^+_1$ by blowing up $(s,e_3)$ while $\mathscr{B}_{1,2}$ is obtained  from $\mathscr{B}_1$ by blowing up $(s,e_2)$. 
In the same way, $\mathscr{T}^+_{1+}$ is obtained from   $\mathscr{T}^+_1$  by blowing up $(y,e_3)$ while blowing up $(y,e_2)$ from $\mathscr{B}_1$ would give $\mathscr{B}_{1,3}$. 
It follows that $\mathscr{T}^+_{1+}$ is isomorphic to $\mathscr{B}_{1,3}$ and that  $\mathscr{T}^+_{1-}$ is isomorphic to $\mathscr{B}_{1,2}$. One can further show that $\mathscr{T}^+_{2+}$ is isomorphic to $\mathscr{T}^+_{1+}$. We conclude that
 \begin{align}
 &\mathscr{T}^+_{1+}\cong \mathscr{B}_{1,3} \cong \mathscr{T}^+_{2+},~~~\\
 &\mathscr{T}^+_{1-}\cong \mathscr{B}_{1,2}.
 \end{align} 
 Note that the resolution in the first line is the toric type I studied in the previous sections. Their isomorphisms can be most easily seen from the toric diagrams in Figure \ref{typeI:fan1}, \ref{typeI:fan2}, and \ref{typeI:fan3}.
 
Similarly one can show that
\begin{align}
\mathscr{T}^+_{3+} \cong \mathscr{T}^+_{2-}.
\end{align}
We therefore assign the same notation $\mathscr{B}_{1,3}^1$ for both of them in Figure \ref{network}. 

Finally, we also have the isomorphisms between their counterparts under the Mordell-Weil involution.

\newpage
\section{Discussion and conclusion}

\begin{figure}
\includegraphics[scale=1]{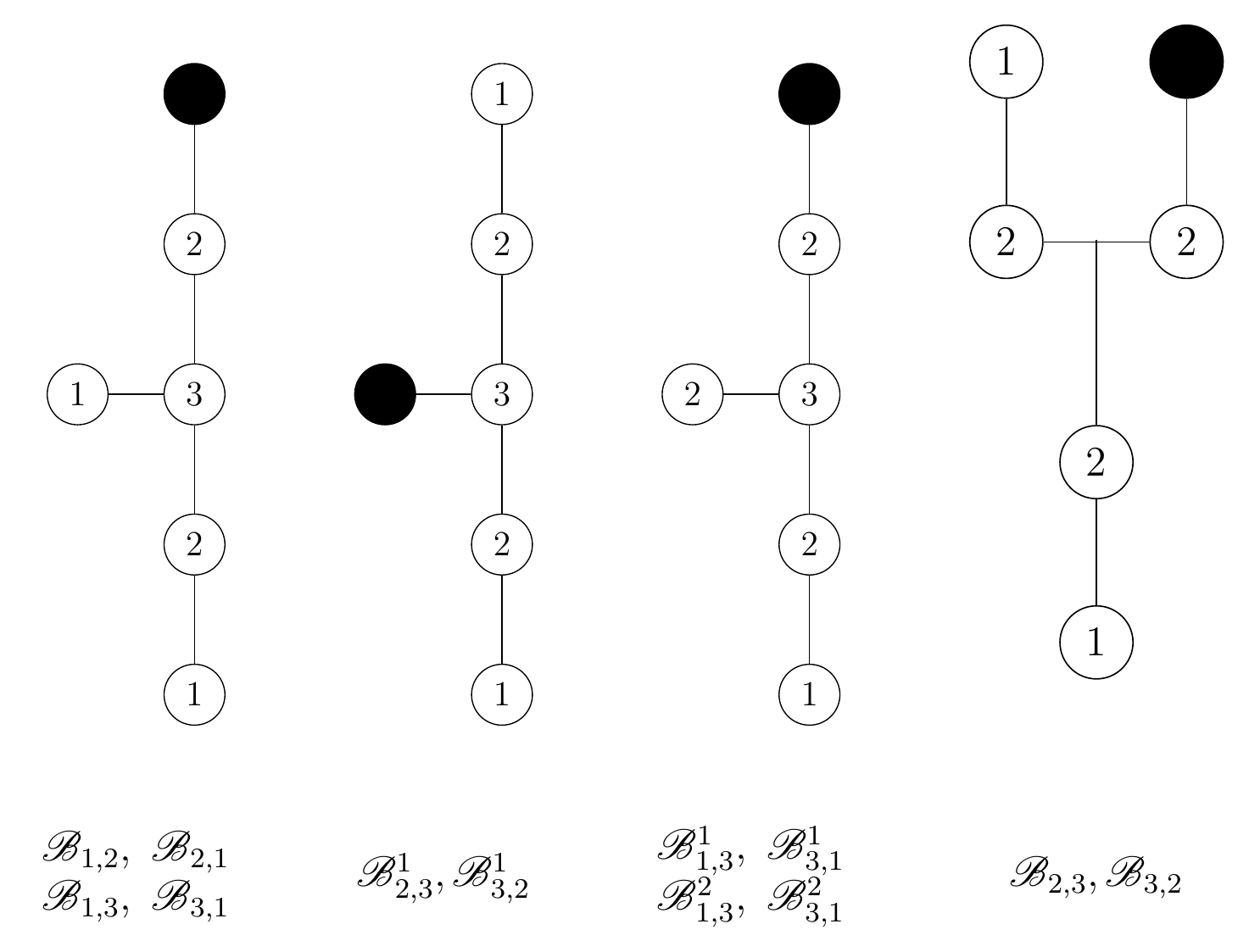}
\caption{The fibers over the codimension three locus  $w=a_1=a_{2,1}=0$ for all twelve resolutions. The numbers of the nodes are the multiplicities and the black node stands for the $e_0=0$ node that passes through the section $z=0$. Here $w=e_0e_1e_2e_3e_4=0$.}\label{fig:codim3fibers}
\end{figure}

\begin{itemize}

\item We present twelve  resolutions for the Tate form with general coefficients $a_{i,j}$ of the I$_5^s$ type over a base $B$ of complex dimension two or three. Ten of them can be obtained by sequences of blow ups summarized in Figure \ref{network}, while another two resolutions $\mathscr{B}_{2,3}^1$, $\mathscr{B}_{3,2}^1$ are obtained by sequences of  weighted blow ups.

\item
Six of the twelve resolutions are $\mathscr{B}_{i,j}$ with $(i\neq j)$ obtained in \cite{EY}, corresponding to the hexagon in Figure \ref{SU5Res}. The new resolutions $\mathscr{B}_{1,3}^1$, $\mathscr{B}_{1,3}^2$ and their Mordell-Weil duals $\mathscr{B}_{3,1}^1$ and $\mathscr{B}_{3,1}^2$ can also be obtained from the network in Figure \ref{network}. 
Finally, the resolution  $\mathscr{B}_{2,3}^1$ and its Mordell-Weil dual $\mathscr{B}_{3,2}^1$ can be obtained by sequences of weighted blow ups.

\item The twelve resolutions correspond to the twelve subchambers in the $SU(5)$ Coulomb branch with fundamental $\bf5$ and antisymmetric representations $\bf10$. We have also identified the partial resolutions in the network in Figure \ref{network} as interior walls, planes, lines, and the origin on the Coulomb branch coming from the intersections of interior walls. It would be interesting to find all the partial resolutions corresponding to the full incidence geometry $(A_4,{\bf 5\oplus 10})$ in all codimensions.

\item We should emphasize that the correspondence between the network with the incidence geometry holds for Weierstrass models on general bases $B$ of complex dimension two or three, regardless of the Calabi-Yau condition. In this sense it is more general than the string/M-theory context.

\item All twelve resolutions share the same fibers in all codimensional loci (but not necessarily the same splitting for each node), except for the codimension three locus $w=a_1=a_{2,1}=0$. The (non-Kodaira) fibers over the codimension three locus $w=a_1=a_{2,1}=0$ for all twelve resolutions are summarized in Figure \ref{fig:codim3fibers}.

\item The network of resolutions in Figure \ref{network} provides a unified framework for all the known resolutions of the model. 
In particular, we have shown that the toric resolution  of type I, type II, and type III are identified as  $\mathscr{B}_{1,3}\cong\mathscr{T}^+_{1+}\cong\mathscr{T}^+_{2+}$,  $\mathscr{B}_{2,3}$, and $\mathscr{B}_{2,3}^1$ in our network, respectively.

\item One important aspect of this work is that all the resolutions we obtained are manifestly projective varieties as they are  constructed by sequences of blow ups or projective flops (in the case of 
 $\mathscr{B}_{2,3}^1$ and  $\mathscr{B}_{3,2}^1$). 
Since they are projective crepant resolutions of the same space, they all share many common topological invariants. 
For example, in the case of elliptic  Calabi-Yau varieties, they share the same  Hodge diamond by a famous theorem of Batyrev on projective crepant resolutions of Calabi-Yau spaces \cite{Batyrev}.

\item It would be interesting to consider similar results for more general groups and for enhancements that are not necessary of rank one. 
In such cases it is not clear if the geometry will match the description of the Coulomb branch.

\end{itemize}

\textbf{Acknowledgments}

We are grateful to Paolo Aluffi, David Morrison,  Patrick Jefferson, Sakura Sch\"afer-Nameki, and Washington Taylor for  discussions. 
ME and SHS are grateful to Princeton University and the Institute of Advanced Studies for their hospitality during the final stage of this work.

%

\end{document}